\documentclass[10pt]{amsart}
\usepackage{latexsym,amssymb,amsmath,amscd,amsthm}
\numberwithin{equation}{section}
\theoremstyle{remark}
\newtheorem{theorem}{{\bf THEOREM}}[section]

\newtheorem{corollary}{{\bf COROLLARY}}[section]

\newtheorem{proposition}{{\bf PROPOSITION}}[section]
\newcommand{\bq}{\begin{equation}}
\newcommand{\bea}{\begin{array}}
\newcommand{\eea}{\end{array}}

\newcommand{\ga}{\alpha}
\newcommand{\gep}{\epsilon}
\newcommand{\gD}{\Delta}
\newcommand{\gl}{\lambda}
\newcommand{\gL}{\Lambda}
\newcommand{\gb}{\beta}
\newcommand{\ot}{\otimes}
\newcommand{\mf}{\mathfrak}
\newcommand{\mc}{\mathcal}

\newcommand{\dg}{\dagger}
\newcommand{\wg}{\wedge}

\newcommand{\ci}{\circ}
\newcommand{\ul}[1]{\underline{#1}}
\newcommand{\ol}[1]{\overline{#1}}

\newcommand{\go}{\omega}
\newcommand{\gO}{\Omega}
\newcommand{\gG}{\Gamma}
\newcommand{\gt}{\theta}
\newcommand{\gs}{\sigma}

\newcommand{\gag}{\gamma}
\newcommand{\gd}{\delta}
\newcommand{\pp}{\partial}

\newcommand{\olra}{\overleftrightarrow}

\newcommand{\tl}{\tilde}
\newcommand{\na}{\nabla}
\newcommand{\gk}{\kappa}

\newcommand{\bs}{\blacksquare}

\newcommand{\gT}{\Theta}

\newcommand{\gS}{\Sigma}

\newcommand{{\DDD}}{D\!\!\!\!\!\!-}

\newcommand{\bx}{\Box}
\newcommand{\mpt}{\mapsto}

\title{FLUCTUATIONS, GRAVITY, AND THE QUANTUM POTENTIAL}
\author{Robert Carroll\\University of Illinois, Urbana, IL 61801}

\date{January, 2005\thanks{email: rcarroll@math.uiuc.edu}}

\begin{document}

\bibliographystyle{plain}

\begin{abstract}
We show how the quantum potential arises in various ways and trace its connection
to quantum fluctuations and Fisher information along with its realization in terms
of Weyl curvature.  It represents a genuine quantization factor for certain
classical systems as well as an expression for quantum matter in gravity theories
of Weyl-Dirac type.  Many of the facts and examples are extracted from the
literature (with references cited) and we mainly provide connections and
interpretation, with a few new observations.  We deliberately avoid ontological and
epistemological discussion and resort to a collection of contexts where the
quantum potential plays a visibly significant role.  In particular we sketch some recent
results of F. and A. Shojai on Dirac-Weyl action and Bohmian mechanics which connects
quantum mass to the Weyl geometry.  
Connections \`a la Santamato of the
quantum potential with Weyl curvature arising from a stochastic geometry, are also indicated
for the Schr\"odinger equation (SE) and Klein-Gordon (KG) equation.  Quantum fluctuations
and quantum geometry are linked with the quantum potential via Fisher information.
Derivations of SE and KG from Nottale's scale relativity are sketched along with a variety of
approaches to the KG equation.  Finally
connections of geometry and mass generation via Weyl-Dirac geometry
with many cosmological implications are indicated, following M. Israelit and N. Rosen.
\end{abstract}

\maketitle

\tableofcontents

\section{THE SCHR\"ODINGER EQUATION}
\renewcommand{\theequation}{1.\arabic{equation}}
\setcounter{equation}{0}

The quantum potential seems to have achieved prominence via the work of L.
deBroglie and D. Bohm plus many others on what is often now called Bohmian
mechanics.  There have been many significant contributions here and we refer to
\cite{c1,c2,c3,c4,ch,c10,c12,cqm} for a reasonably complete list of references.
A good picture of the current theory can be obtained from the papers by 
an American-German-Italian (AGI) group of Allori, Barut, Berndl, Daumer, D\"urr,
Georgi, Goldstein, Lebowitz, Teufel, Tumulka, and Zanghi (cf.
\cite{al,ar,b3,bl,bd,bn,be,bbl,dm,d4,d5,d42,d15,dl,dr,dd,g1,g50,g1,gs,gt,
gd,gll,te,tt,t16}).  We refer also to Holland \cite{h99,h98,hd,hl}, Nikoli\'c 
\cite{n57,n61,n62,nk,ni}, Floyd \cite{f7,f8}, and Bertoldi, Faraggi, and Matone
\cite{b9,f2,f3} for other approaches and summaries.  Other specific references
will arise as we go along but we emphasize with apologies that there are many more
interesting papers omitted here which hopefully are covered in \cite{cqm}.
\\[3mm]\indent
First in a simple minded way one can look at a Schr\"odinger equation (SE)
\bq\label{1}
-\frac{\hbar^2}{2m}\psi''+V\psi=i\hbar\psi_t;\,\,\psi=Re^{iS/\hbar}
\end{equation}
leading to ($'\sim\pp_x$)
\bq\label{2}
S_t+\frac{S_x^2}{2m}+V-\frac{\hbar^2R''}{R}=0;\,\,\pp_t(R^2)+\frac{1}{m}(R^2S')'=0
\end{equation}
\bq\label{3}
P=R^2\sim |\psi|^2;\,\,Q=-\frac{\hbar^2}{2m}\frac{R''}{R}\Rightarrow
S_t+\frac{(S')^2}{2m}+Q+V=0;\,\,P_t+\frac{1}{m}(PS')'=0
\end{equation}
Here Q is the quantum potential and in 3-dimensions for example one expresses this
as $Q=-(\hbar^2/2m)(\gD\sqrt{\rho})/\sqrt{\rho}$ ($R=\sqrt{P},\,\,P\sim\rho$).
\\[3mm]\indent
In a hydrodynamic mode one can write (1-dimension for simplicity and with the
proviso that $S\ne const.$)
$p=S'=m\dot{q}=mv$
($v$ a velocity or collective velocity) and $\rho=mP$ ($\rho$ an unspecified mass
density) to obtain an Euler type hydrodynamic equation ($\pp\sim\pp_x$)
\bq\label{4}
\pp_t(\rho v)+\pp(\rho v^2)+\frac{\rho}{m}\pp V+\frac{\rho}{m}\pp Q=0
\end{equation}
\indent
{\bf REMARK 1.1.}
Given a wave function $\psi$ with $|\psi|^2$ representing a probability density as
in conventional quantum mechanics (QM) it is not unrealistic to imagine an
ensemble picture emerging here (as a ``cloud" of particles for example).  This will
be analogous to diffusion or fluid flow of course but can also be modeled on a
Bohmian particle picture and this will be discussed later in more detail.  We note
also that Q appears in the Hamilton-Jacobi (HJ) type equation \eqref{3} but is not
present in the SE \eqref{1}.  If one were to interpret $\pp V$ as a hydrodynamical
pressure term $-(1/\rho)\pp{\mf P}$ then the SE would be unchanged and the
hydrodynamical equation (with no Q term) would be meaningful in the form 
\bq\label{5}
\pp_t(\rho v)+\pp(\rho v^2)=\frac{1}{m}\pp{\mf P}
\end{equation}
Thinking of Q as a quantization of \eqref{5} yielding \eqref{4} leads then to the
SE \eqref{1}.$\hfill\bs$
\\[3mm]\indent
{\bf REMARK 1.2.}
The development of the AGI school involves now
\bq\label{6}
\dot{q}=v=\frac{\hbar}{m}\Im\frac{\psi^*\psi'}{|\psi|^2}
\end{equation}
and this is derived as the simplest Galilean and time reversal invariant form for
velocity transforming correctly under velocity boosts.  This is a nice argument
and seems to avoid any recourse to Floydian time (cf. \cite{c12,cqm}).$\hfill\bs$
\\[3mm]\indent
Next we consider relations of diffusion to QM following Nagasawa, Nelson, et al
(cf. \cite{n9,n15,n6} - see also e.g. \cite{c15,d99,g10,k8,k9,k93}) and
sketch some formulas for a simple
Euclidean metric where
$\gD=\sum(\pp/\pp x^i)^2$. Then 
$\psi(t,x)=exp[R(t,x)+iS(t,x)]$ satisfies a SE
$i\pp_t\psi+(1/2)\gD\psi+ia(t,x)\cdot\na\psi-V(t,x)\psi=0$ 
($\hbar=m=1$)) if and only if
\bq\label{7}
V=-\frac{\pp S}{\pp t}+\frac{1}{2}\gD R+\frac{1}{2}(\na R)^2-\frac{1}{2}(\na
S)^2-a\cdot\na S;
\end{equation}
$$0=\frac{\pp R}{\pp t}+\frac{1}{2}\gD S+(\na S)\cdot(\na R)+a\cdot\na R$$
in the region $D=\{(s,x):\,\psi(s,x)\ne 0\}$.
Solutions are often referred to as weak or
distributional but we do not belabor this point.  From \cite{n9} there results
\begin{theorem}
Let $\psi(t,x)=exp[R(t,x)+iS(t,x)]$ be a solution of the SE above; then 
$\phi(t,x)=exp[R(t,x)+S(t,x)]$ and $\hat{\phi}=exp[R(t,x)-S(t,x)]$ are
solutions of 
\bq\label{8}
\frac{\pp\phi}{\pp t}+\frac{1}{2}\gD\phi+a(t,x)\cdot\na\phi+c(t,x,\phi)\phi=0;
\end{equation}
$$-\frac{\pp\hat{\phi}}{\pp t}+\frac{1}{2}\gD\hat{\phi}-a(t,x)\cdot\na\hat{\phi}
+c(t,x,\phi)\hat{\phi}=0$$
where the creation and annihilation term $c(t,x,\phi)$ is given via
\bq\label{9}
c(t,x,\phi)=-V(t,x)-2\frac{\pp S}{\pp t}(t,x)-(\na S)^2(t,x)-2a\cdot\na S(t,x)
\end{equation}
Conversely given $(\phi,\hat{\phi})$ as above satisfying \eqref{8}
it follows that $\psi$ satisfies the SE with V as in
\eqref{9} (note $R=(1/2)log(\hat{\phi}\phi)$ and
$S=(1/2)log(\phi/\hat{\phi})$ with $exp(R)=(\hat{\phi}\phi)^{1/2}$).$\hfill\bs$
\end{theorem}
\indent
From this one can conclude that nonrelativistic
QM is diffusion theory in terms of Schr\"odinger processes (described by
$(\phi,\hat{\phi}$) - more details later).  Further it is shown that key postulates in 
Nelson's stochastic mechanics or Zambrini's Euclidean  QM (cf. \cite{z3}) can both be
avoided in connecting the SE to diffusion processes (since they are automatically
valid).  
Look now at Theorem 1.1 for one dimension and write $T=\hbar t$ with
$X=(\hbar/\sqrt{m})x$; then some simple calculation leads to
\begin{corollary}
Equation \eqref{8}, written in the $(X,\,T)$ variables becomes
\bq\label{10}
\hbar\phi_T+\frac{\hbar^2}{2m}\phi_{XX}+A\phi_X+\tl{c}\phi=0;\,\,-\hbar\hat{\phi}_T+
\frac{\hbar^2}{2m}\hat{\phi}_{XX}-A\hat{\phi}_X+\tl{c}\hat{\phi}=0;
\end{equation}
$$\tl{c}=-\tl{V}(X,T)-2\hbar S_T-\frac{\hbar^2}{m}S_X^2-2AS_X$$
Thus the diffusion processes pick up factors of $\hbar$ and $\hbar/\sqrt{m}$.$\hfill\bs$
\end{corollary}
\indent
Next we sketch a derivation of the SE following scale relativity \`a la Nottale
(cf. \cite{c25,n5,n12,n14,n13,n30} and \cite{c29,c13,c31,c32} for some refinements
and variations); this material is expanded in \cite{c1,cqm}.
\\[3mm]\indent
{\bf REMARK 1.3.}
One considers quantum paths \`a la Feynman so that $lim_{t\to
t'}[X(t)-X(t')]^2/(t-t')$ exists.  This implies $X(t)\in H^{1/2}$ where $H^{\ga}$
means $c\gep^{\ga}\leq |X(t)-X(t')|\leq C\gep^{\ga}$ and from \cite{f15} for example
this means $dim_HX[a,b]=1/2$.  Now one ``knows" (see e.g. \cite{a1}) that quantum and
Brownian motion paths (in the plane) have H-dimension 2 and some clarification is
needed here. 
We refer to
\cite{mb} where there is a paper on Wiener Brownian motion (WBM), random walks,
etc. discussing Hausdorff and other dimensions of various sets.  Thus given
$0<\gl<1/2$ with probability 1 a Browian sample function $X$ satisfies
$|X(t+h)-X(t)|\leq b|h|^{\gl}$ for $|h|\leq h_0$ where $b=b(\gl)$.  This leads to
the result that with probability 1 the graph of a Brownian sample function has
Hausdorff and box dimension
$3/2$.  On the other hand a Browian trail (or path) in 2 dimensions has Hausdorff and
box dimension 2 (note a quantum path can have self intersections, etc.).$\hfill\bs$
\\[3mm]\indent
Now fractal spacetime here will mean some kind of continuous nonsmooth pathspace so that a 
bivelocity structure is defined.  One
defines first 
\bq\label{11}
\frac{d_{+}}{dt}y(t)=lim_{\gD t\to 0_{+}}\left<\frac{y(t+\gD t)-y(t)}{\gD
t}\right>;
\end{equation}
$$\frac{d_{-}}{dt}y(t)=lim_{\gD t\to 0_{+}}\left<\frac{y(t)-y(t-\gD t)}{\gD
t}\right>$$
Applied to the position vector x this yields forward and backward mean velocities,
namely
$(d_{+}/dt)x(t)=b_{+}$ and $(d_{-}/dt)x(t)=b_{-}$.  Here these
velocities are defined as the average at a point q and time t of the respective
velocities of the outgoing and incoming fractal trajectories; in stochastic QM
this corresponds to an average on the quantum state.  The position vector $x(t)$
is thus ``assimilated" to a stochastic process which satisfies respectively
after ($dt>0$) and before ($dt<0$) the instant t a relation 
$dx(t)=b_{+}[x(t)]dt+d\xi_{+}(t)=b_{-}[x(t)]dt+d\xi_{-}(t)$ where $\xi(t)$ is a
Wiener process (cf. \cite{n6}).  It is in the description of $\xi$ that the $D=2$
fractal character of trajectories is inserted; indeed that $\xi$ is a Wiener
process means that the $d\xi$'s are assumed to be Gaussian with mean 0, mutually
independent, and such that
\bq\label{12}
<d\xi_{+i}(t)d\xi_{+j}(t)>=2{\mc
D}\gd_{ij}dt;\,\,<d\xi_{-i}(t)d\xi_{-j}(t)>=-2{\mc D}\gd_{ij}dt
\end{equation}
where $<\,\,>$ denotes averaging (${\mc D}$ is now the diffusion coefficient).
Nelson's postulate (cf. \cite{n6}) is that ${\mc D}=\hbar/2m$ and this has
considerable justification (cf. \cite{n5}).  Note also that \eqref{12} is indeed
a consequence of fractal (Hausdorff) dimension 2 of trajectories follows from
$<d\xi^2>/dt^2=dt^{-1}$, i.e. precisely Feynman's result $<v^2>^{1/2}\sim \gd
t^{-1/2}$.  Note
that Brownian motion (used in Nelson's postulate) is known to be of fractal
(Hausdorff)
dimension 2. Note also that any value of ${\mc D}$ may lead to QM and for ${\mc D}\to 0$
the theory becomes equivalent to the Bohm theory.
Now expand any function $f(x,t)$ in a Taylor series up to order 2, take averages,
and use properties of the Wiener process $\xi$ to get
\bq\label{13}
\frac{d_{+}f}{dt}=(\pp_t+b_{+}\cdot\na+{\mc D}\gD)f;\,\,\frac{d_{-}f}{dt}=(\pp_t
+b_{-}\cdot\na-{\mc D}\gD)f
\end{equation}
Let $\rho(x,t)$ be the probability density of $x(t)$; it is known that for any
Markov (hence Wiener) process one has $\pp_t\rho+div(\rho
b_{+})={\mc D}\gD\rho$ (forward equation) and $\pp_t\rho+div(\rho
b_{-})=-{\mc D}\gD\rho$ (backward equation).  These are called Fokker-Planck
equations and one defines two new average velocities 
$V=(1/2)[b_{+}+b_{-}]$ and $U=(1/2)[b_{+}-b_{-}]$.  Consequently adding and
subtracting one obtains $\rho_t+div(\rho V)=0$ (continuity equation)
and $div(\rho U)-{\mc D}\gD\rho=0$ which is equivalent to
$div[\rho(U-{\mc D}\na log(\rho))]=0$.
One can show, using \eqref{13} that the term in square brackets in the last equation
is zero leading to $U={\mc D}\na log(\rho)$.  Now place oneself in
the $(U,V)$ plane and write ${\mc V}=V-iU$.  Then write
$(d_{{\mc V}}/dt)=(1/2)(d_{+}+d_{-})/dt$ and $(d_{{\mc U}}/dt)=
(1/2)(d_{+}-d_{-})/dt$.    
Combining the equations in \eqref{13} one defines
$(d_{{\mc V}}/dt)=\pp_t+V\cdot\na$ and $(d_{{\mc U}}/dt)={\mc
D}\gD+U\cdot \na$; then define
a complex operator $(d'/dt)=(d_{{\mc V}}/dt)-i(d_{{\mc U}}/dt)$
which becomes 
\bq\label{14}
\frac{d'}{dt}=\left(\frac{\pp}{\pp t}-i{\mc D}\gD\right)+{\mc V}\cdot\na
\end{equation}
\indent
One now postulates that the passage from classical mechanics to a new
nondifferentiable process considered here can be implemented by the unique
prescription of replacing the standard $d/dt$ by $d'/dt$.  Thus
consider ${\mf S}=\left<\int_{t_1}^{t_2}{\mc L}(x,{\mc
V},t)dt\right>$ yielding by least action $(d'/dt)(\pp{\mc
L}/\pp{\mc V}_i)=
\pp{\mc L}/\pp x_i$.  Define then ${\mc P}_i=\pp{\mc L}/\pp{\mc V}_i$ leading to
${\mc P}=\na {\mf S}$ (recall the classical action principle with $dS=pdq-Hdt$).
Now for Newtonian mechanics write $L(x,v,t)=(1/2)mv^2-{\bf U}$ which becomes ${\mc L}(x,{\mc
V},t)=(1/2)m{\mc V}^2-{\mf U}$ leading to $-\na{\mf U}=m(d'/dt){\mc
V}$. One separates real and imaginary parts of the complex acceleration
$\gag=(d'{\mc V}/dt$ to get
\bq\label{15}
d'{\mc V}=(d_{{\mc V}}-id_{{\mc U}})(V-iU)=(d_{{\mc V}}V-d_{{\mc U}}U)-
i(d_{{\mc U}}V+d_{{\mc V}}U)
\end{equation}
The force $F=-\na{\mf U}$ is real so the imaginary part of the complex
acceleration vanishes; hence
\bq\label{16}
\frac{d_{{\mc U}}}{dt}V+\frac{d_{{\mc V}}}{dt}U=\frac{\pp U}{\pp t}+U\cdot\na
V+V\cdot\na U+{\mc D}\gD V=0
\end{equation}
from which $\pp U/\pp t$ may be obtained.  
This is a weak point in the derivation since one has to assume e.g. that $U(x,t)$ has certain
smoothness properties.  Now considerable calculation leads to
the SE $i\hbar\psi_t=-(\hbar^2/2m)\gD\psi
+{\mf U}\psi$ and this suggests an interpretation of QM as mechanics in a
nondifferentiable (fractal) space. 
In fact (using one space dimension for convenience) we see that if ${\mf U}=0$ then
the free motion
$m(d'/dt){\mc V}=0$ yields the SE $i\hbar\psi_t=-(\hbar^2/2m)\psi_{xx}$ as a geodesic
equation in ``fractal" space.  Further from $U=(\hbar/m)(\pp\sqrt{\rho}/\sqrt{\rho})$
and $Q=-(\hbar^2/2m)(\gD\sqrt{\rho}/\sqrt{\rho})$ one arrives at a lovely relation,
namely
\begin{proposition}
The quantum potential Q can be written in the form $Q=-(m/2)U^2-(\hbar/2)\pp U$. 
Hence the quantum potential arises directly from the fractal nonsmooth nature of the
quantum paths.  Since Q can be thought of as a quantization of a classical motion
we see that the quantization corresponds exactly to the existence of nonsmooth
paths.  Consequently smooth paths imply no quantum mechanics.
\end{proposition}
\indent
{\bf REMARK 1.4.}
In \cite{a16} one writes again $\psi=Rexp(iS/\hbar)$ with field
equations in the hydrodynamical picture (1-D for convenience)
\bq\label{17}
d_t(m_0\rho v)=\pp_t(m_0\rho v)+\na(m_0\rho v)=-\rho\na(u+Q);\,\,\pp_t\rho+\na\cdot(\rho v)=0
\end{equation}
where $Q=-(\hbar^2/2m_0)(\gD\sqrt{\rho}/\sqrt{\rho})$. The Nottale approach is used
as above with $d_v\sim d_{{\mc V}}$ and $d_u\sim d_{{\mc U}}$.  One assumes
that the velocity field from the hydrodynamical model agrees with the real part $v$ of the
complex velocity $V=v-iu$ so $v=(1/m_0)\na s\sim 2{\mc D}\pp s$ and $u=-(1/m_0)
\na\gs\sim {\mc D}\pp log(\rho)$ where ${\mc D}=\hbar/2m_0$.  In this context the quantum potential
$Q=-(\hbar^2/2m_0)\gD{\mc D}\sqrt{\rho}/\sqrt{\rho}$ becomes 
\bq\label{18}
Q=-m_0{\mc D}\na\cdot
u-(1/2)m_0u^2\sim -(\hbar/2)\pp u-(1/2)m_0u^2
\end{equation}
Consequently Q arises from the fractal
derivative and the nondifferentiability of spacetime again, as in Proposition 1.1.  Further
one can relate
$u$ (and hence Q) to an internal stress tensor whereas the $v$ equations correspond
to systems of Navier-Stokes type.
\\[3mm]\indent
{\bf REMARK 1.5.}
We note that it is the presence of $\pm$ derivatives that makes possible the introduction
of a complex plane to describe velocities and hence QM; one can think of this as the
motivation for a complex valued wave function and the nature of the SE.  
$\hfill\bs$
\\[3mm]\indent
{\bf REMARK 1.6.}
In \cite{c29} one extends ideas
of Nottale and Ord (cf. \cite{o2,o5,o6,o7}) in order to derive an interesting nonlinear
Schr\"odinger equation (NLSE) using a complex diffusion coefficient and a hydrodynamic
model.

\subsection{THE SCHR\"ODINGER EQUATION IN WEYL SPACE}

We go now to Santamato \cite{sa} and derive the SE from classical mechanics in Weyl space
(i.e. from Weyl geometry - cf. also \cite{au,c3,c4,c16,i2,sb,w3}).
The idea is to relate the quantum force (arising from the quantum potential) to geometrical
properties of spacetime; the Klein-Gordon (KG) equation is also treated in this spirit
in \cite{c16,sb}. 
One wants to show how geometry acts as a guidance field for matter (as in general relativity).
Initial positions are assumed random (as in the Madelung approach) and thus the theory is 
statistical and is really
describing the motion of an ensemble.  Thus assume that the particle motion is given by some
random process $q^i(t,\go)$ in a manifold M (where $\go$ is the sample space tag) whose
probability density $\rho(q,t)$ exists and is properly normalizable.  Assume that the process
$q^i(t,\go)$ is the solution of differential equations 
\bq\label{19}
\dot{q}^i(t,\go)=
(dq^i/dt)(t,\go)=v^i(q(t,\go),t)
\end{equation}
with random initial conditions $q^i(t_0,\go)=q_0^i(\go)$.
Once the joint distribution of the random variables $q_0^i(\go)$ is given the process
$q^i(t,\go)$ is uniquely determined by \eqref{19}.  One knows that in this situation 
$\pp_t\rho+\pp_i(\rho v^i)=0$ (continuity equation) with initial Cauchy data
$\rho(q,t)=\rho_0(q)$. The natural origin of $v^i$ arises via a least action principle based
on a Lagrangian
$L(q,\dot{q},t)$ with
\bq\label{20}
L^*(q,\dot{q},t)=L(q,\dot{q},t)-\Phi(q,\dot{q},t);\,\,\Phi=\frac{dS}{dt}=\pp_tS+\dot{q}^i\pp_iS
\end{equation}
Then $v^i(q,t)$ arises by minimizing 
\bq\label{21}
I(t_0,t_1)=E[\int_{t_0}^{t_1}L^*(q(t,\go),\dot{q}(t,\go),t)dt]
\end{equation}
where
$t_0,\,t_1$ are arbitrary and E denotes the expectation (cf. \cite{c1,c2,n9,n15,n6} for
stochastic ideas).  The minimum is to be achieved over the class of all
random motions $q^i(t,\go)$ obeying \eqref{20} with arbitrarily varied velocity field
$v^i(q,t)$ but having common initial values.  One proves first
\bq\label{22}
\pp_tS+H(q,\na S,t)=0;\,\,v^i(q,t)=\frac{\pp H}{\pp p_i}(q,\na S(q,t),t)
\end{equation}
Thus the value of I in \eqref{21} along the random curve $q^i(t,q_0(\go))$ is
\bq\label{23}
I(t_1,t_0,\go)=\int_{t_0}^{t_1}L^*(q(,q_0(\go)),\dot{q}(t,q_0(\go)),t)dt
\end{equation}
Let
$\mu(q_0)$ denote the joint probability density of the random variables $q^i_0(\go)$ and then the
expectation value of the random integral is
\bq\label{24}
I(t_1,t_0)=E[I(t_1,t_0,\go)]=\int_{{\bf
R}^n}\int_{t_0}^{t_1}\mu(q_0)L^*(q(t,q_0),\dot{q}(t,q_0),t)d^nq_0dt
\end{equation}
Standard variational methods give then
\bq\label{25}
\gd I=\int_{{\bf R}^n}d^nq_0\mu(_0)\left[\frac{\pp L^*}{\pp\dot{q}^i}(q(t_1,q_0),\pp_tq(t_1,q_0),
t)\gd q^i(t_1,q_0)-\right.
\end{equation}
$$-\left.\int_{t_0}^{t_1}dt\left(\frac{\pp}{\pp t}\frac{\pp
L^*}{\pp\dot{q}^i}(q(t,q_0),\pp_tq)t,q_0),t)-\frac{\pp L^*}{\pp
q^i}(q(t,q_0),\pp_tq(t,q_0),t)\right)\gd q^i(t,q_0)\right]$$
where one uses the fact that $\mu(q_0)$ is independent of time and $\gd q^i(t_0,q_0)=0$ (recall
common initial data is assumed).  Therefore 
\bq\label{26}
{\bf (A)}\,\,\,(\pp L^*/\pp
\dot{q}^i)(q(t,q_0),\pp_tq(t,q_0),t)=0;
\end{equation}
$${\bf (B)}\,\,\,\frac{\pp}{\pp t}\frac{\pp
L^*}{\pp\dot{q}^i}(q(t,q_0),\pp_tq(t,q_0,t)-\frac{\pp L^*}{\pp q^i}
(q(t,q_0),\pp_tq(t,q_0),t)=0$$
are the necessary conditions for obtaining a minimum of I.  Conditions {\bf (B)} are the usual
Euler-Lagrange (EL) equations whereas {\bf (A)} is a consequence of the fact that in the
most general case one must retain varied motions with $\gd q^i(t_1,q_0)$ different from zero
at the final time $t_1$.  Note that since $L^*$ differs from L by a total time derivative one
can safely replace $L^*$ by L in {\bf (B)} and putting \eqref{20} into {\bf (A)} one
obtains the classical equations 
\bq\label{27}
p_i=(\pp
L/\pp\dot{q}^i)(q(t,q_0),\dot{q}(t,q_0),t)=\pp_i S (q(t,q_0),t)
\end{equation}
It is known now that if
$det[(\pp^2L/\pp\dot{q}^i\pp\dot{q}^j]\ne 0$ then the second equation in
\eqref{22} is a consequence of the gradient condition \eqref{27} and of the definition
of the Hamiltonian function $H(q,p,t)=p_i\dot{q}^i-L$.  Moreover {\bf (B)} in \eqref{26}
and
\eqref{27} entrain the HJ equation in \eqref{63.33}.  In order to show that the average
action integral
\eqref{24} actually gives a minimum one needs $\gd^2I>0$ but this is not necessary for
Lagrangians whose Hamiltonian H has the form
\bq\label{28}
H_C(q,p,t)=\frac{1}{2m}g^{ik}(p_i-A_i)(p_k-A_k)+V
\end{equation}
with arbitrary fields $A_i$ and V (particle of mass m in an EM field A) which is the form for
nonrelativistic applications; given positive definite $g_{ik}$ such Hamiltonians involve
sufficiency conditions $det[\pp^2L/\pp\dot{q}^i\pp\dot{q}^k]=mg>0$.  Finally
{\bf (B)} in \eqref{26} with $L^*$ replaced by L) shows that along particle trajectories
the EL equations are satisfied, i.e. the particle undergoes a classical motion with
probability one.  Notice here that in \eqref{22} no explicit mention of generalized
momenta is made; one is dealing with a random motion entirely based on position.  Moreover
the minimum principle \eqref{21} defines a 1-1 correspondence between solutions $S(q,t)$
in
\eqref{22} and minimizing random motions $q^i(t,\go)$.  Provided $v^i$ is given via
\eqref{22} the particle undergoes a classical motion with probability one.  Thus once the
Lagrangian L or equivalently the Hamiltonian H is given, $\pp_t\rho+\pp_i(\rho v^i)=0$  and
\eqref{22} uniquely determine the stochastic process
$q^i(t,\go)$.  Now suppose that some geometric structure is given on M so that the notion
of scalar curvature $R(q,t)$ of M is meaningful.  Then we assume (ad hoc) that the actual
Lagrangian is 
\bq\label{29}
L(q,\dot{q},t)=L_C(q,\dot{q},t)+\gag(\hbar^2/m)R(q,t)
\end{equation}
where
$\gag=(1/6)(n-2)/ (n-1)$ with $n=dim(M)$.  Since both $L_C$ and R are independent
of $\hbar$ we have $L\to L_C$ as $\hbar\to 0$.
\\[3mm]\indent
Now for a differential manifold with $ds^2=g_{ik}(q)dq^idq^k$ it is standard that in
a transplantation
$q^i\to q^i+\gd q^i$ one has $\gd A^i=\gG^i_{\,k\ell}A^{\ell}dq^k$ with $\gG^i_{\,k\ell}$
general affine connection coefficients on M (Riemannian structure is not assumed).  In
\cite{sa} it is assumed that for $\ell=(g_{ik}A^iA^k)^{1/2}$ one has 
$\gd\ell=\ell\phi_kdq^k$ where the $\phi_k$ are covariant components of an arbitrary
vector (Weyl geometry).  Then the actual affine connections $\gG^i_{k\ell}$ can be found by
comparing this with $\gd\ell^2=\gd(g_{ik}A^iA^k)$ and using $\gd
A^i=\gG^i_{\,k\ell}A^{\ell}dq^k$.  A little linear algebra gives then 
\bq\label{30}
\gG^i_{k\ell}=-\left\{\begin{array}{c}
i\\
k\,\,\ell\end{array}\right\}+g^{im}(g_{mk}\phi_{\ell}+g_{m\ell}\phi_k-g_{k\ell}\phi_m)
\end{equation}
Thus we may prescribe the metric tensor $g_{ik}$ and $\phi_i$ and determine via
\eqref{30} the connection coefficients.  Note that $\gG^i_{\,k\ell}=\gG^i_{\,\ell k}$
and for $\phi_i=0$ one has Riemannian geometry.  Covariant derivatives are defined via
\bq\label{31}
A^k_{,\i}=\pp_iA^k-\gG^{k\ell}A^{\ell};\,\,
A_{k,i}=\pp_iA_k+\gG^{\ell}_{ki}A_{\ell}
\end{equation}
for covariant and contravariant vectors respectively (where $S_{,i}=\pp_iS$).  Note Ricci's
lemma  no longer holds (i.e. $g_{ik,\ell}\ne 0$) so covariant differentiation and operations of
raising or lowering indices do not commute.  The curvature tensor $R^i_{\,k\ell m}$ in Weyl
geometry is introduced via $A^i_{,k,\ell}-A^i_{,\ell,k}=F^i_{mk\ell}A^m$
from which arises the standard formula of Riemannian geometry 
\bq\label{32}
R^i_{mk\ell}=
-\pp_{\ell}\gG^i_{mk}+\pp_k\gG^i_{m\ell}+\gG^i_{n\ell}\gG^n_{mk}-\gG^i_{nk}\gG^n_{m\ell}
\end{equation}
where \eqref{30} is used in place of the Christoffel symbols.  The tensor
$R^i_{\,mk\ell}$ obeys the same symmetry relations as the curvature tensor of Riemann geometry
as well as the  Bianchi identity.  The Ricci symmetric tensor $R_{ik}$ and the scalar
curvature R are defined by the same formulas also, viz. $R_{ik}=R^{\ell}_{\,i\ell k}$ and
$R=g^{ik}R_{ik}$.  For completeness one derives here
\bq\label{33}
R=\dot{R}+(n-1)[(n-2)\phi_i\phi^i-2(1/\sqrt{g})\pp_i(\sqrt{g}\phi^i)]
\end{equation}
where 
$\dot{R}$ is the Riemannian curvature built by the Christoffel symbols.  Thus from
\eqref{30} one obtains
\bq\label{34}
g^{k\ell}\gG^i_{k\ell}=-g^{k\ell}\left\{\begin{array}{c}
i\\
k\,\,\ell\end{array}\right\}-(n-2)\phi^i;\,\,\gG^i_{k\ell}=-\left\{\begin{array}{c}
i\\
k\,\,\ell\end{array}\right\}+n\phi_k
\end{equation}
Since the form of a scalar is independent of the coordinate system used one may compute R in a
geodesic system where the Christoffel symbols and all $\pp_{\ell}g_{ik}$ vanish; 
then \eqref{30}
reduces to $\gG^i_{k\ell}=\phi_k\gk^i_{\ell}+\phi_{\ell}\gd^i_k-g_{k\ell}\phi^i$ and
hence 
\bq\label{35}
R=-g^{km}\pp_m\gG^i_{k\ell}+\pp_i(g^{k\ell}\gG^i_{k\ell})+g^{\ell
m}\gG^i_{n\ell}\gG^n_{mi}-g^{m\ell}\gG^i_{n\ell}\gG^n_{m\ell}
\end{equation}
Further 
one has $g^{\ell m}\gG^i_{n\ell}\gG^n_{mi}=-(n-2)(\phi_k\phi^k)$ at the point in
consideration.  Putting all this in \eqref{35} one arrives at 
\bq\label{36}
R=\dot{R}+(n-1)(n-2)(\phi_k\phi^k)-2(n-1)\pp_k\phi^k
\end{equation}
which becomes \eqref{33} in covariant
form. Now the geometry is to be derived from physical principles so the $\phi_i$ cannot be
arbitrary but must be obtained by the same averaged least action principle \eqref{21}
giving the motion of the particle.  The minimum in \eqref{21} is to be evaluated now with
respect to the class of all Weyl geometries having arbitrarily varied gauge vectors but
fixed metric tensor.  Note that once  \eqref{29} is inserted in \eqref{20} the only
term in
\eqref{21} containing the gauge vector is the curvature term.  Then observing that $\gag>0$
when $n\geq 3$ the minimum principle {\eqref{21} may be reduced to the simpler form 
$E[R(q(t,\go),t)]=min$ where only the gauge vectors $\phi_i$ are varied.  Using 
\eqref{33} this is easily done.  First a little argument shows that
$\hat{\rho}(q,t)=\rho(q,t)/\sqrt{g}$ transforms as a scalar in a coordinate change and this
will be called the scalar probability density of the random motion of the particle
(statistical determination of geometry).  Starting
from $\pp_t\rho+\pp_i(\rho v^i)=0$ a manifestly covariant equation for $\hat{\rho}$ is found to
be
$\pp_t\hat{\rho}+(1/\sqrt{g})\pp_i(\sqrt{g}v^i\hat{\rho})=0$.  Now return to the
minimum problem $E[R(q(t,\go),t)]=min$; from \eqref{33} and $\hat{\rho}=\rho/\sqrt{g}$
one obtains
\bq\label{37}
E[R(q(t,\go),t)]=E[\dot{R}(q(t,\go),t)]+
\end{equation}
$$+(n-1)\int_M[(n-2)\phi_i\phi^i-2(1/\sqrt{g})
\pp_i(\sqrt{g}\phi^i)]\hat{\rho}(q,t)\sqrt{g}d^nq$$
Assuming fields go to 0 rapidly enough on $\pp M$ and integrating by parts one gets then
\bq\label{38}
E[R]=E[\dot{R}]-\frac{n-1}{n-2}E[g^{ik}\pp_i(log(\hat{\rho})\pp_k(log(\hat{\rho})]+
\end{equation}
$$+\frac{n-1}{n-2}E\{g^{ik}[(n-2)\phi_i+\pp_i(log(\hat{\rho})][(n-2)\phi_k+\pp_k(log(\hat{\rho})]\}
$$
Since the first two terms on the right are independent of the gauge vector and $g^{ik}$ is
positive definite $E[R]$ will be a minimum when 
\bq\label{39}
\phi_i(q,t)=-[1/(n-2)]\pp_i[log(\hat{\rho})(q,t)]
\end{equation}
This shows that the
geometric properties of space are indeed affected by the presence of the particle and in turn
the alteration of geometry acts on the particle through the quantum force
$f_i=\gag(\hbar^2/m)\pp_iR$ which according to \eqref{33} depends on the gauge vector and
its derivatives.  It is this peculiar feedback between the geometry of space and the motion of
the particle which produces quantum effects.
\\[3mm]\indent
In this spirit one goes now to a geometrical derivation of the SE.  Thus inserting
\eqref{39} into \eqref{33} one gets
\bq\label{40}
R=\dot{R}+(1/2\gag\sqrt{\hat{\rho}})[1/\sqrt{g})\pp_i(\sqrt{g}g^{ik}\pp_k\sqrt{\hat{\rho}})]
\end{equation}
where the value $(n-2)/6(n-1)$ for $\gag$ is used.  On the other hand the HJ equation 
\eqref{20} can be written as 
\bq\label{41}
\pp_tS+H_C(q,\na S,t)-\gag(\hbar^2/m)R=0
\end{equation}
where \eqref{29} has been used.  When \eqref{40} is introduced into \eqref{41} the
HJ equation and the continuity equation
$\pp_t\hat{\rho}+(1/\sqrt{g})(\sqrt{g}v^i\hat{\rho})=0$, with velocity field given by
\eqref{22}, form a set of two nonlinear PDE which are coupled by the curvature of space. 
Therefore self consistent random motions of the particle (i.e. random motions compatible with
\eqref{35}) are obtained by solving \eqref{41} and the continuity equation
simultaneously.  
For every pair of solutions
$S(q,t,\hat{\rho}(q,t))$ one gets a possible random motion for the particle whose invariant
probability density is $\hat{\rho}$.  The present approach is so different from traditional QM
that a proof of equivalence is needed and this is only done for Hamiltonians of the form 
\eqref{28} (which is not very restrictive).  The HJ equation corresponding to
\eqref{28} is
\bq\label{42}
\pp_tS+\frac{1}{2m}g^{ik}(\pp_iS-A_i)(\pp_kS-A_k)+V-\gag\frac{\hbar^2}{m}R=0
\end{equation}
with R given by \eqref{40}.  Moreover using \eqref{22} as well as \eqref{33} the
continuity equation becomes 
\bq\label{43}
\pp_t\hat{\rho}+(1/m\sqrt{g})\pp_i[\hat{\rho}\sqrt{g}g^{ik}(\pp_kS-A_k)]=0
\end{equation}
Owing to \eqref{40}, \eqref{42} and \eqref{43} form a set of two nonlinear PDE
which must be solved for the unknown functions S and $\hat{\rho}$.  Now a straightforward
calculations shows that, setting 
\bq\label{44}
\psi(q,t)=\sqrt{\hat{\rho}(q,t)}exp](i/\hbar)S(q,t)],
\end{equation}
the quantity
$\psi$ obeys a linear PDE (corrected from \cite{sa})
\bq\label{45}
i\hbar\pp_t\psi=\frac{1}{2m}\left\{\left[\frac{i\hbar\pp_i\sqrt{g}}{\sqrt{g}}+A_i
\right]g^{ik}(i\hbar\pp_k+A_k)\right\}\psi+\left[V-\gag\frac{\hbar^2}{m}\dot{R}\right]\psi
\end{equation}
where only the Riemannian curvature $\dot{R}$ is present (any explicit reference to the gauge
vector $\phi_i$ having disappeared).  \eqref{45} is of course the SE in curvilinear
coordinates whose invariance under point transformations is well known.  Moreover
\eqref{44} shows that
$|\psi|^2=\hat{\rho}(q,t)$ is the invariant probability density of finding the particle in the
volume element $d^nq$ at time t.  Then following Nelson's arguments that the SE together with
the density formula contains QM the present theory is physically equivalent to traditional
nonrelativistic QM.  One sees also from \eqref{44} and \eqref{45} that the time
independent SE is obtained via $S=S_0(q)-Et$ with constant E and
$\hat{\rho}(q)$.  In this case the scalar curvature of space becomes time independent; since
starting data at $t_0$ is  meaningless one replaces the continuity equation with a condition
$\int_M\hat{\rho}(q)\sqrt{g}d^nq=1$.
\\[3mm]\indent
{\bf REMARK 1.6.}
We recall that in the nonrelativistic context the quantum potential has the
form
$Q=-(\hbar^2/2m)(\pp^2\sqrt{\rho}/\sqrt{\rho})\,\,(\rho\sim\hat{\rho}$ here) and in more
dimensions this corresponds to $Q=-(\hbar^2/2m)(\gD\sqrt{\rho}/\sqrt{\rho})$.  Here we
have a SE involving $\psi=\sqrt{\rho}exp[(i/\hbar)S]$ with corresponding HJ equation 
\eqref{42}
which corresponds to the flat space 1-D $S_t+(s')^2/2m+V+Q=0$ with continuity
equation $\pp_t\rho+\pp(\rho S'/m)=0$ (take $A_k=0$ here).  The continuity equation in 
\eqref{43}
corresponds to $\pp_t\rho+(1/m\sqrt{g})\pp_i[\rho\sqrt{g}g^{ik}(\pp_kS)]=0$.  For $A_k=0$
\eqref{42} becomes 
\bq\label{46}
\pp_tS+(1/2m)g^{ik}\pp_iS\pp_kS+V-\gag(\hbar^2/m)R=0
\end{equation}
This leads to an identification $Q\sim-\gag(\hbar^2/m)R$ where R is the Ricci
scalar in the Weyl geometry (related to the Riemannian curvature built on standard Christoffel
symbols via \eqref{33}).  Here $\gag=(1/6)[(n-2)(n-2)]$ as above which for $n=3$
becomes $\gag=1/12$; further the Weyl field $\phi_i=-\pp_i log(\rho)$.
Consequently (see below).
\begin{proposition}
For the SE \eqref{45} in Weyl space the quantum potential is $Q=-(\hbar^2/12m)R$ where R
is the Weyl-Ricci scalar curvature.  For Riemannian flat space $\dot{R}=0$ this becomes
via \eqref{40}
\bq\label{47}
R=\frac{1}{2\gag\sqrt{\rho}}\pp_ig^{ik}\pp_k\sqrt{\rho}\sim\frac{1}{2\gag}
\frac{\gD\sqrt{\rho}}{\sqrt{\rho}}\Rightarrow
Q=-\frac{\hbar^2}{2m}\frac{\gD\sqrt{\rho}}{\sqrt{\rho}}
\end{equation}
as is should and the SE \eqref{45} reduces to the standard SE in the form
$i\hbar\pp_t\psi= -(\hbar^2/2m)\gD\psi+V\psi$ ($A_k=0$).
$\hfill\bs$
\end{proposition}
\indent
{\bf REMARK 1.7.}
In \cite{sb} (first paper) one begins with a generic 4-dimensional manifold with torsion
free connections and a metric tensor $g_{\mu\nu}$ ($\hbar=c=1$ for convenience).  Then
working with an average action principle based on \cite{hk} the particle motion and (Weyl)
spacetime geometry are derived in a gauge invariant manner (cf. Section 3.2).  Thus an
integrable Weyl geometry is produced from a stochastic background via an extremization
procedure (see Section 3).   An effective particle mass is taken as $m^2-(R/6)\sim
m^2(1+Q)\approx m^2exp(Q)$ corresponding to $R/6=-m^2Q=-\bx\sqrt{\rho}/\sqrt{\rho}$ (here
$\hbar=c=1$ and one has signature $(-,+,+,+)$ while the term $exp(Q)$ arises from
\cite{s36}).  We refer to
\cite{c3,c4,cqm,c16,sb} and Section 2 for details (for various other approaches see
\cite{au,w3}).$\hfill\bs$

\subsection{FISHER INFORMATION REVISITED}

We recall first
that the
classical Fisher information associated with translations of a 1-D observable X with
probability density
$P(x)$ (related to a quantum geometry probability measure $ds^2=\sum [(dp_j)^2/p_j]$) is
\bq\label{48}
F_X=\int dx\,P(x)([log(P(x)]')^2>0
\end{equation}
(cf. \cite{c1,c4,f1,h6,h3,h4,h5,r3,r4}).
One has a well known Cramer-Rao inequality $Var(X)\geq F_X^{-1}$
where $Var(X)\sim$ variance of X.  A Fisher length for X is defined via 
$\gd X=F_X^{-1/2}$ and this quantifies the length scale over which $p(x)$
(or better $log(p(x))$) varies appreciably.  Then the root mean square deviation
$\gD X$ satisfies $\gD X\geq \gd X$.  Let now 
P be the momentum observable conjugate to X, and $P_{cl}$ a classical
momentum observable corresponding to the state $\psi$ given via 
$p_{cl}(x)=(\hbar/2i)[(\psi'/\psi)-(\bar{\psi}'/\bar{\psi})]$.   
One has then the
identity $<p>_{\psi}=<p_{cl}>_{\psi}$ following via integration by parts. 
Now define the nonclassical momentum by $p_{nc}=p-p_{cl}$ 
and one shows then 
\bq\label{49}
\gD X\gD p\geq \gd X\gD p\geq \gd X\gD p_{nc}=\hbar/2
\end{equation}
Then consider a
classical ensemble of n-dimensional particles of mass m moving under a potential V.  The
motion can be described via the HJ and continuity equations
\bq\label{50}
\frac{\pp s}{\pp t}+\frac{1}{2m}|\na s|^2+V=0;\,\,\frac{\pp P}{\pp t}+
\na\cdot\left[P\frac{\na s}{m}\right]=0
\end{equation}
for the momentum potential $s$ and the position probability density P
(note that there is no quantum
potential and this will be supplied by the information term).   These
equations follow from the variational principle $\gd L=0$ with Lagrangian
\bq\label{50}
L=\int dt\,d^nx\,P\left[(\pp s/\pp t)+(1/2m)|\na s|^2+V\right]
\end{equation}
It is now assumed that the classical Lagrangian must be modified due to the existence
of random momentum fluctuations.  The nature of such fluctuations is immateria and one can
assume that the momentum associated with position x is given by $p=\na s + N$ where the
fluctuation term N vanishes on average at each point x.  Thus s changes to being an average
momentum potential.  It follows that the average kinetic energy $<|\na s|^2>/2m$ appearing in
the Lagrangian above should be replaced by $<|\na s+N|^2>/2m$ giving rise to
\bq\label{51}
L'=L+(2m)^{-1}\int dt<N\cdot N>=L+(2m)^{-1}\int dt(\gD N)^2
\end{equation}
where $\gD N=<N\cdot N>^{1/2}$ is a measure of the strength of the quantum fluctuations. 
The additional term is specified uniquely, up to a multiplicative constant, by the
three assumptions 
\begin{enumerate}
\item
Action principle:  $L'$ is a scalar Lagrangian with respect to the fields P and s
where the principle $\gd L'=0$ yields causal equations of motion.  Thus 
$$(\gD N)^2=\int
d^nx\,pf(P,\na P,\pp P/\pp t,s,\na s,\pp s/\pp t,x,t)$$
for some scalar function $f$.
\item
Additivity:  If the system comprises two independent noninteracting subsystems with 
$P=P_1P_2$ then the Lagrangian decomposes into additive subsystem contributions; thus
$f=f_1+f_2$ for $P=P_1P_2$.
\item
Exact uncertainty:  The strength of the momentum fluctuation at any given time is
determined by and scales inversely with the uncertainty in position at that time.  
Thus $\gD N\to k\gD N$ for $x\to x/k$.  Moreover since position
uncertainty is entirely characterized by the probability density P at any given time
the function $f$ cannot depend on $s$, nor explicitly on $t$, nor on $\pp P/\pp t$.
\end{enumerate}
This leads to the result that (cf. \cite{c1,cqm,h6})
\bq\label{53}
(\gD N)^2=c\int d^nx\,P|\na
log(P)|^2
\end{equation}
where c is a positive universal constant.  Further for
$\hbar=2\sqrt{c}$ and $\psi=\sqrt{P}exp(is/\hbar)$ the equations of motion for p and s
arising from $\gd L'=0$ are 
$i\hbar\frac{\pp\psi}{\pp t}=-\frac{\hbar^2}{2m}\na^2\psi+V\psi$.
\\[3mm]\indent
A second derivation is given in \cite{r4,r4}.  Thus let
$P(y^i)$ be a probability density and $P(y^i+\gD y^i)$ be the density resulting from a
small change in the $y^i$.  Calculate the cross entropy via
\bq\label{54}
J(P(y^i+\gD y^i):P(y^i))=\int P(y^i+\gD y^i)log\frac{P(y^i+\gD y^i)}{P(y^i)}d^ny\simeq
\end{equation}
$$\simeq\left[\frac{1}{2}\int \frac{1}{P(y^i)}\frac{\pp P(y^i)}{\pp y^i}\frac
{\pp P(y^i)}{\pp y^k)}d^ny\right]\gD y^i\gD y^k=I_{jk}\gD y^i\gD y^k$$
The $I_{jk}$ are the elements of the Fisher information matrix.  The most general
expression has the form
\bq\label{55}
I_{jk}(\gt^i)=\frac{1}{2}\int\frac{1}{P(x^i|\gt^i)}\frac{\pp P(x^i|\gt^i)}{\pp \gt^j}
\frac{\pp P(x^i|\gt^i)}{\pp \gt^k}d^nx
\end{equation}
where $P(x^i|\gt^i)$ is a probability distribution depending on parameters $\gt^i$ in
addition to the $x^i$.  For $P(x^i|\gt^i)=P(x^i+\gt^i)$ one recovers
\eqref{54}.  If P is defined over an n-dimensional
manifold with positive inverse metric $g^{ik}$ one obtains a natural definition of the
information associated with P via
\bq\label{56}
I=g^{ik}I_{ik}=\frac{g^{ik}}{2}\int\frac{1}{P}\frac{\pp P}{\pp y^i}\frac{\pp P}{\pp
y^k}d^ny
\end{equation}
Now in the HJ formulation of classical mechanics the equation of motion takes the form
\bq\label{57}
\frac{\pp S}{\pp t}+\frac{1}{2}g^{\mu\nu}\frac{\pp S}{\pp x^{\mu}}\frac{\pp S}
{\pp x^{\nu}}+V=0
\end{equation}
where $g^{\mu\nu}=diag(1/m,\cdots,1/m)$.  The velocity field $u^{\mu}$ is given by
$u^{\mu}=g^{\mu\nu}(\pp S/\pp x^{\nu})$.  When the exact coordinates
are unknown one can describe the system by means of a probability density
$P(t,x^{\mu})$ with $\int Pd^nx=1$ and 
\bq\label{58}
(\pp P/\pp t)+(\pp/\pp x^{\mu})(Pg^{\mu\nu}(\pp S/\pp x^{\nu})=0
\end{equation} 
These equations
completely describe the motion and can be derived from the Lagrangian
\bq\label{59}
L_{CL}=\int P\left\{(\pp S/\pp t)+(1/2)g^{\mu\nu}(\pp S/\pp x^{\mu})
(\pp S/\pp x^{\nu})+V\right\}dtd^nx
\end{equation}
using fixed endpoint variation in S and P.  Quantization is obtained by adding a term
proportional to the information I defined in \eqref{56}.  This leads to
\bq\label{60}
L_{QM}=L_{CL}+\gl I=\int P\left\{\frac{\pp S}{\pp t}+\frac{1}{2}g^{\mu\nu}\left[
\frac{\pp S}{\pp x^{\mu}}\frac{\pp S}{\pp x^{\nu}}+\frac{\gl}{P^2}\frac{\pp P}
{\pp x^{\mu}}\frac{\pp P}{\pp x^{\nu}}\right]+V\right\}dtd^nx
\end{equation}
Fixed endpoint variation in S leads again to \eqref{58} while variation in P leads to
\bq\label{61}
\frac{\pp S}{\pp t}+\frac{1}{2}g^{\mu\nu}\left[\frac{\pp S}{\pp x^{\mu}}\frac{\pp S}
{\pp x^{\nu}}+\gl\left(\frac{1}{P^2}\frac{\pp P}{\pp x^{\mu}}\frac{\pp P}{\pp x^{\nu}}
-\frac{2}{P}\frac{\pp^2P}{\pp x^{\mu}\pp x^{\nu}}\right)\right]+V=0
\end{equation}
These equations are equivalent to the SE if 
$\psi=\sqrt{P} exp(iS/\hbar)$ with $\gl=(2\hbar)^2$.
\\[3mm]\indent
{\bf	REMARK 1.8.}
Following ideas in \cite{c16,c29,n5} we note in \eqref{60} for
$\phi_{\mu}\sim A_{\mu}=\pp_{\mu}log(P)$ (cf. \eqref{39}) and
$P_{\mu}=\pp_{\,u}S$, a complex velocity can be envisioned leading to
\bq\label{62}
|P_{\mu}+i\sqrt{\gl}A_{\mu}|^2=P_{\mu}^2+\gl A_{\mu}^2\sim g^{\mu\nu}\left(\frac{\pp S}{\pp
x^{\mu}}
\frac{\pp S}{\pp x^{\nu}}+\frac{\gl}{P^2}\frac{\pp P}{\pp x^{\mu}}\frac{\pp P}{\pp
x^{\nu}}\right)
\end{equation}
Further I in \eqref{56} is exactly known from $\phi_{\mu}$ so one has a direct connection
between Fisher information and the Weyl field $\phi_{\mu}$, along with motivation for a
complex velocity. $\hfill\bs$
\\[3mm]\indent
{\bf REMARK 1.9.}
Comparing now with \cite{c4} and quantum geometry in the form $ds^2=\sum(dp_j^2/p_j)$ on a
space of probability distributions we can  define
\eqref{56} as a Fisher information metric in the present context.  This should be
positive definite in view of its relation to
$(\gD N)^2$ in
\eqref{53} for example. Now for $\psi=Rexp(iS/\hbar)$ one has ($\rho\sim\hat{\rho}$ here)
\bq\label{63}
-\frac{\hbar^2}{2m}\frac{R''}{R}\equiv
-\frac{\hbar^2}{2m}\frac{\pp^2\sqrt{\rho}}{\sqrt{\rho}}=-\frac{\hbar^2}{8m}\left[
\frac{2\rho''}{\rho}-\left(\frac{\rho'}{\rho}\right)^2\right]
\end{equation}
in 1-D while in more dimensions we have a form ($\rho\sim P$)
\bq\label{64}
Q\sim -2\hbar^2g^{\mu\nu}\left[\frac{1}{P^2}\frac{\pp P}{\pp x^{\mu}}\frac{\pp P}{\pp
x^{\nu}} -\frac{2}{P}\frac{\pp^2P}{\pp x^{\mu}\pp x^{\nu}}\right]
\end{equation}
as in \eqref{63} (arising from the Fisher metric I of \eqref{56} upon variation in P
in the Lagrangian).  It can also be related to an osmotic velocity field $u=D\na
log(\rho)$ via $Q=(1/2)u^2+D\na\cdot u$ connected to Brownian
motion where D is a diffusion coefficient (cf. \cite{c29,c15,g10,n5}).
For $\phi_{\mu}=-\pp_{\mu}log(P)$ we have then $u=-D{\bf \phi}$ with
$Q=D^2((1/2)(|u|^2-\na\cdot {\bf \phi})$, expressing Q directly in terms of the Weyl
vector.  This enforces the idea that QM is built into Weyl geometry!
$\hfill\bs$

\section{BOHMIAN MECHANICS AND WEYL GEOMETRY}
\renewcommand{\theequation}{2.\arabic{equation}}
\setcounter{equation}{0}

From
Chapters 1 and 2 we know something about Bohmian mechanics and the quantum
potential and we go now to the papers \cite{s33,s34,ss5,s35,s36,s37,s71,sss} 
by A. and F. Shojai to begin the
present discussion (cf. also
\cite{a21,b10,b81,ddd,m14,m8,s52,s53,s55,s57,s2,s4,s5,s6,s51,s31,ss3,ss4,s32,s38,
sf,s1s}).
for related work from the Tehran school and \cite{c4,c16,lo,or,sa,sb,s35} for
linking of dBB theory with Weyl geometry).  In nonrelativistic deBroglie-Bohm
theory the quantum potential is 
$Q=-(\hbar^2/2m)(\na^2|\Psi|/|\Psi|)$.  The
particles trajectory can be derived from Newton's law of motion in
which the quantum force $-\na Q$ is present in addition to the
classical force $-\na V$.  The enigmatic quantum behavior is attributed
here to the quantum force or quantum potential (with $\Psi$
determining a ``pilot wave" which guides the particle motion).  Setting 
$\Psi=\sqrt{\rho}exp[iS/\hbar]$ one has 
\bq\label{65}
\frac{\pp S}{\pp t}+\frac{|\na S|^2}{2m}+V+Q=0;\,\,\frac{\pp \rho}{\pp
t}+\na\cdot\left(\rho\frac{\na S}{m}\right)=0
\end{equation}
The first equation in \eqref{65} is a Hamilton-Jacobi (HJ) equation
which is identical to Newton's law and represents an energy condition
$E=(|p|^2/2m)+V+Q$ (recall from HJ theory $-(\pp S/\pp
t)=E (=H)$ and $\na S=p$).  The second equation
represents a continuity equation for a hypothetical ensemble related to
the particle in question.  For the relativistic extension one could simply try to
generalize the relativistic energy equation 
$\eta_{\mu\nu}P^{\mu}P^{\nu}=m^2c^2$ to the form 
\bq\label{66}
\eta_{\mu\nu}P^{\mu}P^{\nu}=m^2c^2(1+{\mc Q})={\mc M}^2c^2;\,\,
{\mc Q}=
(\hbar^2/m^2c^2)(\bx |\Psi|/|\Psi|)
\end{equation}
\bq\label{67}
{\mc
M}^2=m^2\left(1+\ga\frac{\bx|\Psi|}{|\Psi|}\right);\,\,\,\ga=\frac{\hbar^2}{m^2c^2}
\end{equation}
This could be derived e.g. by setting $\Psi=\sqrt{\rho}exp(iS/\hbar)$ in the
Klein-Gordon (KG) equation and separating the real and imaginary parts, leading to
the relativistic HJ equation $\eta_{\mu\nu}\pp^{\mu}S\pp^{\nu}S={\mf
M}^2c^2$ (as in \eqref{65} - note $P^{\mu}=-\pp^{\mu}S$) and the continuity
equation is $\pp_{\mu}(\rho\pp^{\mu}S)=0$.  The problem of ${\mc M}^2$
not being positive definite here (i.e. tachyons) is serious however and in fact
\eqref{66} is not the correct equation (see e.g. \cite{ss5,s36,s71}).  One must
use the covariant derivatives $\na_{\mu}$ in place of $\pp_{\mu}$ and for spin
zero in a curved background there results
\bq\label{68} 
\na_{\mu}(\rho\na^{\mu}S)=0;\,\,g^{\mu\nu}\na_{\mu}S\na_{\nu}S=
{\mf M}^2c^2;
\end{equation}
To see this one must require that a correct relativistic equation of motion should
not only be Poincar\'e invariant but also it should have the correct
nonrelativistic limit.  Thus for a relativistic particle of mass ${\mf M}$
(which is a Lorentz invariant quantity)  
\bq\label{69}
{\mf A}=\int d\gl (1/2){\mf M}(r)(dr_{\mu}/d\gl)(dr^{\nu}/d\gl)
\end{equation}
is
the action functional where
$\gl$ is any scalar parameter parametrizing the path $r_{\mu}(\gl)$ (it could e.g.
be the proper time $\tau$).  Varying the path via $r_{\mu}\to
r'_{\mu}=r_{\mu}+\gep_{\mu}$ one gets
\bq\label{70}
{\mf A}\to {\mf A}'={\mf A}+\gd{\mf A}={\mf A}+\int d\gl\left[\frac{dr_{\mu}}{d\gl}
\frac{d\gep^{\mu}}{d\gl}+\frac{1}{2}\frac{dr_{\mu}}{d\gl}\frac{dr^{\mu}}{d\gl}
\gep_{\nu}\pp^{\nu}{\mf M}\right]
\end{equation}
By least action the correct path satisfies $\gd{\mf A}=0$ with fixed
boundaries so the equation of motion is 
\bq\label{71}
(d/d\gl)({\mf
M}u_{\mu})=(1/2)u_{\nu}u^{\nu}\pp_{\mu}{\mf M};
\end{equation}
$${\mf
M}(du_{\mu}/d\gl)=((1/2)\eta_{\mu\nu}u_{\ga}u^{\ga}-u_{\mu}u_{\nu})\pp^{\nu}{\mf
M}$$
where $u_{\mu}=dr_{\mu}/d\gl$.  Now look at the symmetries of the action
functional via $\gl\to \gl+\gd$.  The conserved current is then
the Hamiltonian ${\mf H}=-{\mf L}+u_{\mu}(\pp{\mf L}/\pp u_{\mu})=(1/2){\mf
M}u_{\mu}u^{\mu}=E$.  This can be seen by setting $\gd{\mf A}=0$ where
\bq\label{72}
0=\gd{\mf A}={\mf A}'-{\mf A}=\int
d\gl\left[\frac{1}{2}u_{\mu}u^{\mu}u^{\nu}\pp_{\nu}{\mf M}+{\mf M}u_{\mu}\frac{d
u^{\mu}}{d\gl}\right]\gd
\end{equation}
which means that the integrand is zero, i.e. $(d/d\gl)[(1/2){\mf M}
u_{\mu}u^{\mu}]=0$.  Since the proper time is defined as $c^2d\tau^2=dr_{\mu}
dr^{\mu}$ this leads to $(d\tau/d\gl)=\sqrt{(2E/{\mf M}c^2)}$ and
the equation of motion becomes 
\bq\label{73}
{\mf M}(dv_{\mu}/d\tau)=(1/2)(c^2\eta_{\mu\nu}-v_{\mu}v_{\nu})\pp^{\nu}{\mf M}
\end{equation} 
where
$v_{\mu}=dr_{\mu}/d\tau$.  The nonrelativistic limit can be derived by letting the
particles velocity be ignorable with respect to light velocity.  In this limit the
proper time is identical to the time coordinate $\tau=t$ and the result is that
the $\mu=0$ component is satisfied identically via ($r\sim\vec{r}$)
\bq\label{74}
{\mf M}\frac{d^2r}{dt^2}=-\frac{1}{2}c^2\na{\mf M}\Rightarrow
m\left(\frac{d^2r}{dt^2}\right)=-\na\left[\frac{mc^2}{2}log\left(\frac{{\mf
M}}{\mu}\right)\right]
\end{equation}
where $\mu$ is an arbitrary
mass scale.  In order to have the correct limit the term in parenthesis on the
right side should be equal to the quantum potential so 
\bq\label{75}
{\mf M}=\mu
exp[-(\hbar^2/m^2c^2)(\na^2|\Psi|/|\Psi|)]
\end{equation}
The relativistic quantum mass field
(manifestly invariant) is ${\mf M}=\mu exp[(\hbar^2/2m)(\bx|\Psi|/|\Psi|)]$ and
setting $\mu=m$ we get 
\bq\label{76}
{\mf M}=m exp[(\hbar^2/m^2c^2)(\bx|\Psi|/|\Psi|)]
\end{equation}
If
one starts with the standard relativistic theory and goes to the nonrelativistic limit one does
not get the correct nonrelativistic equations; this is a result of 
an improper decomposition of the
wave function into its phase and norm in the KG equation (cf. also \cite{b9} for
related procedures).  One notes here also that \eqref{76} leads to a positive
definite mass squared.  Also from \cite{ss5} this can be extended to a many
particle version and to a curved spacetime.  In summary, for a particle in a
curved background one has (cf. \cite{s36} which we continue to follow) 
\bq\label{77}
\na_{\mu}(\rho\na^{\mu}S)=0;\,\,g^{\mu\nu}\na_{\mu}S\na_{\nu}S={\mf M}^2c^2;
\,\,{\mf M}^2=m^2e^{{\mf Q}};\,\,{\mf
Q}=\frac{\hbar^2}{m^2c^2}\frac{\bx_g|\Psi|}{|\Psi|}
\end{equation}
Since, following deBroglie, the quantum HJ equation (QHJE) in \eqref{77} can be
written in the form $(m^2/{\mf M}^2)g^{\mu\nu}\na_{\mu}S\na_{\nu}S=m^2c^2$
the quantum effects are identical to a change of spacetime metric 
\bq\label{78}
g_{\mu\nu}\to\tl{g}_{\mu\nu}=({\mf M}^2/m^2)g_{\mu\nu}
\end{equation}
which is a
conformal transformation.  The QHJE becomes then
$\tl{g}^{\mu\nu}\tl{\na}_{\mu}S\tl{\na}_{\nu}S=m^2c^2$ where $\tl{\na}_{\mu}$
represents covariant differentiation with respect to the metric
$\tl{g}_{\mu\nu}$ and the continuity equation is then 
$\tl{g}_{\mu\nu}\tl{\na}_{\mu}(\rho\tl{\na}_{\nu}S)=0$.
The important conclusion here is that the presence of the quantum potential is
equivalent to a curved spacetime with its metric given by \eqref{78}.  This is a
geometrization of the quantum aspects of matter and it seems that there is a dual
aspect to the role of geometry in physics.  The spacetime geometry sometimes looks
like ``gravity" and sometimes reveals quantum behavior.  The curvature due to the
quantum potential may have a large influence on the classical contribution to the
curvature of spacetime.  The particle trajectory can now be derived from the
guidance relation via differentiation of \eqref{77} leading to the Newton
equations of motion
\bq\label{79}
{\mf M}\frac{d^2x^{\mu}}{d\tau^2}+{\mf M}\gG^{\mu}_{\nu\gk}u^{\nu}u^{\gk}=
(c^2g^{\mu\nu}-u^{\mu}u^{\nu})\na_{\nu}{\mf M}
\end{equation}
Using the conformal transformation above \eqref{79} reduces to the standard
geodesic equation.
\\[3mm]\indent
Now a general ``canonical" relativistic system consisting of gravity and classical matter 
(no quantum effects) is determined by the action
\bq\label{80}
{\mc A}=\frac{1}{2\gk}\int d^4x\sqrt{-g}{\mc R}+\int
d^4x\sqrt{-g}\frac{\hbar^2}{2m}\left(\frac{\rho}{\hbar^2}{\mc D}_{\mu}S{\mc
D}^{\mu}S-\frac{m^2}{\hbar^2}\rho\right)
\end{equation}
where $\gk=8\pi G$ and $c=1$ for convenience.  It was seen above that via
deBroglie the introduction of a quantum potential is equivalent to introducing a
conformal factor $\gO^2={\mf M}^2/m^2$ in the metric.  Hence in order to introduce
quantum effects of matter into the action \eqref{80} one uses this conformal
transformation to get ($1+Q\sim exp(Q)$)
\bq\label{81}
{\mf A}=\frac{1}{2\gk}\int d^4x\sqrt{-\bar{g}}(\bar{{\mc R}}\gO^2-6\bar{\na}_{\mu}
\gO\bar{\na}^{\mu}\gO)+
\end{equation}
$$+\int
d^4x\sqrt{-\bar{g}}\left(\frac{\rho}{m}\gO^2\bar{\na}_{\mu}S\bar{\na}^{\mu}S-
m\rho\gO^4\right)+\int
d^4x\sqrt{-\bar{g}}\gl\left[\gO^2-\left(1+\frac{\hbar^2}{m^2}
\frac{\bar{\bx}\sqrt{\rho}}{\sqrt{\rho}}\right)\right]$$
where a bar over any quantity means that it corresponds to the nonquantum regime.
Here only the first two terms of the expansion of ${\mf M}^2=m^2exp({\mf Q})$ 
in \eqref{77} have been used, namely ${\mf M}^2\sim m^2(1+{\mf Q})$.  No
physical change is involved in considering all the terms.  $\gl$ is a Lagrange
multiplier introduced to identify the conformal factor with its Bohmian value. 
One uses here $\bar{g}_{\mu\nu}$ to raise of lower indices and to evaluate the
covariant derivatives; the physical metric (containing the quantum effects of
matter) is
$g_{\mu\nu}=\gO^2\bar{g}_{\mu\nu}$.  By variation of the action with respect to
$\bar{g}_{\mu\nu},\,\gO,\,\rho,\,S,$ and $\gl$ one arrives at the following
quantum equations of motion:
\begin{enumerate}
\item
The equation of motion for $\gO$
\bq\label{82}
\bar{{\mc
R}}\gO+6\bar{\bx}\gO+\frac{2\gk}{m}\rho\gO(\bar{\na}_{\mu}S\bar{\na}^{\mu}S-
2m^2\gO^2)+2\gk\gl\gO=0
\end{equation}
\item
The continuity equation for particles 
$\bar{\na}_{\mu}(\rho\gO^2\bar{\na}^{\mu}S)=0$
\item
The equations of motion for particles (here $a'\equiv\bar{a}$)
\bq\label{83}
(\bar{\na}_{\mu}S\bar{\na}^{\mu}S-m^2\gO^2)\gO^2\sqrt{\rho}+\frac{\hbar^2}{2m}
\left[\bx'\left(\frac{\gl}{\sqrt{\rho}}\right)-\gl\frac{\bx'\sqrt{\rho}}{\rho}\right]
=0
\end{equation}
\item
The modified Einstein equations for $\bar{g}_{\mu\nu}$
\bq\label{84}
\gO^2\left[\bar{{\mc R}}_{\mu\nu}-\frac{1}{2}\bar{g}_{\mu\nu}\bar{{\mc R}}\right]
-[\bar{g}_{\mu\nu}\bx'-\bar{\na}_{\mu}\bar{\na}_{\nu}]\gO^2-6\bar{\na}_{\mu}
\gO\bar{\na}_{\nu}\gO+3\bar{g}_{\mu\nu}\bar{\na}_{\ga}\gO\bar{\na}^{\ga}\gO+
\end{equation}
$$+\frac{2\gk}{m}\rho\gO^2\bar{\na}_{\mu}S\bar{\na}_{\nu}S-\frac{\gk}{m}
\rho\gO^2\bar{g}_{\mu\nu}\bar{\na}_{\ga}S\bar{\na}^{\ga}S+\gk
m\rho\gO^4\bar{g}_{\mu\nu}+$$
$$+\frac{\gk\hbar^2}{m^2}\left[\bar{\na}_{\mu}\sqrt{\rho}\bar{\na}_{\nu}\left(\frac
{\gl}{\sqrt{\rho}}\right)+\bar{\na}_{\nu}\sqrt{\rho}\bar{\na}_{\mu}\left(\frac{\gl}{\sqrt{\rho}}
\right)\right]-\frac{\gk\hbar^2}{m^2}\bar{g}_{\mu\nu}\bar{\na}_{\ga}\left[
\gl\frac{\bar{\na}^{\ga}\sqrt{\rho}}{\sqrt{\rho}}\right]=0$$
\item
The constraint equation
$\gO^2=1+(\hbar^2/m^2)[(\bar{\bx}\sqrt{\rho})/\sqrt{\rho}]$
\end{enumerate}
Thus the back reaction effects of the quantum factor on the background metric are
contained in these highly coupled equations.  A simpler form of \eqref{72} can
be obtained by taking the trace of \eqref{84} and using \eqref{82} which
produces 
$\gl=(\hbar^2/m^2)\bar{\na}_{\mu}[\gl(\bar{\na}^{\mu}\sqrt{\rho})/
\sqrt{\rho}]$.  A solution of this via perturbation methods using the small
parameter $\ga=\hbar^2/m^2$ yields the trivial solution $\gl=0$ so the above
equations reduce to
\bq\label{85}
\bar{\na}_{\mu}(\rho\gO^2\bar{\na}^{\mu}S)=0;\,\,\bar{\na}_{\mu}S\bar{\na}^{\mu}
S=m^2\gO^2;\,\,{\mf G}_{\mu\nu}=-\gk{\mf T}^{(m)}_{\mu\nu}-\gk{\mf
T}^{(\gO)}_{\mu\nu}
\end{equation}
where ${\mf T}_{\mu\nu}^{(m)}$ is the matter energy-momentum (EM) tensor and 
\bq\label{86}
\gk{\mf T}_{\mu\nu}^{(\gO)}=\frac{[g_{\mu\nu}\bx-\na_{\mu}\na_{\nu}]\gO^2}{\gO^2}
+6\frac{\na_{\mu}\gO\na_{\nu}\gO}{\go^2}-2g_{\mu\nu}\frac{\na_{\ga}\gO\na^{\ga}
\gO}{\gO^2}
\end{equation}
with $\gO^2=1+\ga(\bar{\bx}\sqrt{\rho}/\sqrt{\rho})$.  Note that
the second relation in \eqref{85} is the Bohmian equation of motion and written
in terms of $g_{\mu\nu}$ it becomes $\na_{\mu}S\na^{\mu}S=m^2c^2$.
\\[3mm]\indent
{\bf REMARK 2.1.}
In the preceeding one has tacitly assumed that there is an ensemble of quantum
particles so what about a single particle?  One translates now the quantum
potential into purely geometrical terms without reference to matter parameters 
so that the original form of the quantum potential can only be deduced after using
the field equations.  Thus the theory will work for a single particle or an
ensemble.
One notes that the use of $\psi\psi^*$ automatically suggests or involves an ensemble if 
(or its square root) it is to
be interpreted as a probability density.  Thus the idea that a particle has only a probability of
being at or near x seems to mean that some paths take it there but others don't and this is
consistent with Feynman's use of path integrals for example.  This seems also to say that there
is no such thing as a particle, only a collection of versions or cloud connected to the particle
idea.  Bohmian theory on the other hand for a fixed energy gives a one parameter family of
trajectories associated to $\psi$ (see here 
\cite{c10,c12,cqm} for details).  This is because the
trajectory arises from a third order differential while fixing the solution $\psi$ of the second
order stationary Schr\"odinger equation involves only two ``boundary" conditions.  As was shown in
\cite{c12} this automatically generates a Heisenberg inequality $\gD x\gD p\geq c\hbar$; i.e.
the uncertainty is built in when using the wave function $\psi$ and amazingly can be expressed by
the operator theoretical framework of quantum mechanics.  Thus a one parameter family of paths
can be associated with the use of $\psi\psi^*$ and this generates the cloud or ensemble
automatically associated with the use of $\psi$.
In fact (cf. Remark 3.2) one might conjecture that upon using a wave function
discription of quantum particle motion, one opens the door to a cloud of particles, all of whose
motions are incompletely governed by the SE, since one determining condition for particle 
motion is ignored.  Thus automatically the quantum potential will give rise to a force acting
on any such particular trajectory and the ``ensemble" idea naturally applies to a cloud of
identical particles. 
$\hfill\bs$
\\[3mm]\indent
{\bf REMARK 2.2.}
Now first ignore gravity and look at the geometrical properties of the
conformal factor given via
\bq\label{87}
g_{\mu\nu}=e^{4\gS}\eta_{\mu\nu};\,\,e^{4\gS}=\frac{{\mf M}^2}{m^2}=
exp\left(\ga\frac{\bx_{\eta}\sqrt{\rho}}{\sqrt{\rho}}\right)=exp\left(\ga\frac
{\bx_{\eta}\sqrt{|{\mf T}|}}{\sqrt{|{\mf T}|}}\right)
\end{equation}
where ${\mf T}$ is the trace of the EM tensor and is substituted for $\rho$ 
(true for dust).  The Einstein tensor for this metric is
\bq\label{88}
{\mf
G}_{\mu\nu}=4g_{\mu\nu}\bx_{\eta}exp(-\gS)+2exp(-2\gS)\pp_{\mu}\pp_{\nu}exp(2\gS);
\,\,\gS=\frac{\ga}{4}\frac{\bx_{\eta}\sqrt{\rho}}{\sqrt{\rho}}
\end{equation}
Hence as an Ansatz one can suppose that in the presence of gravitational effects
the field equation would have a form
\bq\label{89}
{\mc R}_{\mu\nu}-\frac{1}{2}{\mc R}g_{\mu\nu}=\gk{\mf
T}_{\mu\nu}+4g_{\mu\nu}e^{\gS}\bx e^{-\gS}+2e^{-2\gS}\na_{\mu}\na_{\nu}e^{2\gS}
\end{equation}
This is written in a manner such that in the limit ${\mf T}_{\mu\nu}\to 0$ one
will obtain \eqref{87}.  Taking the trace of the last equation one gets
$-{\mc R}=\gk{\mf T}-12\bx\gS+24(\na\gS)^2$ which has the iterative
solution $\gk{\mf T}=-{\mc R}+12\ga\bx[(\bx\sqrt{{\mc R}})/
\sqrt{{\mc R}}]$ leading to 
\bq\label{90}
\gS\propto\ga[(\bx\sqrt{|{\mf
T}|}/\sqrt{|{\mf T}|})]\simeq \ga[(\bx\sqrt{|{\mc R}|})/\sqrt{|{\mc R}|})]
\end{equation}
to first order in $\ga$.
One goes now to the field equations for this toy model.  First from the above one
sees that ${\mf T}$ can be replaced by ${\mc R}$ in the expression for the quantum
potential or for the conformal factor of the metric.  This is important since the
explicit reference to ensemble density is removed and the theory works for a
single particle or an ensemble.  So from \eqref{32.24} for a toy quantum gravity
theory one assumes the following field equations
\bq\label{91}
{\mf G}_{\mu\nu}-\gk{\mf T}_{\mu\nu}-{\mf
Z}_{\mu\nu\ga\gb}exp\left(\frac{\ga}{2}\Phi\right)\na^{\ga}\na^{\gb}exp
\left(-\frac{\ga}{2}\Phi\right)=0
\end{equation}
where ${\mf
Z}_{\mu\nu\ga\gb}=2[g_{\mu\nu}g_{\ga\gb}-g_{\mu\ga}g_{\nu\gb}]$ and $\Phi=(\bx
\sqrt{|{\mc R}|}/\sqrt{|{\mc R}|})$.  The number 2 and the minus sign of the
second term are chosen so that the energy equation derived later
will be correct.  Note that the trace of \eqref{91} is
\bq\label{92}
{\mc R}+\gk{\mf T}+6exp(\ga\Phi/2)\bx exp(-\ga\Phi/2)=0
\end{equation}
and this represents 
the connection of the Ricci scalar curvature of space time and the trace of the
matter EM tensor.  If a perturbative solution is admitted one can expand in powers
of $\ga$ to find ${\mc R}^{(0)}=-\gk{\mf T}$ and ${\mc R}^{(1)}=
-\gk{\mf T}-6exp(\ga\Phi^0/2)\bx exp(-\ga\Phi^0/2)$ where $\Phi^{(0)}=\bx
\sqrt{|{\mf T}|}/\sqrt{|{\mf T}|}$.  The energy relation can be obtained by taking
the four divergence of the field equations and since the divergence of the
Einstein tensor is zero one obtains 
\bq\label{93}
\gk\na^{\nu}{\mf T}_{\mu\nu}=\ga{\mc R}_{\mu\nu}\na^{\nu}\Phi-\frac{\ga^2}{4}
\na_{\mu}(\na\Phi)^2+\frac{\ga^2}{2}\na_{\mu}\Phi\bx\Phi
\end{equation}
For a dust with ${\mf T}_{\mu\nu}=\rho u_{\mu}u_{\nu}$ and
$u_{\mu}$ the velocity field, the conservation of mass law is 
$\na^{\nu}(\rho{\mf M} u_{\nu})=0$ so one gets to first order in $\ga$ 
$\na_{\mu}{\mf M}/{\mf M}=-(\ga/2)\na_{\mu}\Phi$ or ${\mf
M}^2=m^2exp(-\ga\Phi)$ where $m$ is an integration constant.  This is the correct
relation of mass and quantum potential.$\hfill\bs$
\\[3mm]\indent
In \cite{s36} there is then some discussion about making the conformal factor dynamical 
via a general scalar tensor action (cf. also
\cite{ss4}) and subsequently one makes both the conformal factor and the quantum
potential into dynamical fields and creates a scalar tensor theory with two scalar fields.  Thus
first start with a general action
\bq\label{94}
{\mf A}=\int d^4x\sqrt{-g}\left[\phi{\mc R}-\go\frac{\na_{\mu}\phi\na^{\mu}\phi}{\phi}
-\frac{\na_{\mu}Q\na^{\mu} Q}{\phi}+2\gL\phi +{\mf L}_m\right]
\end{equation}
The cosmological constant generally has an interaction term with the scalar field and here
one uses an ad hoc matter Lagrangian
\bq\label{95}
{\mf L}_m=\frac{\rho}{m}\phi^a\na_{\mu}S\na^{\mu}S-m\rho\phi^b-\gL(1+Q)^c+\ga\rho(e^{\ell
Q}-1)
\end{equation}
(only the first two terms $1+Q$ from $exp(Q)$ are used for simplicity in the third term).
Here $a,b,c$ are constants to be fixed later and the last term is chosen 
(heuristically) in such a manner
as to have an interaction between the quantum potential field and the ensemble density (via 
the equations of motion); further the interaction is chosen so that it
vanishes in the classical limit but this is ad hoc.  Variation of the above action yields
\begin{enumerate}
\item
The scalar fields equation of motion
\bq\label{96}
{\mc R}+\frac{2\go}{\phi}\bx\phi-\frac{\go}{\phi^2}\na^{\mu}\phi\na_{\mu}\phi+2\gL+
\end{equation}
$$+\frac{1}{\phi^2}\na^{\mu}Q\na_{\mu}Q+\frac{a}{m}\rho\phi^{a-1}\na^{\mu}S
\na_{\mu}S-mb\rho\phi^{b-1}=0$$
\item
The quantum potential equations of motion
\bq\label{97}
(\bx Q/\phi)-(\na_{\mu}Q\na^{\mu}\phi/\phi^2)-\gL c(1+Q)^{c-1}+\ga\ell\rho
exp(\ell Q)=0
\end{equation}
\item
The generalized Einstein equations
\bq\label{98}
{\mf G}^{\mu\nu}-\gL g^{\mu\nu}=-\frac{1}{\phi}{\mf
T}^{\mu\nu}-\frac{1}{\phi}[\na^{\mu}\na^{\nu}-g^{\mu\nu}\bx]\phi+\frac{\go}{\phi^2}\na^{\mu}\phi
\na^{\nu}\phi-
\end{equation}
$$-\frac{\go}{2\phi^2}g^{\mu\nu}\na^{\ga}\phi\na_{\ga}\phi+\frac{1}{\phi^2}\na^{\mu}Q\na^{\nu}Q
-\frac{1}{2\phi^2}g^{\mu\nu}\na^{\ga}Q\na_{\ga}Q$$
\item
The continuity equation $\na_{\mu}(\rho\phi^a\na^{\mu}S)=0$
\item
The quantum Hamilton Jacobi equation 
\bq\label{99}
\na^{\mu}S\na_{\mu}S=m^2\phi^{b-a}
-\ga m\phi^{-a}(e^{\ell Q}-1)
\end{equation}
\end{enumerate}
In \eqref{96} the scalar curvature and the term $\na^{\mu}S\na_{\mu}S$ can be eliminated
using \eqref{98} and \eqref{99}; further on using the matter Lagrangian and the
definition of the EM tensor one has
\bq\label{100}
(2\go -3)\bx\phi =(a+1)\rho\ga(e^{\ell Q}-1)-2\gL(1+Q)^c+2\gL\phi-\frac{2}{\phi}\na_{\mu}Q
\na^{\mu}Q
\end{equation}
(where $b=a+1$).  Solving \eqref{97} and \eqref{100} with a perturbation expansion in
$\ga$ one finds
\bq\label{101}
Q=Q_0+\ga Q_1+\cdots;\,\,\phi=1+\ga Q_1+\cdots;\,\,\sqrt{\rho}=\sqrt{\rho_0}+
\ga\sqrt{\rho_1}+\cdots
\end{equation}
where the conformal factor is chosen to be unity at zeroth order so that as $\ga\to 0$
\eqref{99} goes to the classical HJ equation.  Further since by \eqref{99} the quantum
mass is $m^2\phi+\cdots$ the first order term in $\phi$ is chosen to be $Q_1$  (cf.
\eqref{77}). Also we will see that $Q_1\sim \bx\sqrt{\rho}/\sqrt{\rho}$ plus corrections
which is in accord with Q as a quantum potential field.  In any case after some computation one 
obtains $a=2\go k,\,\,b=a+1,$ and $\ell=(1/4)(2\go k+1)=(1/4)(a+1)=b/4$ 
with
$Q_0=[1/c(2c-3)]
\{[-(2\go k+1)/2\gL]k\sqrt{\rho_0}-(2c^2-c+1)\}$ while $\rho_0$ can be
determined (cf. \cite{s36} for details).  Thus heuristically the quantum potential can be regarded as
a dynamical field and perturbatively one gets the correct dependence of quantum potential upon
density, modulo some corrective terms.
\\[3mm]\indent
{\bf REMARK 2.3.}
The gravitational effects determine
the causal structure of spacetime as long as quantum effects give its conformal structure.
This does not mean that quantum effects have nothing to do with the causal structure; they
can act on the causal structure through back reaction terms appearing in the metric field
equations.  The conformal factor of the metric is a function of the quantum potential and
the mass of a relativistic particle is a field produced by quantum corrections to the
classical mass.  One has shown that the presence of the quantum potential is equivalent to a
conformal mapping of the metric.  Thus in different conformally related frames one feels
different quantum masses and different curvatures.  In particular there are two frames with
one containing the quantum mass field and the classical metric while the other contains
the classical mass and the quantum metric.  In general frames both the spacetime metric and
the mass field have quantum properties so one can state that different conformal frames are
identical pictures of the gravitational and quantum phenomena.  We feel different quantum
forces in different conformal frames.  The question then arises of whether the
geometrization of quantum effects implies conformal invariance just as gravitational effects
imply general coordinate invariance.  One sees here that Weyl geometry provides 
additional degrees of freedom which can be identified with quantum effects and seems
to create a unified geometric framework for understanding both gravitational and quantum
forces.  Some features here are: (i) Quantum effects appear independent of any preferred
length scale.  (ii) The quantum mass of a particle is a field.  (iii) The gravitational
constant is also a field depending on the matter distribution via the quantum potential
(cf. \cite{ss4, s37}).  (iv)  A local variation of matter field distribution changes the
quantum potential acting on the geometry and alters it globally; the nonlocal character is
forced by the quantum potential (cf. \cite{s32}).$\hfill\bs$

\subsection{DIRAC-WEYL ACTION}

Next (still following \cite{s36}) one goes to Weyl geometry based on the Weyl-Dirac action
\bq\label{102}
{\mf A}=\int d^4x\sqrt{-g}(F_{\mu\nu}F^{\mu\nu}-\gb^2\,\,{}^W{\mc
R}+(\gs+6)\gb_{;\mu}\gb^{;\mu}+ {\mf L}_{matter}
\end{equation}
Here $F_{\mu\nu}$ is the curl of the Weyl 4-vector $\phi_{\mu}$, $\gs$ is an arbitrary
constant and $\gb$ is a scalar field of weight $-1$.  The symbol
``;" represents a covariant derivative under general coordinate and conformal transformations
(Weyl covariant derivative) defined as $X_{;\mu}={}^W\na_{\mu}X-{\mc
N}\phi_{\mu}X$ where 
${\mc N}$ is the Weyl weight of X.  The equations of motion are then
\bq\label{103}
{\mf G}^{\mu\nu}=-\frac{8\pi}{\gb^2}({\mf
T}^{\mu\nu}+M^{\mu\nu})+\frac{2}{\gb}(g^{\mu\nu}{}^W\na^{\ga}{}^W\na_{\ga}\gb-{}^W\na^{\mu}{}^W\na^{\nu}
\gb)+
\end{equation}
$$+\frac{1}{\gb^2}(4\na^{\mu}\gb\na^{\nu}\gb-g^{\mu\nu}\na^{\ga}\gb\na_{\ga}\gb)+\frac{\gs}
{\gb^2}(\gb^{;\mu}\gb^{;\nu}-\frac{1}{2}g^{\mu\nu}\gb^{;\ga}\gb_{;\ga});$$
$${}^W\na_{\mu}F^{\mu\nu}=\frac{1}{2}\gs(\gb^2\phi^{\mu}+\gb\na^{\mu}\gb)+4\pi J^{\mu};$$
$${\mc
R}=-(\gs+6)\frac{{}^W\bx\gb}{\gb}+\gs\phi_{\ga}\phi^{\ga}-\gs{}^W\na^{\ga}\phi_{\ga}+\frac
{\psi}{2\gb}$$
where 
\bq\label{104}
M^{\mu\nu}=(1/4\pi)[(1/4)g^{\mu\nu}F^{\ga\gb}F_{\ga\gb}-F^{\mu}_{\ga}F^{\nu\ga}
\end{equation} 
and 
\bq\label{105}
8\pi{\mf T}^{\mu\nu}=\frac{1}{\sqrt{-g}}\frac{\gd\sqrt{-g}{\mf L}_{matter}}{\gd
g_{\mu\nu}};\,\,
16\pi J^{\mu}=\frac{\gd {\mf L}_{matter}}{\gd\phi_{\mu}};\,\,\psi=\frac{\gd{\mf
L}_{matter}}{\gd \gb}
\end{equation}
For the equations of motion of matter and the trace of the EM tensor one uses invariance of
the action under coordinate and gauge transformations, leading to
\bq\label{106}
{}^W\na_{\nu}{\mf T}^{\mu\nu}-{\mf
T}\frac{\na^{\mu}\gb}{\gb}=J_{\ga}\phi^{\ga\mu}-
\left(\phi^{\mu}+\frac{\na^{\mu}\gb}{\gb}\right){}^W\na_{\ga}J^{\ga};
\end{equation}
$$16\pi{\mf T}-16\pi{}^W\na_{\mu}J^{\mu}-\gb\psi=0$$
The first relation is a geometrical identity (Bianchi identity) and the second shows the
mutual dependence of the field equations.  Note that in the Weyl-Dirac theory the Weyl
vector does not couple to spinors so $\phi_{\mu}$ cannot be interpreted as the EM potential;
the Weyl vector is used as part of the spacetime geometry and the auxillary field (gauge field)
$\gb$ represents the quantum mass field.  The gravity fields $g_{\mu\nu}$ and $\phi_{\mu}$ and
the quantum mass field determine the spacetime geometry.   Now one constructs a Bohmian 
quantum gravity which is conformally invariant in the framework of Weyl geometry.  If the
model has mass this must be a field (since mass has non-zero Weyl weight).  The Weyl-Dirac 
action is a general Weyl invariant action as above and for simplicity now assume the matter
Lagrangian does not depend on the Weyl vector so that $J_{\mu}=0$.  The equations of motion
are then
\bq\label{107}
{\mf G}^{\mu\nu}=-\frac{8\pi}{\gb^2}({\mf
T}^{\mu\nu}+M^{\mu\nu})+\frac{2}{\gb}(g^{\mu\nu}{}^W\na^{\ga}{}^W\na_{\ga}\gb-{}^W\na^{\mu}{}^W\na^{\nu}
\gb)+
\end{equation}
$$+\frac{1}{\gb^2}(4\na^{\mu}\gb\na^{\nu}\gb-g^{\mu\nu}\na^{\ga}\gb\na_{\ga}\gb)+\frac{\gs}
{\gb^2}\left(\gb^{;\mu}\gb^{;\nu}-\frac{1}{2}g^{\mu\nu}\gb^{;\ga}\gb_{;\ga}\right);$$
$${}^W\na_{\nu}F^{\mu\nu}=\frac{1}{2}\gs(\gb^2\phi^{\mu}+\gb\na^{\mu}\gb);\,\,{\mc R}=
-(\gs+6)\frac{{}^W\bx\gb}{\gb}+\gs\phi_{\ga}\phi^{\ga}-\gs{}^W\na^{\ga}\phi_{\ga}
+\frac{\psi}{2\gb}$$
The symmetry conditions are 
\bq\label{108}
{}^W\na_{\nu}{\mf T}^{\mu\nu}-{\mf T}
(\na^{\mu}\gb/\gb)=0;\,\,16\pi{\mf T}-\gb\psi=0
\end{equation}
(recall ${\mf T}={\mf
T}_{\mu\nu}^{\mu\nu}$).   One notes that from \eqref{107}
results ${}^W\na_{\mu}(\gb^2\phi^{\mu}+\gb\na^{\mu}\gb)=0$ 
so $\phi_{\mu}$ is
not independent of $\gb$.  To see how this is related to the Bohmian quantum theory one
introduces a quantum mass field and shows it is proportional to the Dirac field.  Thus
using \eqref{107} and \eqref{108} one has
\bq\label{109}
\bx\gb+\frac{1}{6}\gb{\mc R}=\frac{4\pi}{3}\frac{{\mf
T}}{\gb}+\gs\gb\phi_{\ga}\phi^{\ga}+2(\gs-6)\phi^{\gag}\na_{\gag}\gb+\frac{\gs}{\gb}
\na^{\mu}\gb\na_{\mu}\gb
\end{equation}
This can be solved iteratively via
\bq\label{110}
\gb^2=(8\pi{\mf T}/{\mc R})-\{1/[({\mc
R}/6)-\gs\phi_{\ga}\phi^{\ga}]\}\gb\bx\gb+\cdots
\end{equation}
Now assuming ${\mf
T}^{\mu\nu}=\rho u^{\mu}u^{\nu}$ (dust with ${\mf T}=\rho$) we multiply \eqref{108} by
$u_{\mu}$ and sum to get 
\bq\label{111}
{}^W\na_{\nu}(\rho u^{\nu})-\rho(u_{\mu}\na^{\mu}\gb/\gb)=0
\end{equation}
Then put \eqref{108}
into \eqref{111} which yields
\bq\label{112}
u^{\nu}{}^W\na_{\nu}u^{\mu}=(1/\gb)(g^{\mu\nu}-u^{\mu}u^{\nu})\na_{\nu}\gb
\end{equation}
To see this write (assuming $g^{\mu\nu}\na_{\nu}\gb=\na^{\mu}\gb$)
\bq\label{113}
{}^W\na_{\nu}(\rho
u^{\mu}u^{\nu})=u^{\mu}{}^W\na_{\nu}\rho u^{\mu}+\rho u^{\nu}{}^W\na_{\nu}u^{\mu}\Rightarrow
\end{equation}
$$\Rightarrow
u^{\mu}\left(\frac{u_{\mu}\na^{\mu}\gb}{\gb}\right)+u^{\nu}{}^W\na_{\nu}u^{\mu}-
\frac{\na^{\mu}\gb}{\gb}=0\Rightarrow u^{\nu}{}^W\na_{\nu}u^{\mu}=(1-u^{\mu}u_{\mu})\frac
{\na^{\mu}\gb}{\gb}=$$
$$(g^{\mu\nu}-u^{\mu}u_{\mu}g^{\mu\nu})\frac{\na_{\nu}\gb}{\gb}=(g^{\mu\nu}-
u^{\mu}u^{\nu})\frac{\na_{\nu}\gb}{\gb}$$
which is \eqref{111}.
Then from \eqref{110}
\bq\label{114}
\gb^{2(1)}=\frac{8\pi{\mf T}}{{\mc R}};\,\,\gb^{2(2)}=\frac{8\pi{\mf T}}{{\mc R}}
\left(1-\frac{1}{({\mc R}/6)-\gs\phi_{\ga}\phi^{\ga}}\frac{\bx\sqrt{{\mf T}}}{\sqrt{{\mf
T}}}\right);\cdots
\end{equation}
Comparing with
\eqref{79} and \eqref{67} shows that we have the correct equations for the Bohmian theory
provided one identifies
\bq\label{115}
\gb\sim{\mf M};\,\,\frac{8\pi{\mf T}}{{\mc R}}\sim
m^2;\,\,\frac{1}{\gs\phi_{\ga}\phi^{\ga}-({\mc R}/6)}\sim \ga
\end{equation}
Thus $\gb$ is the Bohmian quantum mass field and the coupling constant $\ga$ (which depends
on $\hbar$) is also a field, related to geometrical properties of spacetime.  One notes that
the quantum effects and the length scale of the spacetime are related.  To see this suppose
one is in a gauge in which the Dirac field is constant; apply a gauge transformation to
change this to a general spacetime dependent function, i.e.
\bq\label{116}
\gb=\gb_0\to\gb(x)=
\gb_0exp(-\Xi(x));\,\,\phi_{\mu}\to\phi_{\mu}+\pp_{\mu}\Xi
\end{equation}
Thus the gauge in which the
quantum mass is constant (and the quantum force is zero) and the gauge in which the quantum
mass is spacetime dependent are related to one another via a scale change.  In particular
$\phi_{\mu}$ in the two gauges differ by $-\na_{\mu}(\gb/\gb_0)$ and since $\phi_{\mu}$ is a
part of Weyl geometry and the Dirac field represents the quantum mass one concludes that the
quantum effects are geometrized (cf. also \eqref{107} which shows that $\phi_{\mu}$ is not
independent of $\gb$ so the Weyl vector is determined by the quantum mass and thus the
geometrical aspect of the manifold is related to quantum effects).

\section{MORE ON KLEIN GORDON EQUATIONS}
\renewcommand{\theequation}{3.\arabic{equation}}
\setcounter{equation}{0}

We give several approaches here, from various points of view.

\subsection{BERTOLDI-FARAGGI-MATONE THEORY}

The equivalence principle (EP) of Faraggi-Matone (cf. \cite{b9,ch,c8,c9,f3}) is based on
the idea that all physical systems can be connected by a coordinate transformation to
the free situation with vanishing energy (i.e. all potentials are equivalent under
coordinate transformations).  This automatically leads to the quantum stationary
Hamilton-Jacobi equation (QSHJE) which is a third order nonlinear differential
equation providing a trajectory representation of quantum mechanics (QM).  The theory
transcends in several respects the Bohm theory and in particular utilizes a Floydian
time (cf. \cite{f7,f8})
leading to $\dot{q}=p/m_Q\ne p/m$ where
$m_Q=m(1-\partial_EQ)$ is the ``quantum mass" and Q the ``quantum potential". 
Thus the EP is reminscient of the Einstein equivalence of relativity theory.  This
latter served as a midwife to the birth of relativity but was somewhat inaccurate in its
original form.  It is better put as saying that all laws of physics should be invariant
under general coordinate transformations (cf. \cite{o1}).  This demands that not only
the form but also the content of the equations be unchanged.  More precisely the
equations should be covariant and all absolute constants in the equations are to be left
unchanged (e.g. $c,\,\hbar,\,e,\,m$ and $\eta_{\mu\nu}=$ Minkowski tensor).  
Now for the EP, the classical picture with $S^{cl}(q,Q^0,t)$ the Hamilton
principal function ($p=\partial S^{cl}/\partial q$) and $P^0,\,Q^0$ playing the role of
initial conditions involves the classical HJ equation (CHJE) $H(q,p=
(\partial S^{cl}/\partial q),t)+(\partial S^{cl}/\partial t)=0$.  For time independent V one
writes
$S^{cl}=S_0^{cl}(q,Q^0)-Et$ and arrives at the classical stationary HJ equation
(CSHJE) $(1/2m)(\partial S_0^{cl}/\partial q)^2+{\mathfrak W}=0$ where
${\mathfrak W}=V(q)-E$.  In the Bohm theory one looked at Schr\"odinger equations 
$i\hbar\psi_t=-(\hbar^2/2m)\psi''+V\psi$ with $\psi=\psi(q)exp(-iEt/\hbar)$
and
$\psi(q)=R(qexp(i\hat{W}/\hbar)$ leading to 
\bq\label{2.25}
\left(\frac{1}{2m}\right)(\hat{W}')^2+V-E-\frac{\hbar^2R''}{2mR}=0;\,\,
(R^2\hat{W}')'=0
\end{equation}
where $\hat{Q}=-\hbar^2R''/2mR$ was called the quantum potential; this
can be written in the Schwartzian form 
$\hat{Q}=(\hbar^2/4m)\{\hat{W};q\}$ (via $R^2\hat{W}'=c$).  Here 
$\{f;q\}=(f'''/f')-(3/2)(f''/f')^2$.  Writing ${\mathfrak W}=V(q)-E$ as in 
above we have the quantum stationary HJ equation (QSHJE)
\bq\label{3.3a}
(1/2m)(\partial\hat{W}'/\partial q)^2+{\mathfrak W}(q)+\hat{Q}(q)=0
\equiv {\mathfrak W}=-(\hbar^2/4m)\{exp(2iS_0/\hbar);q\}
\end{equation}
This was 
worked out in the Bohm school (without the Schwarzian connections) but $\psi=Rexp(i\hat{W}/\hbar)$
is not appropriate for all situations and care must be taken ($\hat{W}=constant$ must be excluded
for example - cf. \cite{f3,f7,f8}).
The technique of Faraggi-Matone (FM) is completely general and with only the EP as
guide one exploits the relations between Schwarzians, Legendre duality, and the
geometry of a second order differential operator $D_x^2+V(x)$ (M\"obius
transformations play an important role here) to arrive at the QSHJE in the form
\bq\label{2.26}
\frac{1}{2m}\left(\frac{\partial S_0^v(q^v)}{\partial q^v}\right)^2+{\mathfrak W}(q^v)+
{\mathfrak Q}^v(q^v)=0
\end{equation}
where $v:\,q\to q^v$ represents an arbitrary locally invertible coordinate
transformation.  Note in this direction for example that the Schwarzian derivative
of the the ratio of two linearly independent elements in $ker(D^2_x+V(x))$ is twice
$V(x)$.  In particular given an arbitrary system with coordinate $q$ and reduced
action $S_0(q)$ the system with coordinate $q^0$ corresponding to $V-E=0$ involves 
${\mathfrak W}(q)=(q^0;q)$ where $(q^0,q)$ is a cocycle term which has
the form $(q^a;q^b)=-(\hbar^2/4m)\{q^a;q^b\}$.  In fact it can be said
that the essence of the EP is the cocycle conditon 
\bq\label{3.4a}
(q^a;q^c)=(\partial_{q^c}q^b)^2[(q^a;q^b)-(q^c;q^b)]
\end{equation}
In addition FM developed a theory of $(x,\psi)$ duality (cf. \cite{f2})) which
related the space coordinate and the wave function via a prepotential (free energy)
in the form ${\mathfrak F}=(1/2)\psi\bar{\psi}+iX/\epsilon$ for example.
A number of interesting philosophical points arise (e.g. the emergence of space from
the wave function) and we connected this to various features of
dispersionless KdV in \cite{ch,c9} in a sort of extended WKB spirit.  One should
note here that although a form $\psi=Rexp(i\hat{W}/\hbar)$ is not generally appropriate it is
correct when one is dealing with two independent solutions of the Schr\"odinger equation
$\psi$ and $\bar{\psi}$ which are not proportional.  In this context we utilized
some interplay between various geometric properties of KdV which involve the Lax
operator $L^2=D_x^2+V(x)$ and of course this is all related to Schwartzians,
Virasoro algebras, and vector fields on $S^1$ (see e.g. \cite{ch,c7,c9,c38,c39}).
Thus the simple presence of the
Schr\"odinger equation (SE) in QM automatically incorporates a host of geometrical
properties of $D_x=d/dx$ and the circle $S^1$.  In fact since the FM theory exhibits the
fundamental nature of the  SE via its geometrical properties connected to the QSHJE one
could speculate about trivializing QM (for 1-D) to a study of $S^1$ and $\partial_x$.
\\[3mm]\indent
We import here some comments based on \cite{b9} concerning the Klein-Gordon (KG)
equation and the equivalence principle (EP)  
(details are in \cite{b9} and cf. also \cite{d20}).
One starts with the relativistic classical Hamilton-Jacobi equation (RCHJE) with a
potential $V(q,t)$ given as 
\bq\label{2.27}
\frac{1}{2m}\sum_1^D(\pp_kS^{cl}(q,t))^2+{\mf W}_{rel}(q,t)=0;
\end{equation}
$${\mf W}_{rel}(q,t)=
\frac{1}{2mc^2}[m^2c^4-(V(q,t)+\pp_tS^{cl}(q,t))^2]$$
In the time-independent case one has $S^{cl}(q,t)=S_0^{cl}(q)-Et$ and \eqref{2.26}
becomes 
\bq\label{2.28}
\frac{1}{2m}\sum_1^D(\pp_kS_0^{cl})^2+{\mf W}_{rel}=0;\,\,{\mf W}_{rel}(q)=
\frac{1}{2mc^2}[m^2c^4-(V(q)-E)^2]
\end{equation}
In the latter case one can go through the same steps as in the nonrelativistic case and 
the relativistic quantum HJ equation (RQHJE) becomes 
\bq\label{3.6a}
(1/2m)(\na
S_0)^2+{\mf W}_{rel}-(\hbar^2/2m)(\gD R/R)=0;\,\, \na\cdot(R^2\na S_0)=0
\end{equation}
these
equations imply the stationary KG equation 
\bq\label{3.6b}
-\hbar^2c^2\gD
\psi+(m^2c^4-V^2+2EV-E^2)\psi=0
\end{equation}
where
$\psi=Rexp(iS_0/\hbar)$.  Now in the time dependent case the (D+1)-dimensional RCHJE
is ($\eta^{\mu\nu}=diag(-1,1,\cdots,1)$
\bq\label{3.6c}
(1/2m)\eta^{\mu\nu}\pp_{\mu}S^{cl}\pp_{\nu}S^{cl}
+{\mf W}'_{rel}=0;
\end{equation}
$${\mf W}'_{rel}=(1/2mc^2)[m^2c^4-V^2(q)-2cV(q)\pp_0S^{cl}(q)]$$
with $q=(ct,q_1,\cdots,q_D)$.  Thus \eqref{3.6c} has the same structure as \eqref{2.28}
with Euclidean metric replaced by the Minkowskian one.  We know how to implement the EP
by adding Q via $(1/2m)(\pp S)^2+{\mf W}_{rel}+Q=0$ (cf. \cite{f3} and
remarks above).  Note now that ${\mf W}'_{rel}$ depends on $S^{cl}$ requires an
identification
\bq\label{3.6d}
{\mf W}_{rel}=(1/2mc^2)[m^2c^4-V^2(q)-2cV(q)\pp_0S(q)]
\end{equation}
($S$ replacing $S^{cl}$) and
implementation of the EP requires that for an arbitrary ${\mf W}^a$ state ($q\sim q^a$)
one must have 
\bq\label{3.6e}
{\mf W}_{rel}^b(q^b)=(p^b|p^a){\mf W}_{rel}^a(q^a)
+(q^q;q^b);\,\,Q^b(q^b)=(p^b|p^a)Q(q^a)-(q^a;q^b)
\end{equation}
where 
\bq\label{3.6f}
(p^b|p)=[\eta^{\mu\nu}p_{\mu}^bp_{\nu}^b/\eta^{\mu\nu}p_{\mu}p_{\nu}]=
p^TJ\eta J^Tp/p^T\eta p;\,\,J_{\nu}^{\mu}=\pp q^{\mu}/\pp q^{b^{\nu}}
\end{equation}
(J is a Jacobian and these formulas are the natural multidimensional generalization
- see \cite{b9} for details).  Furthermore there is a cocycle condition
$(q^a;q^c)=(p^c|p^b)[(q^a;q^b)-(q^c;q^b)]$.
\\[3mm]\indent
Next one shows that ${\mf W}_{rel}=(\hbar^2/2m)[\bx
(Rexp(iS/\hbar))/Rexp(iS/\hbar)]$ and hence the corresponding quantum potential
is $Q_{rel}=-(\hbar^2/2m) [\bx R/R]$.  Then the RQHJE becomes
$(1/2m)(\pp S)^2+{\mf W}_{rel} +Q=0$ with $\pp\cdot (R^2\pp S)=0$ (here $\bx
R=\pp_{\mu}\pp^{\mu}R$) and this reduces to the standard SE in the classical limit $c\to
\infty$
(note $\pp\sim (\pp_0,\pp_1,\cdots,\pp_D)$ with $q_0=ct,$ etc. - cf. \eqref{3.6c}).
To see how the EP
is simply implemented one considers the so called minimal coupling prescription for an
interaction with an electromagnetic four vector
$A_{\mu}$.  Thus set
$P_{\mu}^{cl}=p_{\mu}^{cl}+eA_{\mu}$ where $p_{\mu}^{cl}$ is a particle momentum and
$P_{\mu}^{cl}=\pp_{\mu} S^{cl}$ is the generalized momentum.  Then the RCHJE reads
as $(1/2m)(\pp S^{cl}-eA)^2+(1/2)mc^2=0$ where $A_0=-V/ec$.  Then
${\mf W}=(1/2)mc^2$ and the critical case ${\mf W}=0$ corresponds to the
limit situation where $m=0$.  One adds the standard Q correction for implementation of
the EP to get
$(1/2m)(\pp S-eA)^2+(1/2)mc^2+Q=0$ and there are transformation
properties (here $(\pp S - eA)^2\sim \sum (\pp_{\mu}S-eA_{\mu})^2$)
\bq\label{2.29}
{\mf W}(q^b)=(p^b|p^a){\mf W}^a(q^a)+(q^a;q^b);\,\,Q^b(q^b)=(p^q|p^a)Q^a(q^a)-
(q^a;q^b)
\end{equation}
$$(p^b|p)=\frac{(p^b-eA^b)^2}{(p-eA)^2}=\frac{(p-eA)^TJ\eta J^T(p-eA)}{(p-eA)^T
\eta (p-eA)}$$
Here J is a Jacobian $J_{\nu}^{\mu}=\pp q^{\mu}/\pp q^{b^{\nu}}$ and
this all implies the cocycle condition again.  One finds now that
(recall $\pp\cdot (R^2(\pp S-eA))=0$ - continuity equation)
\bq\label{2.30}
(\pp S-eA)^2=\hbar^2\left(\frac{\bx R}{R}-\frac{D^2(Re^{iS/\hbar})}
{Re^{iS/\hbar}}\right);\,\,D_{\mu}=\pp_{\mu}-\frac{i}{\hbar}eA_{\mu}
\end{equation}
and it follows that
\bq\label{2.31}
{\mf W}=\frac{\hbar^2}{2m}\frac{D^2(Re^{iS/\hbar})}{Re^{iS/\hbar}};\,\,Q=-\frac
{\hbar^2}{2m}\frac{\bx R}{R};\,\,D^2=\bx-\frac{2ieA\pp}{\hbar}-\frac{e^2A^2}{\hbar^2}
-\frac{ie\pp A}{\hbar}
\end{equation}
\bq\label{2.32}
(\pp S-eA)^2+m^2c^2-\hbar^2\frac{\bx R}{R}=0;\,\,\pp\cdot (R^2(\pp S-eA))=0
\end{equation}
Note also that \eqref{3.6c} agrees with $(1/2m)(\pp S^{cl}-eA)^2+(1/2)mc^2=0$ after setting ${\mf
W}_{rel} =mc^2/2$ and replacing $\pp_{\mu}S^{cl}$ by $\pp_{\mu}S^{cl}-eA_{\mu}$.  One can check
that \eqref{2.32} implies the KG equation $(i\hbar\pp+eA)^2\psi+
m^2c^2\psi=0$ with $\psi=Rexp(iS/\hbar)$.  
\\[3mm]\indent
{\bf REMARK 3.1.}
We extract now a remark about mass generation and the EP from \cite{b3}.  Thus a
special property of the EP is that it cannot be implemented in classical mechanics
(CM) because of the fixed point corresponding to ${\mf W}=0$.  One is forced to
introduce a uniquely determined piece to the classical HJ equation (namely a quantum
potential Q). In the case of the RCHJE the fixed point ${\mf W}(q^0)=0$
corresponds to
$m=0$ and the EP then implies that all the other masses can be generated by a
coordinate transformation.  Consequently one concludes that masses correspond to
the inhomogeneous term in the transformation properties of the ${\mf W}^0$ state, i.e.
$(1/2)mc^2=(q^0;q)$.  Furthermore by \eqref{2.29} masses are expressed
in terms of the quantum potential $(1/2)mc^2=(p|p^0)Q^0(q^0)-Q(q)$.
In particular in \cite{f3} the role of the quantum potential was seen as a sort of
intrinsic self energy which is reminiscent of the relativistic self energy and
this provides a more explicit evidence of such an interpretation.$\hfill\bs$
\\[3mm]\indent
{\bf REMARK 3.2.}
In a previous paper \cite{c10} (working with stationary states and $\psi$
satisfying the Schr\"odinger equation (SE) 
$-(\hbar^2/2m)\psi''+V\psi=E\psi$) we suggested that the notion of
uncertainty in quantum mechanics (QM) could be phrased as incomplete information. 
The background theory here is taken to be the trajectory theory of
Bertoldi-Faraggi-Matone-Floyd as above and the idea in \cite{c10} goes as follows.
First recall that Floydian microstates satisfy a third order quantum stationary
Hamilton-Jacobi equation (QSHJE)
\bq\label{01.1}
\frac{1}{2m}(S_0')^2+{\mf W}(q)+Q(q)=0;\,\,Q(q)=\frac{\hbar^2}{4m}\{S_0;q\};
\end{equation}
$${\mf W}(q)=-\frac{\hbar^2}{4m}\{exp(2iS_0/\hbar);q\}\sim V(q)-E$$
where $\{f;q\}=(f'''/f')-(3/2)(f''/f')^2$ is the Schwarzian and
$S_0$ is the Hamilton principle function.  Also one recalls that the EP of 
Faraggi-Matone can only be implemented when $S_0\ne const$; thus consider
$\psi=Rexp(iS_0/\hbar)$ with $Q=-\hbar^2 R''/2mR$ and $(R^2S_0')'=0$
where $S_0'=p$ and $m_Q\dot{q}=p$ with $m_Q=m(1-\pp_EQ)$ and 
$t\sim\pp_ES_0$.  Thus microstates require three initial or boundary conditions
in general to determine $S_0$ whereas the SE involves only two such conditions.
Hence in dealing with the SE in the standard QM Hilbert space formulation one is
not using complete information about the ``particles" described by microstate
trajectories.  The price of underdetermination is then uncertainty in $q,p,t$ for
example.  In the present note we will make this more precise and add further
discussion.  Following
\cite{c12} we now make this more precise and add further
discussion.
For the stationary SE $-(\hbar^2/2m)\psi''+V\psi=E\psi$ it is shown in \cite{f3} that one has
a general formula
\bq\label{02.1}
e^{2iS_0(\gd)/\hbar}=e^{i\ga}\frac{w+i\bar{\ell}}{w-i\ell}
\end{equation} 
($\gd\sim (\ga,\ell)$) with three integration constants, $\ga,\ell_1,\ell_2$ where
$\ell=\ell_1+i\ell_2$ and $w\sim \psi^D/\psi\in{\bf R}$.
Note $\psi$ and $\psi^D$ are linearly independent solutions of the SE and one
can arrange that $\psi^D/\psi\in{\bf R}$ in describing any situation.
Here $p$ is determined by the two constants in $\ell$ and has a form
\bq\label{02.2}
p=\frac{\pm\hbar\gO\ell_1}{|\psi^D-i\ell\psi|^2}
\end{equation}
(where $w\sim \psi^D/\psi$ above and $\gO=\psi'\psi^D-\psi(\psi^D)'$).
Now let p be determined exactly with $p=p(q,E)$ via the Schr\"odinger equation and
$S_0'$.  Then $\dot{q}=(\pp_Ep)^{-1}$ is also exact so $\gD
q=(\pp_Ep)^{-1}(\tau)\gD t$ for some $\tau$ with $0\leq \tau\leq t$ is exact (up to
knowledge of $\tau$).  Thus given the wave function $\psi$
satisfying the stationary SE
with two boundary conditions at $q=0$ say to fix
uniqueness, one can create a probability density $|\psi|^2(q,E)$ and
the function
$S_0'$.  This determines $p$ uniquely
and hence $\dot{q}$.  The additional constant needed for $S_0$ 
appears in \eqref{02.1} and we can write $S_0=S_0(\ga,q,E)$ since
from \eqref{02.1} one has 
\bq\label{02.2a}
S_0-(\hbar/2)\ga=-(i\hbar/2)log(\gb)
\end{equation}
and 
$\gb=(w+i\bar{\ell})/(w-i\ell)$ with $w=\psi^D/\psi$ is to be considered as known
via a determination of suitable $\psi,\,\psi^D$.  Hence $\pp_{\ga}S_0=
-\hbar/2$ and consequently $\gD S_0\sim\pp_{\ga}S_0\gd\ga=-(\hbar/2)
\gD\ga$ measures the indeterminacy in $S_0$.
\\[3mm]\indent
Let us expand upon this as follows.  Note first that the determination of constants 
necessary to fix $S_0$ from the QSHJE is not usually the same as that
involved in fixing $\ell,\,\bar{\ell}$ in \eqref{02.1}.  In paricular differentiating
in q one gets
\bq\label{02.3}
S_0'=-\frac{i\hbar\beta'}{\beta};\,\,\gb'=-\frac{2i\Re\ell w'}{(w-i\ell)^2}
\end{equation}
Since $w'=-\gO/\psi^2$ where $\gO=\psi'\psi^D-\psi(\psi^D)'$ we get 
$\gb'=-2i\ell_1\gO/(\psi^D-i\ell\psi)^2$ and consequently
\bq\label{02.4}
S_0'=-\frac{\hbar\ell_1\gO}{|\psi^D-i\ell\psi|^2}
\end{equation}
which agrees with p in \eqref{02.2} ($\pm\hbar$ simply indicates direction).  We
see that e.g.
$S_0(x_0)=i\hbar\ell_1\gO/|\psi^D(x_0)-i\ell\psi(x_0)|^2=f(\ell_1,\ell_2,x_0)$
and $S_0''=g(\ell_1,\ell_2,x_0)$ determine the relation between $(p(x_0),
p'(x_0))$ and $(\ell_1,\,\ell_2)$ but they are generally different numbers.  In
any case, taking
$\ga$ to be the arbitrary unknown constant in the determination of $S_0$, we have
$S_0=S_0(q,E,\ga)$ with $q=q(S_0,E,\ga)$ and $t=t(S_0,E,\ga)=
\pp_ES_0$ (emergence of time from the wave function).  One can then write e.g.
\bq\label{02.4a}
\gD q=(\pp q/\pp S_0)(\hat{S}_0,E,\ga)\gD S_0=(1/p)(\hat{q},E)\gD
S_0=-(1/p)(\hat{q},E)(\hbar/2)\gD\ga
\end{equation} 
(for intermediate values ($\hat{S}_0,\,\hat{q}$)) leading to
\begin{theorem}
With p determined uniquely by two ``initial" conditions so that $\gD p$ is determined
and q given via \eqref{02.1} we have from \eqref{02.4a} the inequality 
$\gD p\gD q=O(\hbar)$ which resembles the Heisenberg uncertainty relation.
\end{theorem}
\begin{corollary}
Similarly $\gD t=(\pp t/\pp S_0)(\hat{S}_0,E,\ga)\gD S_0$ for some
intermediate value $\hat{S}_0$ and hence as before 
$\gD E\gD t=O(\hbar)$ ($\gD E$ being precise).
\end{corollary}
Note that there is no physical argument here; one is simply looking at the number of
conditions necessary to fix solutions of a differential equation.  In fact (based on some
corresondence with E. Floyd) it seems somewhat difficult to produce a physical argument.
We refer also to Remark 3.1.2 for additional discussion.
$\hfill\bs$
\\[3mm]\indent
{\bf REMARK 3.3.}
In order to get at the time dependent SE from the BFM (Bertoldi-Faraggi-Matone) theory
we proceed as follows.
From the previous discussion on the KG equation one sees that (dropping the A terms)
in the time independent case one has $S^{cl}(q,t)=S_0^{cl}(q)-Et$ 
\bq\label{10.110}
(1/2m)\sum_1^D(\pp_kS_0^{cl})^2+{\mf W}_{rel}=0;\,\, 
{\mf W}_{rel}(q)=(1/2mc^2)[m^2c^4-(V(q)-E)^2]
\end{equation}
leading to a stationary RQHJE 
\bq\label{10.11a}
(1/2m)(\na S_0)^2+{\mf W}_{rel}-(\hbar^2/2m)(\gD R/R)=0;\,\, 
\na\cdot(R^2\na S_0)=0
\end{equation}
This implies also the stationary KG equation
\bq\label{10.11b}
-\hbar^2c^2\gD\psi+(m^2c^4-V^2+2VE-E^2)\psi=0
\end{equation}
Now in the time dependent case one can write $(1/2m)\eta^{\mu\nu}
\pp_{\mu}S^{cl}\pp_{\nu}S^{cl}+{\mf W}'_{rel}=0$ where $\eta\sim diag(-1,1,\cdots,1)$
and 
\bq\label{10.11c}
{\mf W}'_{rel}(q)=(1/2mc^2)[m^2c^4-V^2(q)-2cV(q)\pp_0S^{cl}(q)]
\end{equation}
with $q\equiv (ct,q_1,\cdots,q_D)$.  Thus we have the same structure as
\eqref{10.110} with Euclidean metric replaced by a Minkowskian one.  
To implement the EP we have to modify the classical equation by adding a function
to be determined, namely
$(1/2m)(\pp S)^2+{\mf W}_{rel}+Q=0$ ($(\pp S)^2\sim \sum (\pp_{\mu}S)^2$ etc.).  Observe that
since
${\mf W}'_{rel}$ depends on $S^{cl}$ we have to make the identification ${\mf
W}_{rel}= (1/2mc^2)[m^2c^4-V^2(q)-2cV(q)\pp_0S(q)]$ which differs from ${\mf W}'_{rel}$
since $S$ now appears instead of $S^{cl}$.  Implementation of the EP requires that for
an arbitrary ${\mf W}^a$ state 
\bq\label{10.11d}
{\mf W}^b_{rel}(q^b)=(p^b|p^a){\mf W}^
a_{rel}(q^a)+(q^q;q^b);\,\,Q^b(q^b)=(p^b|p^a)Q^a(q^a)-(q^a;q^b)
\end{equation}
where now
$(p^b|p)=\eta^{\mu\nu}p_{\mu}^bp_{\nu}^b/\eta^{\mu\nu}p_{\mu}p_{\nu}=
p^TJ\eta J^Tp/p^T\eta p$ and $J^{\mu}_{\nu}=\pp q^{\mu}/\pp (q^b)^{\nu}$.  This leads
to the cocycle condition $(q^a;q^c)=(p^c|p^b)[(q^q;q^b)-(q^c;q^b)]$ as before.
Now consider the identity 
\bq\label{10.11e}
\ga^2(\pp S)^2=\Box(Rexp(\ga S))/Rexp(\ga
S)-(\Box R/R)-(\ga\pp\cdot(R^2\pp S)/R^2)
\end{equation}
and if R satisfies the continuity equation
$\pp\cdot (R^2\pp S)=0$ one sets
$\ga=i/\hbar$ to obtain
\bq\label{10.12}
\frac{1}{2m}(\pp S)^2=-\frac{\hbar^2}{2m}\frac{\Box(Re^{iS/\hbar})}{Re^{iS/\hbar}}+
\frac{\hbar^2}{2m}\frac{\Box R}{R}
\end{equation}
Then it is shown that ${\mf
W}_{rel}=(\hbar^2/2m)(\Box(Rexp(iS/\hbar))/Rexp (iS/\hbar)$ so there results
$\,\,Q_{rel}=-(\hbar^2/2m)(\Box R/R)$.  Thus the RQHJE has the form
(cf. \eqref{2.30} - \eqref{2.32})
\bq\label{10.13}
\frac{1}{2m}(\pp S)^2+{\mf W}_{rel}-\frac{\hbar^2}{2m}\frac{\Box R}{R}=0;\,\,
\pp\cdot(R^2\pp S)=0
\end{equation}
\indent
Now for the time dependent SE one takes the nonrelativistic limit of the RQHJE.
For the classical limit one makes the usual substitution $S=S'-mc^2t$ so as $c\to
\infty$ ${\mf W}_{rel}\to (1/2)mc^2+V$ and $-(1/2m)(\pp_0S)^2\to
\pp_tS'-(1/2)mc^2$ with $\pp(R^2\pp S)=0\to m\pp_t(R')^2+\na\cdot((R')^2\na S')=0$.
Therefore (removing the primes) \eqref{10.13} becomes $(1/2m)(\na S)^2
+V+\pp_tS-(\hbar^2/2m)(\gD R/R)=0$ with the time dependent nonrelativistic continuity
equation being $m\pp_tR^2+\na\cdot(R^2\na S)=0$.  This leads then 
(for $\psi\sim Rexp(iS/\hbar)$) to the SE
\bq\label{10.14}
i\hbar\pp_t\psi=\left(-\frac{\hbar^2}{2m}\gD +V\right)\psi
\end{equation}
One sees from all this that the BFM theory is profoundly governed by the equivalence
principle and produces a usable framework for computation.  It is surprising that it has not
attracted more adherents.$\hfill\bs$

\subsection{KLEIN GORDON \`A LA SANTAMATO}

The derivation of the SE in \cite{sa} (treated in Section 1.1) was modified in
\cite{sb} to a derivation of
the Klein-Gordon (KG) equation via a somewhat different average action principle.  Recall
that the
spacetime geometry in \cite{sa} was obtained from the average action principle to obtain Weyl
connections with a gauge field
$\phi_{\mu}$ (thus the geometry had a statistical origin).  The Riemann scalar curvature 
$\dot{R}$ was then related to the Weyl scalar curvature R via an equation 
\bq\label{063.46}
R=\dot{R}-3[(1/2)g^{\mu\nu}\phi_{\mu}\phi_{\nu}+(1/\sqrt{-g})\pp_{\mu}(\sqrt{-g}g^{\mu\nu}
\phi_{\nu})]
\end{equation}
Explicit reference to the underlying Weyl structure disappears in the
resulting SE and we refer to Remark 1.7 for a few comments (cf. also \cite{c16} for an
incisive review).  We recall also here from
\cite{q1,q2,q3,q4} (cf. \cite{c3,c4,cqm}) that in
the conformal geometry the particles do not follow geodesics of the conformal metric alone;
further the work in \cite{q1,q2,q3,q4} is absolutely fundamental in exhibiting a correct
framework for general relativity via the conformal (Weyl) version.
Summarizing from 
\cite{sa} and the second paper in \cite{sb} one can say that traditional QM is equivalent
(in some sense) to classical statistical mechanics in Weyl spaces.
The moral seems to be (loosly) that quantum mechanics in Riemannian spacetime is the same as
classical statistical mechanics in a Weyl space.  In particular one wants to establish that
traditional QM, based on wave equations and ad hoc probability calculus is
merely a convenient mathematical construction to overcome the complications arising from a
nontrivial spacetime geometric structure.  Here one works from first principles and includes
gauge invariance (i.e. invariance with respect to an arbitrary choice of the spacetime
calibration).  The spacetime is supposed to be a generic 4-dimensional differential manifold
with torsion free connections
$\gG^{\gl}_{\,\mu\nu}=\gG^{\gl}_{\,\nu\mu}$ and a metric tensor $g_{\mu\nu}$ with signature
$(+,-,-,-)$ (one takes $\hbar=c=1$).  Here the (restrictive) hypothesis of assuming a Weyl
geometry from the beginning is released, both the particle motion and the spacetime geometric
structure are derived from a single average action principle.  A result of this approach is
that the spacetime connections are forced to be integrable Weyl connections by the 
extremization principle.  
\\[3mm]\indent
The particle is supposed to undergo a motion in spacetime with deterministic trajectories and
random initial conditions taken on an arbitrary spacelike 3-dimensional hypersurface; thus
the theory describes a relativistic Gibbs ensemble of particles (cf. \cite{hk,sb} for all
this in detail and see also \cite{cqm}).  Both the particle motion and the spacetime
connections can be obtained from the average stationary action principle
\bq\label{3.470}
\gd\left[E\left(\int_{\tau_1}^{\tau_2}L(x(\tau,\dot{x}(\tau))d\tau\right)\right]=0
\end{equation}
This action integral must be parameter invariant, coordinate invariant, and gauge invariant.
All of these requirements are met if L is positively homogeneous of the first degree in 
$\dot{x}^{\mu}=dx^{\mu}/d\tau$ and transforms as a scalar of Weyl type $w(L)=0$.  The
underlying probabiity measure must also be gauge invariant.  A suitable Lagrangian is then
\bq\label{3.480}
L(x,dx)=(m^2-(R/6))^{1/2}ds+A_{\mu}dx^{\mu}
\end{equation}
where $ds=(g_{\mu\nu}\dot{x}^{\mu}\dot{x}^{\nu})^{1/2}d\tau$ is the arc length and R is the
space time scalar curvature; m is a parameterlike scalar field of Weyl type (or
weight) $w(m)=-(1/2)$. The factor 6 is essentially arbitrary and has been chosen for future
convenience.  The vector field $A_{\mu}$ can be interpreted as a 4-potential due to an
externally applied EM field and the curvature dependent factor in front of $ds$ is an
effective particle mass. This seems a bit ad hoc but some feeling for the nature of the
Lagrangian can be obtained from Section 1.1 (cf. also \cite{au}).  The Lagrangian will be
gauge invariant provided the $A_{\mu}$ have Weyl type $w(A_{\mu})=0$.  Now one can split
$A_{\mu}$ into its gradient and divergence free parts $A_{\mu}=\bar{A}_{\mu}-\pp_{\mu}S$,
with both $S$ and $\bar{A}_{\mu}$ having Weyl type zero, and with $\bar{A}_{\mu}$ interpreted
as and EM 4-potential in the Lorentz gauge. Due to the nature of the action principle
regarding fixed endpoints in variation one notes that the average action principle is not
invariant under EM gauge transformations $A_{\mu}\to A_{\mu}+\pp_{\mu}S$; but one knows that
QM is also not invariant under EM gauge  transformations (cf. \cite{ah}) so there is no
incompatability with QM here.
\\[3mm]\indent
Now the set of all spacetime trajectories accessible to the particle (the particle path
space) may be obtained from \eqref{3.470} by performing the variation with respect to the
particle trajectory with fixed metric tensor, connections, and an underlying probability
measure.  Thus (cf. \cite{cqm,hk,sb}) the solution is given by the so-called Carath\'eodory
complete figure associated with the Lagrangian
\bq\label{3.490}
\bar{L}(x,dx)=(m^2-(R/6))^{1/2}ds+\bar{A}_{\mu}dx^{\mu}
\end{equation}
(note this leads to the same equations as \eqref{3.480} since the Lagrangians differ by a
total differential $dS$).  The resulting complete figure is a geometric entity formed by a
one parameter family of hypersurfaces $S(x)=const.$ where $S$ satisfies the HJ equation
\bq\label{3.500}
g^{\mu\nu}(\pp_{\mu}S-\bar{A}_{\mu})(\pp_{\nu}S-\bar{A}_{\nu})=m^2-\frac{R}{6}
\end{equation}
and by a congruence of curves intersecting this family given by
\bq\label{3.510}
\frac{dx^{\mu}}{ds}=\frac{g^{\mu\nu}(\pp_{\nu}S-\bar{A}_{\nu})}
{[g^{\rho\gs}(\pp_{\rho}S-\bar{A}_{\rho})(\pp_{\gs}S-\bar{A}_{\gs})]^{1/2}}
\end{equation}
The congruence yields the actual particle path space and the underlying probability measure
on the path space may be defined on an arbitrary 3-dimensional hypersurface intersecting all
of the members of the congruence without tangencies (cf. \cite{hk}).  The measure
will be completely identified by its probability current density $j^{\mu}$ (see
\cite{cqm,sb}).  Moreover, since the measure is independent of the arbitrary
choice of the hypersurface, $j^{\mu}$ must be conserved, i.e. $\pp_{\mu}j^{\mu}=0$.
Since the trajectories are deterministically defined by \eqref{3.510}, $j^{\mu}$ must be
parallel to the particle 4-velocity \eqref{3.510}, and hence
\bq\label{3.520}
j^{\mu}=\rho\sqrt{-g}g^{\mu\nu}(\pp_{\nu}S-\bar{A}_{\nu})
\end{equation}
with some $\rho>0$.  Now gauge invariance of the underlying measure as well as of the
complete figure requires that $j^{\mu}$ transforms as a vector density of Weyl type
$w(j^{\mu})=0$ and S as a scalar of Weyl type $w(S)=0$.  From \eqref{3.520} one sees then
that $\rho$ transforms as a scalar of Weyl type $w(\rho)=-1$ and $\rho$ is called the scalar
probability density of the particle random motion. 
\\[3mm]\indent
The actual spacetime affine connections are obtained from \eqref{3.470} by performing the
variation with respect to the fields $\gG^{\gl}_{\,\mu\nu}$ for a fixed metric tensor,
particle trajectory, and probability measure.  It is expedient to tranform the average action
principle to the form of a 4-volume integral
\bq\label{3.530}
\gd\left[\int_{\gO}d^4x[(m^2-(R/6))(g_{\mu\nu}j^{\mu}j^{\nu}]^{1/2}+A_{\mu}j^{\mu}\right]=0
\end{equation}
where $\gO$ is the spacetime region occupied by the congruence \eqref{3.510} and $j^{\mu}$ is
given by \eqref{3.520} (cf. \cite{cqm,sb} for proofs).  Since the connection
fields $\gG^{\gl}_{\,\mu\nu}$ are contained only in the curvature term R the variational
problem \eqref{3.530} can be further reduced to 
\bq\label{3.540}
\gd\left[\int_{\gO}\rho R\sqrt{-g}d^4x\right]=0
\end{equation}
(here the HJ equation \eqref{3.500} has been used).  This states that the average spacetime
curvature must be stationary under a variation of the fields $\gG^{\gl}_{\mu\nu}$ (principle
of stationary average curvature).  The extremal connections $\gG^{\gl}_{\,\mu\nu}$
arising from \eqref{3.540} are derived in \cite{sb} using standard field theory techniques
and the result is 
\bq\label{3.550}
\gG^{\gl}_{\,\mu\nu}=\left\{\begin{array}{c}
\gl\\
\mu\,\,\nu
\end{array}\right\}+\frac{1}{2}(\phi_{\mu}\gd^{\gl}_{\nu}+\phi_{\nu}\gd_{\mu}^{\gl}-
g_{\mu\nu}g^{\gl\rho}\phi_{\rho});\,\,\phi_{\mu}=\pp_{\mu}log(\rho)
\end{equation}
This shows that the resulting connections are integrable Weyl connections with a gauge field
$\phi_{\mu}$ (cf. \cite{sa} and Sections 1.1-1.2).  The HJ equation \eqref{3.500} and
the continuity equation $\pp_{\mu}j^{\mu}=0$ can be consolidated in a single complex equation
for S, namely
\bq\label{3.560}
e^{iS}g^{\mu\nu}(iD_{\mu}-\bar{A}_{\mu})(iD_{\nu}-\bar{A}_{\nu})e^{-iS}-(m^2-(R/6))=0;\,\,
D_{\mu}\rho=0
\end{equation}
Here $D_{\mu}$ is (doubly covariant - i.e. gauge and coordinate invariant) Weyl derivative
given by (cf. \cite{au})
\bq\label{3.570}
D_{\mu}T^{\ga}_{\,\,\gb}=\pp_{\mu}T^{\ga}_{\,\,\gb}+\gG^{\ga}_{\,\mu\gep}T^{\gep}_{\,\,\gb}-
\gG^{\gep}_{\,\mu\gb}T^{\ga}_{\,\,\gep}+w(T)\phi_{\mu}T^{\ga}_{\,\,\gb}
\end{equation}
It is to be noted that the probability density (but not the rest mass) remains constant
relative to $D_{\mu}$.  When written out \eqref{3.560} for a set of two coupled partial
differential equations for $\rho$ and $S$.  To any solution corresponds a particular random
motion of the particle.
\\[3mm]\indent
Next one notes that \eqref{3.560} can be cast in the familiar KG form, i.e.
\bq\label{3.580}
[(i/\sqrt{-g})\pp_{\mu}\sqrt{-g}-\bar{A}_{\mu}]g^{\mu\nu}(i\pp_{\nu}-\bar{A}_{\nu})\psi
-(m^2-(\dot{R}/6))\psi=0
\end{equation}
where $\psi=\sqrt{\rho}exp(-iS)$ and $\dot{R}$ is the Riemannian scalar curvature built out
of $g_{\mu\nu}$ only.  We have the (by now) familiar formula
\bq\label{3.590}
R=\dot{R}-3[(1/2)g^{\mu\nu}\phi_{\mu}\phi_{\nu}+(1/\sqrt{-g})
\pp_{\mu}(\sqrt{-g}g^{\mu\nu}\phi_{\nu})]
\end{equation}
According to point of view {\bf (A)} above in the KG equation \eqref{3.580} any explicit
reference to the underlying spacetime Weyl structure has disappeared; thus the Weyl
structure is hidden in the KG theory.  However we note that no physical meaning is attributed
to $\psi$ or to the KG equation.  Rather the dynamical and statistical behavior of the
particle, regarded as a classical particle, is determined by \eqref{3.560}, which, although
completely equivalent to the KG equation, is expressed in terms of quantities having a more
direct physical interpretation.
\\[3mm]\indent
{\bf REMARK 3.4.}
The formula \eqref{3.590} goes back to Weyl \cite{wy} and the connection of matter to geometry
arises from \eqref{3.550}.  The time variable is treated in a special manner here related to
a Gibbs ensemble and $\rho>0$ is built into the theory.$\hfill\bs$

\subsection{KLEIN GORDON VIA SCALE RELATIVITY}

In \cite{c1,cqm} and Section 1.1 we sketched a few developments in the theory of
scale relativity.  This is by no means the whole story and we want to give a taste of
some further main ideas while deriving the KG equation in this context
(cf. 
\cite{a9,c25,c13,c31,c32,n5,n12,n14,n13,n55,n30,n37}).  A main idea here is that the
Schr\"odinger, Klein-Gordon, and Dirac equations are all geodesic equations in the
fractal framework.  They have the form $D^2/ds^2=0$ where $D/ds$ represents the
appropriate covariant derivative.  The complex nature of the SE and KG equaton arises
from a discrete time symmetry breaking based on nondifferentiability.  For the Dirac
equation further discrete symmetry breakings are needed on the spacetime variables
in a biquaternionic context (cf. here \cite{c25}).  First we go back to
\cite{n5,n12,n30} and sketch some of the fundamentals of scale relativity.  This is a
very rich and beautiful theory extending in both spirit and generality the relativity
theory of Einstein (cf. also \cite{c40} for variations involving Clifford theory).
The basic idea here is that (following Einstein) the laws of nature apply whatever
the state of the system and hence the relevant variables can only be defined
relative to other states.  Standard scale laws of power-law type correspond to
Galilean scale laws and from them one actually recovers quantum mechanics (QM) in a
nondifferentiable space.  The quantum behavior is a manifestation of the fractal
geometry of spacetime. In particular the quantum
potential is a manifestation of fractality in the same way as the Newton potential is
a manifestation of spacetime curvature.  In this spirit one can also conjecture (cf.
\cite{n30}) that this quantum potential may explain various dynamical effects
presently attributed to dark matter (cf. also \cite{a35}).
Now for the KG
equation via scale relativity the derivation in the first paper of \cite{c25}
seems the most concise and we follow that at first (cf. also \cite{n12}).  All of the
elements of the approach for the SE remain valid in the motion relativistic case with
the time replaced by the proper time s, as the curvilinear parameter along the 
geodesics.  Consider a small increment $dX^{\mu}$ of a nondifferentiable four
coordinate along one of the geodesics of the fractal spacetime.  One can decompose
this in terms of a large scale part $\overline{LS}<dX^{\mu}>=dx^{\mu}=v_{\mu}ds$ and
a fluctuation $d\xi^{\mu}$ such that $\overline{LS}<d\xi^{\mu}>=0$.  One is led to
write the displacement along a geodesic of fractal dimension $D=2$ via
\bq\label{04.8}
dX^{\mu}_{\pm}=d_{\pm}x^{\mu}+d\xi^{\mu}_{\pm}=v_{\pm}^{\mu}ds+
u_{\pm}^{\mu}\sqrt{2{\mc D}}ds^{1/2}
\end{equation}
Here $u_{\pm}^{\mu}$ is a dimensionless
fluctuation andd the length scale $2{\mc D}$ is introduced for dimensional purposes.
The large scale forward and backward derivatives $d/ds_{+}$ and $d/ds_{-}$ are
defined via 
\bq\label{04.9}
\frac{d}{ds_{\pm}}f(s)=lim_{s\to 0_{\pm}}\ol{LS}\left<\frac{f(s+\gd s)-f(s)}
{\gd s}\right>
\end{equation}
Applied to $x^{\mu}$ one obtains the forward and backward large scale four velocities
of the form
\bq\label{04.10}
(d/dx_{+})x^{\mu}(s)=v_{+}^{\mu};\,\,(d/ds_{-})x^{\mu}=v_{-}^{\mu}
\end{equation}
Combining yields 
\bq\label{04.11}
\frac{d'}{ds}=\frac{1}{2}\left(\frac{d}{ds_{+}}+\frac{d}{ds_{-}}\right)-\frac{i}{2}
\left(\frac{d}{ds_{+}}-\frac{d}{ds_{-}}\right);
\end{equation}
$${\mc V}^{\mu}=\frac{d'}{ds}x^{\mu}=
V^{\mu}-iU^{\mu}=\frac{v_{+}^{\mu}+v_{-}^{\mu}}{2}-i\frac{v_{+}^{\mu}-v_{-}^{\mu}}{2}$$
For the fluctuations one has
\bq\label{04.12}
\ol{LS}<d\xi_{\pm}^{\mu}d\xi_{\pm}^{\nu}>=\mp 2{\mc
D}\eta^{\mu\nu}ds
\end{equation}
One chooses here $(+,-,-,-)$ for the Minkowski signature for
$\eta^{\mu\nu}$ and there is a mild problem because the diffusion (Wiener) process makes
sense only for positive definite metrics.  Various solutions have been given and they are all
basically equivalent, amounting to the transformatin a Laplacian into a  D'Alembertian.  Thus
the two forward and backward differentials of
$f(x,s)$ should be written as
\bq\label{04.13}
(df/ds_{\pm})=(\pp_s+v_{\pm}^{\mu}\pp_{\mu}\mp{\mc D}
\pp^{\mu}\pp_{\mu})f
\end{equation}
One considers now only stationary functions f, not depending
explicitly on the proper time s, so that the complex covariant derivative operator
reduces to 
\bq\label{04.130}
(d'/ds)=({\mc V}^{\mu}+i{\mc D}\pp^{\mu})\pp_{\mu}
\end{equation}
\indent
Now assume that the large scale part of any mechanical system can be characterized by
a complex action ${\mf S}$ leading one to write 
\bq\label{04.14}
\gd{\mf
S}=-mc\gd\int_a^bds=0;\,\,ds=\ol{LS}<\sqrt{dX^{\nu}dX_{\nu}}>
\end{equation}
This leads to
$\gd{\mf S}=-mc\int_a^b{\mc V}_{\nu}d(\gd x^{\nu})$ with $\gd x^{\nu}
=\ol{LS}<dX^{\nu}>$.  Integrating by parts yields 
\bq\label{04.15}
\gd{\mf S}=-[mc\gd
x^{\nu}]_a^b+mc\int_a^b\gd x^{\nu}(d{\mc V}_{\mu}/ds)ds
\end{equation}
To get the equations of
motion one has to determine $\gd {\mf S}=0$ between the same two points, i.e.
at the limits $(\gd x^{\nu})_a=(\gd x^{\nu})_b=0$.  From \eqref{04.15} one obtains then
a differential geodesic equation $d{\mc V}/ds=0$.  One can also write
the elementary variation of the action as a functional of the coordinates.  So
consider the point a as fixed so $(\gd x^{\nu})_a=0$ and consider b as variable.
The only admissable solutions are those satisfying the equations of motion so the
integral in \eqref{04.15} vanishes and writing $(\gd x^{\nu})_b$ as $\gd x^{\nu}$ gives
$\gd{\mf S}=-mc{\mc V}_{\nu}\gd x^{\nu}$ (the minus sign comes from
the choice of signature).  The complex momentum is now 
\bq\label{04.16}
{\mc P}_{\nu}=
mc{\mc V}_{\nu}=-\pp_{\nu}{\mf S}
\end{equation}
and the complex action completely characterizes the
dynamical state of the particle.  Hence introduce a wave function $\psi=exp(i{\mf S}/
{\mf S}_0)$ and via \eqref{04.16} one gets 
\bq\label{04.17}
{\mc V}_{\nu}=(i{\mf
S}_0/mc)\pp_{\nu}log(\psi)
\end{equation}
Now for the scale relativistic prescription replace the
derivative in $d/ds$ by its covariant expression $d'/ds$.  Using \eqref{04.17}
one transforms $d{\mc V}/ds=0$ into 
\bq\label{04.18}
-\frac{{\mf S}_0^2}{m^2c^2}\pp^{\mu}log(\psi)\pp_{\mu}\pp_{\nu}log(\psi)-
\frac{{\mf S}_0{\mc D}}{mc}\pp^{\mu}\pp_{\mu}\pp_{\nu}log(\psi)=0
\end{equation}
The choice ${\mf S}_0=\hbar=2mc{\mc D}$ allows a simplification of \eqref{04.18} when
one uses the identity
\bq\label{04.19}
\frac{1}{2}\left(\frac{\pp_{\mu}\pp^{\mu}\psi}{\psi}\right)=\left(\pp_{\mu}log(\psi)
+\frac{1}{2}\pp_{\mu}\right)\pp^{\mu}\pp^{\nu}log(\psi)
\end{equation}
Dividing by ${\mc D}^2$ one obtains the equation of motion for the free particle 
$\pp^{\nu}[\pp^{\mu}\pp_{\mu}\psi/\psi]=0$.  Therefore the KG
equation (no electromagnetic field) is 
\bq\label{04.20}
\pp^{\mu}\pp_{\mu}\psi+(m^2c^2/\hbar^2)\psi=0
\end{equation}
and this becomes an
integral of motion of the free particle provided the integration constant is chosen in
terms of a squared mass term $m^2c^2/\hbar^2$.  Thus the quantum behavior described
by this equation and the probabilistic interpretation given to $\psi$ is reduced
here to the description of a free fall in a fractal spacetime, in analogy with
Einstein's general relativity.  Moreover these equations are covariant since the
relativistic quantum equation written in terms of $d'/ds$ has the same form as the
equation of a relativistic macroscopic and free particle using $d/ds$.  One notes
that the metric form of relativity, namely $V^{\mu}V_{\mu}=1$ is not conserved in
QM and it is shown in \cite{p5} that the free particle KG equation expressed in
terms of ${\mc V}$ leads to a new equality 
\bq\label{04.21}
{\mc V}^{\mu}{\mc
V}_{\mu}+2i{\mc D}\pp^{\mu}{\mc V}_{\mu}=1
\end{equation}
In the scale relativistic framework
this expression defines the metric that is induced by the internal scale structures
of the fractal spacetime.  In the absence of an electromagnetic field ${\mc
V}^{\mu}$ and ${\mf S}$ are related by \eqref{04.16} which can be writen as 
${\mc V}_{\mu}=-(1/mc)\pp_{\mu}{\mf S}$ so \eqref{04.21} becomes 
\bq\label{04.22}
\pp^{\mu}{\mf S}\pp_{\mu}{\mf S}-2imc{\mc D}\pp^{\mu}\pp_{\mu}{\mf
S}=m^2c^2
\end{equation}
which is the new form taken by the Hamilton-Jacobi equation.
\\[3mm]\indent
{\bf REMARK 3.5.}
We go back to \cite{n12,p5} now and repeat some of their steps in a perhaps more
primitive but revealing form.  Thus one omits the $\ol{LS}$ notation and uses
$\gl\sim 2{\mc D}$; equations \eqref{04.8} - \eqref{04.130} and \eqref{04.11} are the
same and one writes now ${\mf d}/ds$ for $d'/ds$.  Then ${\mf d}/ds={\mc
V}^{\mu}\pp_{\mu}+(i\gl/2)\pp^{\mu}\pp_{\mu}$ plays the role of a scale covariant
derivative and one simply takes the equation of motion of a free relativistic
quantum particle to be given as $({\mf d}/ds){\mc V}^{\nu}=0$, which
can be interpreted as the equations of free motion in a fractal spacetime or as
geodesic equations.  In fact now $({\mf d}/ds){\mc V}^{\nu}=0$ leads directly to the KG
equation upon writing $\psi=exp(i{\mf S}/mc\gl)$ and ${\mf P}^{\mu}=-\pp^{\mu}{\mf S}=
mc{\mc V}^{\mu}$ so that $i{\mf S}=mc\gl log(\psi)$ and ${\mc V}^{\mu}=
i\gl\pp^{\mu}log(\psi)$.  Then
\bq\label{04.23}
\left({\mc
V}^{\mu}\pp_{\mu}+\frac{i\gl}{2}\pp^{\mu}\pp_{\mu}\right)\pp^{\nu}log(\psi)=0=
i\gl\left(\frac{\pp^{\mu}\psi}{\psi}\pp_{\mu}+
\frac{1}{2}\pp^{\mu}\pp_{\mu}\right)\pp^{\nu}\log(\psi)
\end{equation}
Now some identities are given in \cite{p5} for aid in calculation here, namely
\bq\label{04.24}
\frac{\pp^{\mu}\psi}{\psi}\pp_{\mu}\frac{\pp^{\nu}\psi}{\psi}=
\frac{\pp^{\mu}\psi}{\psi}\pp^{\nu}\left(\frac{\pp_{\mu}\psi}{\psi}\right)=
\end{equation} 
$$=\frac{1}{2}\pp^{\nu}\left(\frac{\pp^{\mu}\psi}{\psi}\frac{\pp_{\mu}\psi}{\psi}
\right);\,\,\pp_{\mu}
\left(\frac{\pp^{\mu}\psi}{\psi}\right)+\frac{\pp^{\mu}\psi}{\psi}\frac{\pp_{\mu}\psi}
{\psi}=\frac{\pp^{\mu}\pp_{\mu}\psi}{\psi}$$
The first term in the last equation of \eqref{04.23} is then
$(1/2)[(\pp^{\mu}\psi/
\psi)(\pp_{\mu}\psi/\psi)]$ and the second is
\bq\label{04.30}
(1/2)\pp^{\mu}\pp_{\mu}\pp^{\nu}log(\psi)=(1/2)\pp^{\mu}\pp^{\nu}\pp_{\mu}log(\psi)=
\end{equation}
$$=(1/2)\pp^{\nu}\pp^{\mu}\pp_{\mu}log(\psi)=(1/2)\pp^{\nu}
\left(\frac{\pp^{\mu}\pp_{\mu}\psi}{\psi}-\frac{\pp^{\mu}\psi\pp_{\mu}\psi}{\psi^2}
\right)$$
Combining we get $(1/2)\pp^{\nu}(\pp^{\mu}\pp_{\mu}\psi/\psi)=0$
which integrates then to a KG equation 
\bq\label{04.31}
-(\hbar^2/m^2c^2)\pp^{\mu}\pp_{\mu}\psi=\psi
\end{equation}
for suitable choice of
integration constant (note $\hbar/mc$ is the Compton wave length).
\\[3mm]\indent
Now in this context or above we refer back to Section 3.1 for example and write
$Q=-(1/2m) (\bx R/R)$ ($\hbar=c=1$ for
convenience here).  Then recall
$\psi=exp(i{\mf S}/m\gl)$ and ${\mf P}_{\mu}=m{\mc
V}_{\mu}=-\pp_{\mu}{\mf S}$ with $i{\mf S} =m\gl log(\psi)$.  Also 
${\mc V}_{\mu}=-(1/m)\pp_{\mu}{\mf S}=i\gl\pp_{\mu}log(\psi)$ with $\psi=Rexp
(iS/m\gl)$ so $log(\psi)=i{\mf S}/m\gl=log(R) +iS/m\gl$, leading to 
\bq\label{04.32}
{\mc V}_{\mu}=
i\gl[\pp_{\mu}log(R)+(i/m\gl)\pp_{\mu}S]=-\frac{1}{m}\pp_{\mu}S+i\gl\pp_{\mu}log(R)
=V_{\mu}+iU_{\mu}
\end{equation}
Then $\bx=\pp^{\mu}\pp_{\mu}$ and $U_{\mu}=\gl\pp_{\mu}log(R)$ leads to 
\bq\label{04.33}
\pp^{\mu}U_{\mu}=\gl\pp^{\mu}\pp_{\mu}log(R)=\gl\bx log(R)
\end{equation}
Further $\pp^{\mu}\pp_{\nu}log(R)=(\pp^{\mu}\pp_{\nu}R/R)-(R_{\nu}R_{\mu}/R^2)$ so
\bq\label{04.34}
\bx log(R)=\pp^{\mu}\pp_{\mu}log(R)=(\bx R/R)-(\sum
R_{\mu}^2/R^2)=
\end{equation}
$$=(\bx R/R)-\sum (\pp_{\mu}R/R)^2=(\bx R/R)-|U|^2$$
for $|U|^2=
\sum U_{\mu}^2$.  Hence via $\gl=1/2m$ for example one has 
\bq\label{04.35}
Q=-(1/2m)(\bx R/R)=-\frac{1}{2m}\left[
|U|^2+\frac{1}{\gl}\bx log(R)\right]=
\end{equation}
$$=-\frac{1}{2m}\left[|U|^2+\frac{1}{\gl}\pp^{\mu}
U_{\mu}\right]=-\frac{1}{2m}|U|^2-\frac{1}{2}div(\vec{U})$$
(cf. Proposition 1.1).
$\hfill\bs$

\subsection{FIELD THEORETIC METHODS}

In trying to imagine particle trajectories of a fractal nature or in a
fractal medium we are tempted to abandon (or rather relax) the particle
idea and switch to quantum fields (QF).  Let the fields sense the bumps
and fractality; if one can think of fields as operator valued
distributions for example then fractal supports for example are quite reasonable. 
There are other reasons of course since the notion of particle in quantum
field theory (QFT) has a rather fuzzy nature anyway. Then of course there
are problems with QFT itself (cf.
\cite{w12}) as well as arguments that there is no first quantization
(except perhaps in the Bohm theory - cf. \cite{n57}).  
Some aspects of particles arising from QF and QFT methods, especially in
a Bohmian spirit are reviewed in \cite{c2,cqm} and here we only briefly indicate one approach
due to Nikoli\'c for bosonic fields (cf. \cite{n57,n61,n62,nk,ni} (cf. also
\cite{bb,h99,h98,hd,hl,hn} for other field aspects of KG).
We refer also to \cite{h97,w12} for interesting philosophical discussion about
particles and localized objects in a QFT. 
Many details are omitted and standard QFT techniques are assumed to be known (see e.g.
\cite{h92}) and we will concentrate here on derivations of KG type equations.
First note that the papers
\cite{n62} are impressive in producing a local operator describing the
particle density current for scalar and spinor fields in an arbitrary
gravitational and electromagnetic background.  This enables one to
describe particles in a local, general covariant, and gauge invariant
manner and this is reviewed in \cite{cqm}. 
We follow here \cite{n61} concerning Bohmian particle trajectories in relativistic
bosonic and fermionic QFT.
First we recall that there is no objection to a Bohmian type
theory for QFT and no contradiction to Bell's theorems etc. (see e.g. \cite{b37,
d42}).  Without discussing philosophical aspects of such a theory we simply construct
one following Nikolic. 
Thus consider first particle trajectories in relativistic QM
and posit a real scalar field
$\phi(x)$ satisfying the Klein-Gordon equation in a Minkowski metric
$\eta_{\mu\nu}=diag(1,-1,-1,-1)$ written as $(\pp_0^2-\na^2+m^2)\phi=0$. 
Let $\psi=\phi^{+}$ with $\psi^*=\phi^{-}$ correspond to positive and negative frequency
parts of $\phi=\phi^{+}+\phi^{-}$.  The particle current is 
$j_{\mu}=i\psi^*\olra{\pp}\!\!\!_{\mu}\psi$ and 
$N=\int d^3xj_0$ is the positive definite number of particles (not the
charge).  This is most easily seen from the plane wave expansion $\phi^{+}(x)=
\int d^3ka(\gk)exp(-ikx)/\sqrt{(2\pi)^32k_0}$ since then $N=\int d^3ka^{\dg}(\gk)
a(\gk)$ (see above and \cite{n56,n62} where it is shown that the particle current and the
decomposition
$\phi=\phi^{+}+\phi^{-}$ make sense even when a background gravitational field or some
other potential is present).  One can write also $j_0=i(\phi^{-}\pi^{+}-\phi^{+}\pi^{-})$
where $\pi=\pi^{+}+\pi^{-}$ is the canonical momentum (cf. \cite{h99}).  Alternatively
$\phi$ may be interpreted not as a field containg an arbitrary number of particles but
rather as a one particle wave function.  Here we note that contrary to a field a wave
function is not an observable and so doing we normalize $\phi$ here so that $N=1$.  The
current $j_{\mu}$ is conserved via $\pp_{\mu}j^{\mu}=0$ which implies that
$N=\int d^3xj_0$ is also conserved, i.e. $dN/dt=0$.  In the causal interpretation one
postulates that the particle has the trajectory determined by $dx^{\mu}/
d\tau=j^{\mu}/2m\psi^*\psi$.  The affine parameter $\tau$ can be eliminated by writing
the trajectory equation as $d{\bf x}/dt={\bf j}(t,{\bf x})/j_0(t,{\bf x})$
where $t=x^0,\,\,{\bf x}=(x^1,x^2,x^3)$ and ${\bf j}=(j^1,j^2,j^3)$.  By writing $\psi
=Rexp(iS)$ where $R,\,S)$ are real one arrives at a Hamilton-Jacobi (HJ) form 
$dx^{\mu}/d\tau=-(1/m)\pp^{mu}S$ and the KG equation is equivalent to
\bq\label{93.1}
\pp^{\mu}(R^2\pp_{\mu}S)=0;\,\,\frac{(\pp^{\mu}S)(\pp_{\mu}S)}{2m}-\frac{m}{2}+Q=0
\end{equation}
Here $Q=-(1/2m)(\pp^{\mu}\pp_{\mu}R/R$ is the quantum potential ($c=\hbar=1$). 
From the HJ form and \eqref{93.1} plus the identity $d/d\tau=(dx^{\mu}/dt)
\pp_{\mu}$ one arrives at the equations of motion $m(d^2x^{\mu}/d\tau^2)
=\pp^{\mu}Q$.  A typical trajectory arising from $d{\bf x}/dt={\bf j}/j_0$ could be
imagined as an S shaped curve in the $t-x$ plane (with $t$ horizontal) and cut with a
vertical line through the middle of the S.  The velocity may be superluminal and may move
backwards in time (at points where $j_0<0$).  There is no paradox with backwards in time
motion since it is physically indistinguishable from a motion forwards with negative
energy.  One introduces a physical number of particles via $N_{phys}=\int
d^3x|j_0|$. Contrary to $N=\int d^3xj_0$ the physical number of particles is not
conserved.  A pair of particles one with positive and the other with negative energy may be
created or annihilated; this resembles the behavior of virtual particles in convential QFT.
\\[3mm]\indent
Now go to relativistic QFT where in the Heisenberg picture the Hermitian field operator
$\hat{\phi}(x)$ satisfies 
\bq\label{93.2a}
(\pp_0^2-\na^2+m^2)\hat{\phi}=J(\hat{\phi})
\end{equation}
where $J$ is a nonlinear function describing the interaction.  In the Schr\"odinger
picture the time evolution is determined via the Schr\"odinger equation (SE)
$H[\phi,-i\gd/\gd\phi]\Psi[\phi,t]=i\pp_t\Psi[\phi,t]$ where $\Psi$ is a
functional with respect to $\phi({\bf x})$ and a function of $t$.  A normalized solution
of this can be expanded as 
$\Psi[\phi,t]=\sum_{-\infty}^{\infty}
\tl{\Psi}_n[\phi,t]$
where the $\tl{\Psi}_n$ are unnormalized n-particle wave
functionals and the analysis proceeds from there (cf. \cite{n61}).  
In the deBroglie-Bohm (dBB) interpretation the field $\phi(x)$ has a causal evolution
determined by
\bq\label{93.6}
(\pp_0^2-\na^2+m^2)\phi(x)=J(\phi(x))-\left(\frac{\gd Q[\phi,t]}{\gd\phi({\bf
x})}\right)_{\phi({\bf x})=\phi(x)};
\end{equation}
$$Q=-\frac{1}{2|\Psi|}\int d^3x\frac{\gd^2
|\Psi|}{\gd\phi^2({\bf x})}$$
where Q is the quantum potential again.  However the n particles attributed to the wave
function $\psi_n$ also have causal trajectories determined by a generalization of
$d{\bf x}/dt={\bf j}/j_0$ as
\bq\label{93.7}
\frac{d{\bf x}_{n,j}}{dt}=\left(\frac{\psi^*_n(x^{(n)})\olra{\na}\!\!\!_j\psi_n
(x^{(n)})}{\psi_n^*(x^{(n)})\olra{\pp}\!\!\!_{t_j}\psi_n(x^{(n)})}
\right)_{t_1=\cdots=t_n=t}
\end{equation}
where the n-particle wave function is
\bq\label{xxx}
\psi_n({\bf x}^{(n)},t)=<0|\hat{\phi}(t,{\bf x}_1)\cdots\hat{\phi}(t,{\bf x}_n)|\Psi>
\end{equation}
These n-particles have well defined trajectories even when the probability (in the
conventional interpretation of QFT) of the experimental detection is equal to zero.  In
the dBB interpretation of QFT we can introduce a new causally evolving parameter 
$e_n[\phi,t]$ defined as 
\bq\label{93.7a}
e_n[\phi,t]=|\tl{\Psi}_n[\phi,t]|^2/\sum_{n'}^
{\infty}|\tl{\Psi}_{n'}[\phi,t]|^2
\end{equation}
The evolution of this parameter is determined by
the evolution of $\phi$ given via \eqref{93.6} and by the solution 
$\Psi=\sum\tl{\Psi}$ of the SE.  This parameter might be interpreted as a probability that
there are n particles in the system at time t if the field is equal (but not measured!) to
be 
$\phi({\bf x})$ at that time.  However in the dBB theory one does not want a stochastic
interpretation.  {\bf Hence assume that $e_n$ is an actual property of the particles
guided by the wave function $\psi_n$ and call it the effectivity of these n particles.}
This is a nonlocal hidden variable attributed to the particles and it is introduced to
provide a deterministic description of the creation and destruction of particles
(see \cite{c2,cqm,n61} for more on this).
\\[3mm]\indent
{\bf REMARK 3.6.}
In \cite{n57} an analogous fermionic theory is developed but it is even more technical and we
refer to
\cite{cqm} for a sketch.$\hfill\bs$
\\[3mm]\indent
{\bf REMARK 3.7.}
In \cite{ni} one addresses the question of statistical transparency.  Thus the
probabilitistic interpretation of the nonrelativistic SE does not work for the
relativistic KG equation ($\pp^{\mu}\pp_{\mu}+m^2)\psi=0$ (where $x=({\bf x},t)$ and
$\hbar=c=1$) since $|\psi|^2$ does not correspond to a probability density.  There is
a conserved current
$j^{\mu}=i\psi^*\olra{\pp}\!\!^{\mu}\psi$ (where
$a\olra{\pp}\!\!^{\mu}b=a\pp^{\mu}b-b\pp^{\mu}a$) but the time component $j^0$ is not positive
definite.  In \cite{n57,n61} the equations that determine the Bohmain trajectories of
relativistic quantum particles described by many particle wave functions were written in a form
requiring a preferred time coordinate.  However a preferred Lorentz frame is not necessary
(cf. \cite{be}) and this is developed in \cite{ni} following \cite{be,n61}.  First note
that as in \cite{be,n61} it appears that particles may be superluminal and the principle of
Lorentz covariance does not forbid superluminal velocities and conversly superluminal velocities
do not lead to causal paradoxes (cf. \cite{be,ni}).  As noted in \cite{be} the Lorentz-covariant
Bohmian interprtation of the many particle KG equation is not statistically transparent. 
This means that the statistical distribution of particle positions cannot be calculated in a
simple way from the wave function alone without the knowledge of particle trajectories.  One
knows that classcal QM is statistically transparent of course and this perhaps helps to
explain why Bohmian mechanics has not attracted more attention.  However statistical
transparency (ST) may not be a fundamental property of nature as suggested by looking at
standard theories (cf. \cite{ni})
The upshot is that since statistical probabilities can be calculated via Bohmian trajectories
that theory is more powerful than other interpretations of general QM and we refer to
\cite{ni} for discussion on this, on the KG equation, and on Lorentz covariance.
$\hfill\bs$

\subsection{DeDONDER-WEYL AND KG}

We go here to a paper \cite{nk} which gives a manifestly covariant canonical method of field
quantization based on the classical DeDonder-Weyl formulation of field theory.  The Bohmian 
formulation is not postulated for intepretational purposes here but derived from purely
technical requirements, namely covariance and consistency with standard QM.  It arises
automatically as a part of the formalism without which the theory cannot be formulated
consistently.  This together with the results of \cite{n57,ni} suggest that it is Bohmian
mechanics that might be the missing bridge between QM and relativity; further it should play
an important role in cosmology (cf. \cite{cqm,i5,k8,i9,i6,i66,i67,ii,ij,rn,t2,t1,ti}).  The
classical covariant canonical DeDonder-Weyl formalism is given first following 
the excellent development in \cite{kp} and
for simplicity one real scalar field in Minkowski spacetime is used.  Thus (classical
formulation) let
$\phi(x)$ be a real scalar field described by 
\bq\label{19.68}
{\mf A}=\int d^4x{\mf L};\,\,{\mf L}=\frac{1}{2}(\pp^{\mu}\phi)(\pp_{\mu}\phi)-V(\phi)
\end{equation}
As usual one has
\bq\label{19.69}
\pi^{\mu}=\frac{\pp{\mf
L}}{\pp(\pp_{\mu}\phi)}=\pp^{\mu}\phi;\,\,\pp_{\mu}\phi=\frac{\pp{\mf
H}}{\pp\pi^{\mu}};\,\,\pp_{\mu}\pi^{\mu}=-\frac{\pp{\mf H}}{\pp\phi}
\end{equation}
where the scalar DeDonder-Weyl (DDW) Hamilonian (not related to the energy density) is given by
the Legendre transform ${\mf H}(\pi^{\mu},\phi)=\pi^{\mu}\pp_{\mu}\phi-{\mf
L}=(1/2)\pi^{\mu}\pi_{\mu}+V$.  The equations \eqref{19.69} are equivalent to the standard
Euler-Lagrange (EL) equations and by introducing the local vector $S^{\mu}(\phi(x),x)$ the
dynamics can also be described by the covariant DDW HJ equation and equations of motion
\bq\label{19.70}
{\mf H}\left(\frac{\pp S^{\ga}}{\pp\phi},\phi\right)+\pp_{\mu}S^{\mu}=0;\,\,\pp^{\mu}\phi
=\pi^{\mu}=\frac{\pp S^{\mu}}{\pp\phi}
\end{equation}
Note here $\pp_{\mu}$ is the partial derivative acting only on the second argument of
$S^{\mu}(\phi(x),x)$; the corresonding total derivative is $d_{\mu}=\pp_{\mu}+(\pp_{\mu}\phi)
(\pp/\pp\phi)$.  Further the first equation in \eqref{19.70} is a single equation for
four quantities $S^{\mu}$ so there is a lot of freedom in finding solutions.  Nevertheless
the theory is equivalent to other formulations of classical field theory.  Now following
\cite{kt} one considers the relation between the covariant HJ equation and the conventional
HJ equation; the latter can be derived from the former as follows.  Using \eqref{19.69},
\eqref{19.70} takes the form
$(1/2)\pp_{\phi}S_{\mu}\pp_{\phi}S^{\mu}+V+\pp_{\mu}S^{\mu}=0$.  Then using the equation of
motion in \eqref{19.70} write the first term as 
\bq\label{19.71}
\frac{1}{2}\frac{\pp S_{\mu}}{\pp\phi}\frac{\pp S^{\mu}}{\pp\phi}=\frac{1}{2}\frac{\pp
S^0}{\pp\phi}\frac{\pp S^0}{\pp\phi}+\frac{1}{2}(\pp_i\phi)(\pp^i\phi)
\end{equation}
Similarly using \eqref{19.70} the last term is
$\pp_{\mu}S^{\mu}=\pp_0S^0+d_iS^i-(\pp_i\phi)(\pp^i\phi)$.  Now 
introduce the quantity ${\bf S}=\int d^3xS^0$ so that $[\pp S^0(\phi(x),x)/\pp\phi(x)]=[\gd{\bf
S}([\phi({\bf x},t)],t)/\gd\phi({\bf x},t)]$ where $\gd/\gd\phi({\bf x},t)\equiv
[\gd/\gd\phi(x)]_{\phi(x)=\phi({\bf x},t)}$ is the space functional derivative.  Putting this
together gives then
\bq\label{19.72}
\int d^3x\left[\frac{1}{2}\left(\frac{\gd {\bf S}}{\gd\phi({\bf x},t)}\right)^2+\frac{1}{2}(\na
\phi)^2+V(\phi)\right]+\pp_t{\bf S}=0
\end{equation}
which corresponds to the standard noncovariant HJ equation.  The time evolution of $\phi({\bf
x},t)$ is given by $\pp_t\phi({\bf x},t)=\gd{\bf S}/\gd\phi({\bf x},t)$ which arises from the
time component of \eqref{19.70}.  Note that in deriving \eqref{19.72} it was necessary to use
the space part of the equations of motion \eqref{19.70} (this does not play an important role
in classical physics but is important here).  
\\[3mm]\indent
Now for the Bohmian formulation look at the
SE $\hat{H}\Psi=i\hbar\pp_t\Psi$ where we write
\bq\label{19.73}
\hat{H}=\int d^3x\left[-\frac{\hbar^2}{2}\left(\frac{\gd}{\gd\phi({\bf
x})}\right)^2+\frac{1}{2}(\na\phi)^2+V(\phi)\right];
\end{equation}
$$\Psi([\phi({\bf x})],t)={\mf
R}([\phi({\bf x})],t)exp[i{\mf S}([\phi({\bf x})],t)/\hbar]$$
Then the complex SE equation is equivalent to two real equations
\bq\label{19.74}
\int d^3x\left[\frac{1}{2}\left(\frac{\gd {\mf S}}{\gd\phi({\bf
x})}\right)^2+\frac{1}{2}(\na\phi)^2+V(\phi)+Q\right]+\pp_t{\mf S}=0;
\end{equation}
$$\int d^3x\left[\frac{\gd{\mf R}}{\gd\phi({\bf x})}\frac{\gd {\mf S}}{\gd\phi({\bf
x})}+J\right]+\pp_t{\mf R}=0;\,\,Q=-\frac{\hbar^2}{2{\mf R}}\frac{\gd^2{\mf R}}{\gd\phi^2({\bf
x})};\,\,J=\frac{{\mf R}}{2}\frac{\gd^2{\mf S}}{\gd\phi^2({\bf x})}$$
The second equation is also equivalent to
\bq\label{19.75}
\pp_t{\mf R}^2+\int d^3x\frac{\gd}{\gd\phi({\bf x})}\left({\mf R}^2\frac{\gd{\mf
S}}{\gd\phi({\bf x})}\right)=0
\end{equation}
and this exhibits the unitarity of the theory because it provides that the norm
$\int [d\phi({\bf x})]^2\Psi^*\Psi=\int[d\phi({\bf x})]{\mf R}^2$ does not depend on
time.  The quantity ${\mf R}^2([\phi({\bf x})],t)$ represents the probability
density for fields to have the configuration $\phi({\bf x})$ at time t.  One can take
\eqref{19.74} as the starting point for quantization of fields (note $exp(i{\mf
S}/\hbar)$ should be single valued).  Equations \eqref{19.74} and \eqref{19.75}
suggest a Bohmian interpretation with deterministic time evolution given via
$\pp_t\phi$.  Remarkably the statistical predictions of this deterministic
interpretation are equivalent to those of the conventional interpretation.  All
quantum uncertainties are a consequence of the ignorance of the actual initial field
configuration $\phi({\bf x},t_0)$.  The main reason for the consistency of this
interpretation is the fact that \eqref{19.75} with $\pp_t\phi$ as above represents
the continuity equation which provides that the statistical distribution
$\rho([\phi({\bf x})],t)$ of field configurations $\phi({\bf x})$ is given by the
quantum distribution $\rho={\mf R}^2$ at any time t, provided that $\rho$ is given
by ${\mf R}^2$ at some initial time.  The initial distribution is arbitrary in
principle but a quantum H theorem explains why the quantum distribution is the most
probable (cf. \cite{v8}).  Comparing \eqref{19.74} with \eqref{19.72} we see that
the quantum field satisfies an equation similar to the classical one, with the
addition of a term resulting from the nonlocal quantum potential Q.  The quantum
equation of motion then turns out to be 
\bq\label{19.76}
\pp^{\mu}\pp_{\mu}\phi+\frac{\pp V(\phi)}{\pp\phi}+\frac{\gd {\mf Q}}{\gd\phi({\bf
x};t)}=0
\end{equation} 
where ${\mf Q}=\int d^3xQ$.  A priori perhaps the main unattractive feature of the
Bohmian formulation appears to be the lack of covariance, i.e. a preferred Lorentz
frame is needed and this can be remedied with the DDW presentation to follow.
\\[3mm]\indent
Thus one wants a quantum substitute for the classical covariant DDW HJ equation
$(1/2)\pp_{\phi}S_{\mu}\pp_{\phi}S^{\mu}+V+\pp_{\mu}S^{\mu}=0$.  Define then the
derivative
\bq\label{19.77}
\frac{dA([\phi],x)}{d\phi(x)}=\int d^4x'\frac{\gd A([\phi],x')}{\gd\phi(x)}
\end{equation}
where $\gd/\gd\phi(x)$ is the spacetime functional derivative (not the space
functional derivative used before in \eqref{19.72}).  In particular if $A([\phi],x)$
is a local functional, i.e. if $A([\phi],x)=A(\phi(x),x)$ then
\bq\label{19.78}
\frac{dA(\phi(x),x)}{d\phi(x)}=\int d^4x'\frac{\gd
A(\phi(x'),x')}{\gd\phi(x)}=\frac{\pp A(\phi(x),x)}{\pp\phi(x)}
\end{equation}
Thus $d/d\phi$ is a generalization of $\pp/\pp\phi$ such that its action on nonlocal
functionals is also well defined.  An example of interest is a functional nonlocal
in space but local in time so that 
\bq\label{19.79}
\frac{\gd A([\phi],x')}{\gd\phi(x)}=\frac{\gd A([\phi],x')}{\gd\phi({\bf x},x^0)}
\gd((x')^0-x^0)\Rightarrow
\end{equation}
$$\Rightarrow\frac{dA([\phi],x)}{d\phi(x)}=\frac{\gd}{\gd\phi({\bf x},x^0)}\int
d^3x'A([\phi],{\bf x}',x^0)$$
Now the first equation in \eqref{19.70} and the equations of motion become
\bq\label{19.80}
\frac{1}{2}\frac{d S_{\mu}}{d\phi}\frac{dS^{\mu}}{d\phi}+V+\pp_{\mu}S^{\mu}=0;\,\,
\pp^{\mu}\phi=\frac{dS^{\mu}}{d\phi}
\end{equation}
which is appropriate for the quantum modification.  Next one proposes a method of
quantization that combines the classical covariant canonical DDW formalism with the
standard specetime asymmetric canonical quantization of fields.  The starting point
is the relation between the noncovariant classical HJ equation \eqref{19.72} and its
quantum analogue \eqref{19.74}.  Suppressing the time dependence of the field in
\eqref{19.72} we see that they differ only in the existence of the Q term in the
quantum case.  This suggests the following quantum analogue of the classical
covariant equation \eqref{19.80}
\bq\label{19.82}
\frac{1}{2}\frac{dS_{\mu}}{d\phi}\frac{dS^{\mu}}{d\phi}+V+Q+\pp_{\mu}S^{\mu}=0
\end{equation}
Here $S^{\mu}=S^{\mu}([\phi],x)$ is a functional of $\phi(x)$ so $S^{\mu}$ at $x$ may
depend on the field $\phi(x')$ at all points $x'$.
One can also allow for time nonlocalities (cf. \cite{ni}).  Thus \eqref{19.83} is
manifestly covariant provided that Q given by \eqref{19.74} can be written in a
covariant form.  The quantum equation \eqref{19.82} must be consistent with the
conventional quantum equation \eqref{19.74}; indeed by using a similar procedure to
that used in showing that \eqref{19.70} implies \eqref{19.72} one can show that
\eqref{19.82} implies \eqref{19.74} provided that some additional conditions are
fulfilled.  First $S^0$ must be local in time so that \eqref{19.79} can be used.
Second $S^i$ must be completely local so that $dS^i/d\phi=\pp S^i/\pp\phi$, which
implies
\bq\label{19.83}
d_iS^i=\pp_iS^i+(\pp_i\phi)\frac{dS^i}{d\phi}
\end{equation}
However just as in the classical case in this procedure it is necessary to use the
space part of the equations of motion \eqref{19.70}.  Therefore these classical
equations of motion must be valid even in the quantum case.  Since we want a
covariant theory in which space and time play equal roles the validity of the space
part of the \eqref{19.70} implies that its time part should also be valid. 
Consequently in the covariant quantum theory based on the DDW formalism one must
require the validity of the second equation in \eqref{19.80}.  This requirement is nothing
but a covariant version of the Bohmian equation of motion written for an arbitrarily nonlocal
$S^{\mu}$ (this clarifies and generalizes results in \cite{kt}).  The next step is
to find a covariant substitute for the second equation in \eqref{19.74}.  One
introduces a vector $R^{\mu}([\phi],x)$ which will generate a preferred foliation of
spacetime such that the vector $R^{\mu}$ is normal to the leaves of the foliation.
Then define 
\bq\label{19.84}
{\mf R}([\phi],\gS)=\int_{\gS}d\gS_{\mu}R^{\mu};\,\,{\mf
S}([\phi],x)=\int_{\gS}d\gS_{\mu}S^{\mu}
\end{equation}
where $\gS$ is a leaf (a 3-dimensional hypersurface) generated by $R^{\mu}$.
Hence the covariant version of $\Psi={\mf R}exp(i{\mf S})$ is $\Psi([\phi],\gS)={\mf
R}([\phi],\gS)exp(i{\mf S}([\phi],\gS)/\hbar)$.  
For $R^{\mu}$ one postulates the equation
\bq\label{19.85}
\frac{dR^{\mu}}{d\phi}\frac{dS^{\mu}}{d\phi}+J+\pp_{\mu}R^{\mu}=0
\end{equation}
In this way a preferred foliation
emerges dynamically as a foliation generated by the solution $R^{\mu}$ of the
equaitons \eqref{19.85} and \eqref{19.82}.  Note that $R^{\mu}$ does not play any
role in classical physics so the existence of a preferred foliation is a purely
quantum effect.  Now the relation between \eqref{19.85} and \eqref{19.74} is
obtained by assuming that nature has chosen a solution of the form
$R^{\mu}=(R^0,0,0,0)$ where $R^0$ is local in time.  Then integrating \eqref{19.85}
over $d^3x$ and assuming again that $S^0$ is local in time one obtains
\eqref{19.74}.  Thus \eqref{19.85} is a covariant substitute for the second equation
in \eqref{19.74}.  It remains to write covariant versions for Q and J and these are
\bq\label{19.86}
Q=-\frac{\hbar^2}{2{\mf R}}\frac{\gd^2{\mf R}}{\gd_{\gS}\phi^2(x)};\,\,J=\frac{\mf
R}{2}\frac{\gd^2{\mf S}}{\gd_{\gS}\phi^2(x)}
\end{equation}
where $\gd/\gd_{\gS}\phi(x)$ is a version of the space functional derivative in which 
$\gS$ is generated by $R^{\mu}$.  Thus \eqref{19.85} and \eqref{19.82} with
\eqref{19.86} represent a covariant substitute for the functional SE equivalent to
\eqref{19.75}.  The covariant Bohmian equations \eqref{19.80} imply a covariant
version of \eqref{19.76}, namely
\bq\label{19.87}
\pp^{\mu}\pp_{\mu}\phi+\frac{\pp V}{\pp\phi}+\frac{dQ}{d\phi}=0
\end{equation}
Since the last term can also be written as $\gd(\int d^4x  Q)/\gd\phi(x)$ the
equation of motion \eqref{19.87} can be obtained by varying the quantum action
\bq\label{19.88}
{\mf A}_Q=\int d^4x{\mf L}_Q=\int d^4x({\mf L}-Q)
\end{equation}
Thus in summary the covariant canonical quantization of fields is given by equations
\eqref{19.80}, \eqref{19.82}, \eqref{19.85}, and \eqref{19.86}.  The conventional
functional SE corresponds to a special class of solutions for which $R^i=0,\,\,
S^i$ are local, while $R^0$ and $S^0$ are local in time.  In \cite{nk} a multifield
generalization is also spelled out, a toy model is considered, and applications to
quantum gravity are treated.  The main result is that a manifestly covariant method
of field quantization based on the DDW formalism is developed which treats space and
time on an equal footing.  Unlike the conventional canonical quantization it is not
formulated in terms of a single complex SE but in terms of two coupled real
equations.  The need for a Bohmian formulation emerges from the requirement that the
covariant method should be consistent with the conventional noncovariant method.  This
suggests that Bohmian mechanics (BM) might be a part of the formalism without which the
covariant quantum theory cannot be formulated consistently.

\section{DIRAC WEYL GEOMETRY}
\renewcommand{\theequation}{4.\arabic{equation}}
\setcounter{equation}{0}

A sketch of Dirac Weyl geometry following \cite{da} was given in \cite{c3} in connection
with deBroglie-Bohm theory in the spirit of the Tehran school (cf. \cite{b10,
b81,m14,m8,
s52,s53,s55,s57,s2,s4,s5,s6,s51,s31,ss3,ss4,s32,s33,s34,ss5,s35,s36,s37,s38,s71,
sss,sf,s1s}).
We go now to \cite{i5,i8,i9,i6,i66,i67,ii,ij,rn} for a very brief discussion of versions
of the Dirac Weyl theory involved in discussing magnetic monopoles, dark matter, quintessence,
matter creation, etc. (see \cite{cqm} for more in this direction).  Thus
go to \cite{i8} where in particular an integrable Weyl-Dirac theory is
developed (the book \cite{i5} is a lovely exposition but the work in \cite{i8} is
somwhat newer).   Note, as remarked in \cite{ml} (where twistors are used), the integrable
Weyl-Dirac geometry is desirable in order that the natural frequency of an atom at a point
should not depend on the whole world line of the atom.  The first paper in \cite{i8} is
designed to investigate the integrable Weyl-Dirac (Int-W-D) geometry and its ability to create
massive matter.  For example in this theory a spherically symmetric static geometric
formation can be spatially confined and an exterior observer will recognize it as a
massive entity.  This may be either a fundamental particle or a cosmic black hole both
confined by a Schwarzschild surface. Here we only summarize some basic features in order
to establish notation, etc. and sketch the preliminary theory (referring to \cite{cqm}
and the work of Israelit and Rosen for many examples).
Thus in the Weyl geometry one has a metric
$g_{\mu\nu}=g_{\nu\mu}$ and a length connection vector $w_{\mu}$ along with an idea of 
Weyl gauge transformation (WGT) 
\bq\label{45.1}
g_{\mu\nu}\to
\tl{g}_{\mu\nu}=e^{2\gl}g_{\mu\nu};\,\,g^{\mu\nu}\to\tl{g}^{\mu\nu}=e^{-2\gl}g^{\mu\nu}
\end{equation}
where $\gl(x^{\mu})$ is an arbitrary differerentiable function.  One is interested in
covariant quantities satisfying $\psi\to\tl{\psi}=exp(n\gl)\psi$ where
the Weyl power n is described via $\pi(\psi)=n,\,\,\pi(g_{\mu\nu})=2,$ and 
$\pi(g^{\mu\nu})=-2$.  If $n=0$ the quantity $\psi$ is said to be gauge invariant
(in-invariant).  Under parallel displacement one has length changes and for a vector
\bq\label{045.1}
(i)\,\,dB^{\mu}=-B^{\gs}\gG^{\mu}_{\gs\nu}dx^{\nu};\,\,(ii)\,\,
B=(B^{\mu}B^{\nu}g_{\mu\nu})^{1/2};\,\,(iii)\,\,
dB=Bw_{\nu}dx^{\nu}
\end{equation} 
(note $\pi(B)=1$).  In order to have agreement between (i) and (iii) one requires
\bq\label{45.2}
\gG^{\gl}_{\mu\nu}=\left\{\begin{array}{c}
\gl\\
\mu\,\,\,\nu
\end{array}\right\}+g_{\mu\nu}w^{\gl}-\gd_{\nu}^{\gl}w_{\mu}-\gd_{\mu}^{\gl}w_{\nu}
\end{equation}
where $\left\{\begin{array}{c}\
\gl\\
\mu\,\,\nu
\end{array}\right\}$ is the Christoffel symbol based on $g_{\mu\nu}$.
In order for (iii) to hold in any gauge one must have the WGT $w_{\mu}\to
\tl{w}_{\mu}=w_{\mu}+\pp_{\mu}\gl$ and if the vector $B^{\mu}$ is transported by parallel
displacement around an infinitesimal closed parallelogram one finds 
\bq\label{45.3}
\gD B^{\gl}=B^{\gs}K^{\gl}_{\gs\mu\nu}dx^{\mu}\gd x^{\nu};\,\,
\gD B=BW_{\mu\nu}dx^{\mu}\gd x^{\nu};
\end{equation}
$$K^{\gl}_
{\gs\mu\nu}=-\gG^{\gl}_{\,\,\gs\mu,\nu}+\gG^{\gl}_{\,\,\gs\nu,\mu}-\gG^{\ga}_{\,\,\gs\mu}\gG^
{\gl}_{\ga\nu}+\gG^{\ga}_{\gs\nu}\gG^{\gl}_{\ga\mu}$$
is the curvature tensor formed from
\eqref{45.2} and $W_{\mu\nu}=w_{\mu,\nu}-w_{\nu,\mu}$.  Equations for the WGT
$w_{mu}\to\tl{w}_{\mu}$ and the definition of $W_{\mu\nu}$ led Weyl to identify
$w_{\mu}$ with the potential vector and $W_{\mu\nu}$ with the EM field strength; he used a
variational principle $\gd I=0$ with
$I=\int L\sqrt{-g}d^4x$ with L built up from $K^{\gl}_{\gs\mu\nu}$ and $W_{\mu\nu}$.  In order
to have an action invariant under both coordinate transformations and WGT he was forced
to use $R^2$ (R the Riemannian curvature scalar) and this led to the gravitational field.
\\[3mm]\indent
Dirac revised this with a scalar field $\gb(x^{\nu})$ which under WGT changes via
$\gb\to\tl{\gb}=e^{-\gl}\gb$ (i.e. $\pi(\gb)=-1$).  His in-invariant action
integral is then ($f_{,\mu}\equiv\pp_{\mu}f$)
\bq\label{45.5}
I=\int [W^{\gl\gs}W_{\gl\gs}-\gb^2R+\gb^2(k-6)w^{\gs}w_{\gs}+2(k-6)\gb w^{\gs}\gb_{,\gs}+
\end{equation}
$$+k\gb_{,\ul{\gs}}\gb_{,\gs}+2\gL\gb^4+L_M]\sqrt{-g}d^4x$$
Here k is a parameter, $\gL$ is the cosmological constant, $L_M$ is the Lagrangian
density of matter, and an underlined index is to be raised with $g^{\mu\nu}$. 
Now according to \eqref{45.3} this is a nonintegrable geometry but there may be situations
when geometric vector fields are ruled out by physical constraints (e.g. the FRW universe). 
In this case one can preserve the WD character of the spacetime by assuming that $w_{\nu}$ is
the gradient of a scalar function $w$ so that $w_{\nu}=w_{,\nu}=\pp_{\nu}w$.
One has then $W_{\mu\nu}=0$ and from \eqref{45.3} results $\gD B=0$ yielding an
integrable spacetime (Int-W-D spacetime).  To develop this begin with \eqref{45.5} but with
$w_{\nu}$ given by $w_{\nu}=\pp_{\nu}w$ so the first term in \eqref{45.5} vanishes.  The
parameter k is not fixed and the dynamical variables are $g_{\mu\nu},\,w,$ and $\gb$. 
Further it is assumed that $L_M$ depends on $(g_{\mu\nu},\,w,\,\gb)$. For convenience write
\bq\label{45.6} 
b_{\mu}=(log(\gb))_{,\mu}=\gb_{,\mu}/\gb
\end{equation}
and use a modified Weyl connection
vector
$W_{\mu}=w_{\mu}+b_{\mu}$ which is a gauge invariant gradient vector.  Write
also $k-6=16\pi \gk$ and varying w in \eqref{45.5} one gets a field equation
\bq\label{45.7}
2(\gk\gb^2W^{\nu})_{;\nu}=S
\end{equation}
where the semicolon denotes covariant
differentiation with the Christoffel symbols and S is the Weylian scalar charge given by 
$16\pi S=\gd L_M/\gd w$.  Varying $g_{\mu\nu}$ one gets also
\bq\label{45.8}
G_{\mu}^{\nu}=-8\pi\frac{T^{\nu}_{\mu}}{\gb^2}+16\pi\gk\left(W^{\nu}W_{\mu}-\frac{1}{2}\gd_{\mu}^
{\nu}W^{\gs}W_{\gs}\right)+
\end{equation}
$$+2(\gd_{\mu}^{\nu}b^{\gs}_{;\gs}-b^{\nu}_{;\mu})+2b^{\nu}b_{\mu}+\gd_{\mu}^{\nu}b^{\gs}_{\gs}
-\gd_{\mu}^{\nu}\gb^2\gL$$
where $G_{\mu}^{\nu}$ represents the Einstein tensor and the EM density tensor of ordinary
matter is 
\bq\label{45.9}
8\pi\sqrt{-g}T^{\mu\nu}=\gd(\sqrt{-g}L_M)/\gd g_{\mu\nu}
\end{equation}
Finally the variation with respect to $\gb$ gives an equation for the $\gb$ field 
\bq\label{45.10}
R+k(b^{\gs}_{;\gs}+b^{\gs}b_{\gs})=16\pi\gk(w^{\gs}w_{\gs}-w^{\gs}_{;\gs})
+4\gb^2\gL+8\pi\gb^{-1}B
\end{equation}
Note in \eqref{45.10} R is the Riemannian curvature scalar and the
Dirac charge B is a conjugate of the Dirac gauge function $\gb$, namely
$16\pi B=\gd L_M/\gd \gb$.
\\[3mm]\indent
By a simple procedure (cf. \cite{da}) one can derive conservation laws; consider e.g.
$I_M=\int L_M\sqrt{-g}d^4x$.  This is an in-invariant so its variation
due to coordinate transformation or WGT vanishes.  Making use of 
$16\pi S=\gd L_M/\gd w$, \eqref{45.9}, and $16\pi B=\gd L_M/\gd \gb$ 
one can write 
\bq\label{45.11}
\gd I_M=8\pi\int (T^{\mu\nu}\gd g_{\mu\nu}+2S\gd
w+2B\gd\gb)\sqrt{-g}d^4x
\end{equation}
Via $x^{\mu}\to \tl{x}^{\mu}=x^{\mu}+\eta^{\mu}$
for an arbitrary infinitesimal vector $\eta^{\mu}$ one can write 
\bq\label{45.12}
\gd
g_{\mu\nu}=g_{\gl\nu}\eta^{\gl}_{;\mu}+g_{\mu\gl}\eta^{\gl}_{;\nu};\,\,\gd
w=w_{,\nu}\eta^{\nu};\,\,
\gd\gb=\gb_{,\nu}\eta^{\nu}
\end{equation}
Taking into account $x^{\mu}\to\tl{x}^{\mu}$ we have $\gd I_M=0$ and
making use of \eqref{45.12} one gets from \eqref{45.11} the energy momentum relations
\bq\label{45.13}
T^{\gl}_{\mu;\gl}-Sw_{\mu}-\gb B b_{\mu}=0
\end{equation}
Further considering a WGT
with infinitesimal $\gl(x^{\mu})$ one has from \eqref{45.11} the equation 
$S+T-\gb B=0$ with $T=T^{\gs}_{\gs}$.  One can contract \eqref{45.8} and make use of
\eqref{45.7} and $S+T=\gb B$ giving again \eqref{45.10}, so that \eqref{45.10} is a corollary
rather than an independent equation and one is free to choose the gauge function $\gb$ in
accordance with the gauge covariant nature of the theory.  Going back to the energy-momentum
relations one inserts $S+T=\gb B$ into \eqref{45.13} to get
$T^{\gl}_{\mu;\gl}-Tb_{\mu}=SW_{\mu}$.  Now go back to the field equation
\eqref{45.8} and introduce the EM density tensor of the
$W_{\mu}$ field
\bq\label{45.14}
8\pi\gT^{\mu\nu}=16\pi\gk\gb^2[(1/2)g^{\mu\nu}W^{\gl}W_{\gl}-W^{\mu}W^{\nu}]
\end{equation}
Making use of \eqref{45.7} one can prove $\gT^{\gl}_{\mu;\nu}-\gT
b_{\mu}=-SW_{\mu}$ and using $T^{\gl}_{\mu;\gl}-TB_{\mu}=SW_{\mu}$ one has an equation for the
joint energy momentum density 
\bq\label{45.15}
(T^{\gl}_{\mu}+\gT^{\gl}_{\mu})_{;\gl}
-(T+\gT)b_{\mu}=0
\end{equation}
One can derive now the equation of motion of a test particle 
(following \cite{rn}).  Consider matter consisting of identical particles with rest mass m
and Weyl scalar charge $q_s$, being in the stage of a pressureless gas so that the EM density
tensor can be written $T^{\mu\nu}=\rho U^{\mu}U^{\nu}$ where $U^{\mu}$ is the
4-velocity and the scalar mass density $\rho$ is given by $\rho=m\rho_n$ with
$\rho_n$ the particle density.  Taking into account the conservation of particle number one
obtains from $T^{\gl}_{\mu;\gl}-Tb_{\mu}=SW_{\mu}$ the equation of motion
\bq\label{45.16}
\frac{d U^{\mu}}{ds}+\left\{\begin{array}{c}
\mu\\
\gl\,\,\,\gs
\end{array}\right\}U^{\gl}U^{\gs}=\left(b_{\gl}+\frac{q_s}{m}W_{\gl}\right)(g^{\mu\gl}
-U^{\mu}U^{\gl})
\end{equation}
In the Einstein gauge ($\gb=1$) we are then left with 
\bq\label{45.17}
\frac{dU^{\mu}}{ds}+\left\{\begin{array}{c}
\mu\\
\gl\,\,\gs
\end{array}\right\}U^{\gl}U^{\gs}=\frac{q_s}{m}w_{\gl}(g^{\mu\gl}-U^{\mu}U^{\gl})
\end{equation}
\indent
This gives a sketch of a powerful framework capable of treating many problems involving
``mattter" and geometry.  Connections to Section 2 are obvious and we  have supplied
earlier additional relations to fluctuations via Fisher information and quantum geometry
(cf. also \cite{c1,c2,c3,c4,cqm}).
Many 
cosmological questions of great interest including dark matter, quintessence, etc. are also
treated in \cite{i5,i8,i9,i6,i66,i67,ii,ij} and one can
speculate about the original universe from many points of view. 
The inroads into cosmology
here seem to be an inevitable consequence of the presence of  Weyl-Dirac theory in dealing with
quantum fluctuations via the quantum potential.

\section{REMARKS ON QUANTUM GEOMETRY}
\renewcommand{\theequation}{5.\arabic{equation}}
\setcounter{equation}{0}

We gave a ``hands on" sketch of quantum geometry in \cite{c4} and refer to
\cite{a13,a14,a51,a12,a18,b18,b34,c27,c28,c30,c33,c42,g7,h6,h3,h55,i10,k50,m30,m10,
p6,p7,t2,t1,ti,w7} for background and extensive theory.  Here we follow
\cite{c4,c27,c28,c30,c33,c42} and briefly extract from \cite{c4}.
Roughly the idea is that for H the
Hilbert space of a quantum system there is a natural quantum geometry on the
projective space $P(H)$ with inner product
$<\phi|\psi>=(1/2\hbar)g(\phi,\psi)+(i/2\hbar)\go(\phi,\psi)$ where
$g(\phi,\psi) =2\hbar\Re(\phi|\psi)$ is the natural Fubini-Study (FS) metric and
$g(\phi,\psi)=\go(\phi,J\psi)\,\,(J^2=-1)$.  On the other hand the FS metric is
proportional to the Fisher information metric of the form 
$Cos^{-1}|<\phi|\psi>|$.  Moreover (in 1-D for simplicity) ${\mf
F}\propto
\int
\rho Qdx$ is a functional form of Fisher information where Q is the quantum potential
and
$\rho=|\psi|^2$.  Finally one recalls that in a Riemannian flat spacetime (with
quantum matter and Weyl geometry) the Weyl-Ricci scalar curvature is proportional to
Q.  Thus assume H is separable with a complete orthonormal system
$\{u_n\}$ and for any $\psi\in H$ denote by $[\psi]$ the ray generated by $\psi$ while
$\eta_n=(u_n|\psi)$.  Define for $k\in{\bf N}$
\bq\label{05.1}
U_k=\{[\psi]\in P(H);\,\,\eta_k\ne 0\};\,\,\phi_k:\,U_k\to\ell^2({\bf C}):\,\,
\phi_k([\psi])=\left(\frac{\eta_1}{\eta_k},\cdots,\frac{\eta_{k-1}}{\eta_k},
\frac{\eta_{k+1}}{\eta_k},\cdots\right)
\end{equation}
where $\ell^2({\bf C})$ denotes square summable functions.  Evidently $P(H)=\cup_kU_k$
and $\phi_k\ci\phi_j^{-1}$ is biholomorphic.  It is easily shown that the structure is
independent of the choice of complete orthonormal system.  The coordinaes for $[\psi]$
relative to the chart $(U_k,\phi_k)$ are $\{z^k_n\}$ given via 
$z^k_n=(\eta_n/\eta_k)$ for $n<k$ and $z^k_n=(\eta_{n+1}/\eta_k)$ for $n\geq k$.
To convert this to a real manifold one can use $z^k_n=(1/\sqrt{2})(x^k_n+iy^k_n)$
with
\bq\label{05.2}
\frac{\pp}{\pp z^k_n}=\frac{1}{\sqrt{2}}\left(\frac{\pp}{\pp x^k_n}+i\frac{\pp}{\pp
y^k_n}\right);\,\,\frac{\pp}{\pp \bar{z}^k_n}=\frac{1}{\sqrt{2}}\left(\frac{\pp}{\pp
x_n^k}-i\frac{\pp}{\pp y^k_n}\right)
\end{equation}
etc.  Instead of nondegeneracy as a criterion for a symplectic form inducing a bundle
isomorphism between $TM$ and $T^*M$ one assumes here that a symplectic form on M is a
closed 2-form which induces at each point $p\in M$ a toplinear isomorphism between the
tangent and cotangent spaces at p.  For $P(H)$ one can do more than simply exhibit
such a natural symplectic form; in fact one shows that $P(H)$ is a K\"ahler manifold
(meaning that the fundamental 2-form is closed).  Thus one can choose a Hermitian
metric ${\mf G}=\sum g^k_{mn}dz^k_m\ot d\bar{z}^k_n$ with
\bq\label{05.3}
g^k_{mn}=(1+\sum_i z^k_i\bar{z}^k_i)^{-1}\gd_{mn}-(1+\sum_1
z_i^k\bar{z}_i^k)^{-2}\bar{z}^k_mz^k_n
\end{equation}
relative to the chart $U_k,\phi_k)$.  The fundamental 2-form of the metric ${\mf G}$ is
$\go=i\sum_{m,n}g^k_{mn}dz_m^k\wg d\bar{z}_n^k$ and to show that this is
closed note that $\go=i\pp\bar{\pp}f$ where locally $f=log(1+\sum
z_i^k\bar{z}^k_i)$ (the local K\"ahler function).  Note here that $\pp+\bar{\pp}=d$
and $d^2=0$ implies $\pp^2=\bar{\pp}^2=0$ so $d\go=0$ and thus $P(H)$ is a K manifold
(cf. \cite{m10} for K geometry).
\\[3mm]\indent
Now $P(H)$ is the set of one dimensional subspaces or rays of H; for every 
$x\in H/\{0\},\,\,[x]$ is the ray through $x$.  If H is the Hilbert space of a
Schr\"odinger quantum system then H represents the pure states of the system and 
$P(H)$ can be regarded as the state manifold (when provided with the differentiable
structure below).  One defines the K structure as follows.  On $P(H)$ one has an 
atlas $\{(V_h,b_h,C_h)\}$ where $h\in H$ with $\|h\|=1$.  Here $(V_h,b_h,C_h)$ is the
chart with domain $V_h$ and local model the complex Hilbert space $C_h$ where
\bq\label{05.4}
V_h=\{[x]\in P(H);\,(h|x)\ne 0\};\,\,C_h=[h]^{\perp};\,\,b_h:\,V_h\to C_h;\,\,[x]\to
b_h([x])=\frac{x}{(h|x)}-h
\end{equation}
This produces a analytic manifold structure on $P(H)$.  As a real manifold one uses
an atlas $\{(V_h,R\ci b_h,RC_h)\}$ where e.g. $RC_h$ is the realification of $C_h$
(the real Hilbert space with ${\bf R}$ instead of ${\bf C}$ as scalar field)
and $R:\,C_h\to RC_h;\,v\to Rv$ is the canonical bijection (note $Rv\ne\Re v$).
Now consider the form of the K metric relative to a chart $(V_h,R\ci b_h,RC_h)$ where
the metric $g$ is a smooth section of $L_2(TP(H),{\bf R})$ with local expression
$g^h:\,RC_h\to L_2(RC_h,{\bf R});\,Rz\mpt g^h_{Rz}$ where 
\bq\label{05.5}
g^h_{Rz}(Rv,Rw)=2\nu\Re\left(\frac{(v|w)}{1+\|z\|^2}-
\frac{(v|z)(z|w)}{(1+\|z\|^2)^2}\right)
\end{equation}
The fundamental form $\go$ is a section of $L_2(TP(H),{\bf R})$, i.e. 
$\go^h:\,RC_h\to L_2(RC_h,{\bf R});\,\,Rz\to \go^h_{Rz}$, given via
\bq\label{05.6}
\go^h_{Rz}(Rv,Rw)=2\nu\Im\left(\frac{(v|w)}{1+\|z\|^2}-
\frac{(v|z)(z|w)}{(1+\|z\|^2)^2}\right)
\end{equation}
\indent
Then using e.g. \eqref{05.5} for the FS metric in $P(H)$ consider
a Schr\"odinger Hilbert space with dynamics determined via ${\bf
R}\times P(H)\to P(H):\,(t,[x])\mpt [exp(-(i/\hbar)tH)x]$ where H is a (typically
unbounded) self adjoint operator in H.  One thinks then of K\"ahler isomorphisms of
$P(H)$ (i.e. smooth diffeomorphisms $\Phi:\,P(H)\to P(H)$ with the properties
$\Phi^*J=J$ and $\Phi^*g=g$). If U is any unitary operator on H the map $[x]\mpt [Ux]$
is a K isomorphism of $P(H)$.  Conversely (cf. \cite{c3}) any K isomorphism of $P(H)$
is induced by a unitary operator U (unique up to phase factor).  Further for every self
adjoint operator A in H (possibly unbounded) the family of maps $(\Phi_t)_{t\in{\bf
R}}$ given via 
$\Phi_t:\,[x]\to [exp(-itA)x]$ is a continuous one parameter group of K isomorphisms of 
$P(H)$ and vice versa (every K isomorphism of $P(H)$ is induced by a self adjoint
operator where boundedness of A corresponds to smoothness of the $\Phi_t$).  Thus in the
present framework the dynamics of QM is described by a continuous one parameter group
of K isomorphisms, which automatically are symplectic isomorphisms (for the structure
defined by the fundamental form) and one has a Hamiltonian system.  Next ideally one can
suppose that every self adjoint operator represents an observable and these will be
shown to be in $1-1$ correspondence with the real K functions.
\\[3mm]\indent
One defines a (Riemann) metric (statistical distance) on the space of
probability distributions ${\mc P}$ of the form 
\bq\label{05.7}
ds^2_{PD}=
\sum(dp_j^2/p_j)=\sum p_j(dlog(p_j))^2
\end{equation}
Here one thinks of the central limit theorem
and a distance between probability distributions distinguished via a Gaussian
$exp[-(N/2)(\tl{p}_j-p_j)^2/p_j]$ for two nearby distributions (involving N samples
with probabilities $p_j,\,\tl{p}_j$).  This can be generalized to quantum mechanical
pure states via (note $\psi\sim \sqrt{p}exp(i\phi)$ in a generic manner)
\bq\label{05.8}
|\psi>=\sum\sqrt{p_j}e^{i\phi_j}|j>;\,\,|\tl{\psi}>=|\psi>+|d\psi>=\sum \sqrt{p_j+dp_j}
e^{i(\phi_j+d\phi_j)}|j>
\end{equation}
Normalization requires $\Re(<\psi|d\psi>)=-1/2<d\psi|d\psi>$ and measurements described
by the one dimensional projectors $|j><j|$ can distinguish $|\psi>$ and $|\tl{\psi}>$
according to the metric \eqref{05.7}.  The maximum (for optimal disatinguishability) is
given by the Hilbert space angle $cos^{-1}(|<\tl{\psi}|\psi>|)$ and the
corresponding line element ($PS\sim$ pure state)
\bq\label{05.9}
\frac{1}{4}ds^2_{PS}=[cos^{-1}(|<\tl{\psi}|\psi>|)]^2\sim 1-|<\tl{\psi}|\psi>|^2=
<d\psi_{\perp}|d\psi_{\perp}>\sim
\end{equation}
$$\sim \frac{1}{4}\sum\frac{dp_j^2}{p_j}+\left[\sum p_jd\phi_j^2-(\sum
p_jd\phi_j)^2\right]$$
(called the Fubini-Study (FS) metric) is the natural metric on the manifold of Hilbert
space rays.  Here
\bq\label{05.10}
|d\psi_{\perp}>=|d\psi>-|\psi><\psi|d\psi>
\end{equation}
is the
projection of
$|d\psi>$ orthogonal to $|\psi>$.  Note that if $cos^{-1}(|<\tl{\psi}|\psi>|=\gt$ then 
$cos(\gt)=|<\tl{\psi}|\psi>|$ and $cos^2(\gt)=|<\tl{\psi}|\psi>|^2=1-Sin^2(\gt)\sim
1-\gt^2$ for small $\gt$.  Hence $\gt^2\sim 1-cos^2(\gt)=1-|<\tl{\psi}|\psi>|^2$.
The term in square brackets (the
variance of phase changes) is nonnegative and an appropriate choice of basis makes it
zero.  In
\cite{b18} one then goes on to discuss distance formulas in terms of density operators
and Fisher information but we omit this here.  Generally as in \cite{w7} one observes
that the angle in Hilbert space is the only Riemannian metric on the set of rays which
is invariant uder unitary transformations.
In any event $ds^2=\sum(dp_i^2/p_i),\,\,\sum p_i=1$ is referred to as the
Fisher metric (cf. \cite{m10}).  Note in terms of $dp_i=\tl{p}_i-p_i$ one can write
$d\sqrt{p}=(1/2)dp/\sqrt{p}$ with $(d\sqrt{p})^2=(1/4)(dp^2/p)$ and think of 
$\sum(d\sqrt{p_i})$ as a metric.  Alternatively from $cos^{-1}(|<\tl{\psi}|\psi>|$ one
obtains $ds_{12}=cos^{-1}(\sum \sqrt{p_{1i}}\sqrt{p_{2i}})$ as a distance
in ${\mc P}$.  Note from \eqref{05.10} that 
$ds_{12}^2=4cos^{-1}|<\psi_1|\psi_2>|\sim 4(1-|(\psi_1|\psi_2)|^2\equiv
4(<d\psi|d\psi>-<d\psi|\psi><\psi|d\psi>)$ begins to look like a FS metric before passing
to projective coordinates.  In this direction we observe from \cite{m10} that the FS
metric can be expressed also via 
\bq\label{05.11}
\pp\bar{\pp}log(|z|^2)=\phi=\frac{1}{|z|^2}\sum dz_i\wg d\bar{z}_i-\frac{1}{|z|^4}
\left(\sum \bar{z}_idz_i\right)\wg\left(\sum z_id\bar{z}_i\right)
\end{equation}
so for $v\sim \sum v_i\pp_i+\bar{v}_i\bar{\pp}_i$ and $w\sim \sum
w_i\pp_i+\bar{w}_i\bar{\pp}_i$ and $|z|^2=1$ one has 
$\phi(v,w)=(v|w)-(v|z)(z|w)$. 
\\[3mm]\indent
Now recall the material on fisher information in Section 1.2 and the results on
the SE in Weyl space in Section 1.1 to confirm the connection of quantum geometry
as above to Fisher information, Weyl curvature, and the quantum potential. 
Several features arise which deserve emphasis (cf. also \cite{c16})
\begin{itemize}
\item
Philosophically the wave function seems to be inevitably associated to a cloud or
ensemble (cf. Remarks 2.1 and 3.2).  This provides meaning for
$psi=Rexp(iS/\hbar)$ with $R=\sqrt{\rho}$ and $\rho=\psi^*\psi$ representing a
probability density.  Connections to hydrodynamics, diffusion, and kinetic theory
are then natural and meaningful.
\item
From the ensemble point of view or by statistical derivations as in Section 1.1
one sees that spacetime geometry should also be conceived of in statistical terms
at the quantum level.  This is also connected with the relativistic theory and the
quantum potential (in various forms) is exhibited as a fundamental ingredient of
both QM and spacetime geometry.
\item
Bohmian type mechanics plays a fundamental role in providing unification of all
these ideas.  Similarly fractal considerations as in Nottale's scale relativity
lead to important formulas consistent with the pictures obtained via Bohmian
mechanics and the quantum potential.
\item
Quantum geometry in a projective Hilbert space is connected to all these matters
as indicated in this section.
\end{itemize}

\end{document}